\documentclass[autogobble,dvipsnames,acmsmall,screen]{acmart}

\makeatletter

\@ifclasswith{acmart}{review}{
  \newcommand{\reviewmode}{}
}{
}

\usepackage{amsmath}
\usepackage{xparse}

\usepackage{mathpartir}
\usepackage[capitalise]{cleveref}
\usepackage{booktabs}   %
\usepackage{csquotes}
\usepackage{fancyvrb}
\usepackage{stmaryrd}
\usepackage{adjustbox}
\usepackage{xspace}
\usepackage{relsize}
\usepackage{comment}

\usepackage{savesym}
\savesymbol{program}
\usepackage{amsmath}
\usepackage{semantic}
\usepackage{mdframed}

\numberwithin{desiderata}{section}
\crefname{desiderata}{Desideratum}{Desiderata}

\DeclareMathAlphabet{\mymathbb}{U}{bbold}{m}{n}

\newcommand\doubleplus{+\kern-1.3ex+\kern0.8ex}

\NewDocumentCommand\defineMeta{mmmm}{%
    \expandafter\DeclareDocumentCommand\csname#1#2\endcsname{ E{^_}{{}{}} }{%
    #3{#4^{#3{##1}}_{#3{##2}}}%
    }%
}
\NewDocumentCommand\sgMeta{mG{#1}}{
    \defineMeta{g}{#1}{\gradualstyle}{#2}
    \defineMeta{s}{#1}{\staticstyle}{#2}
}

\NewDocumentCommand\defineStaticFun{mG{#1}}{%
    \expandafter\NewDocumentCommand\csname#1\endcsname{od()}{%
    \staticstyle{\mathtt{#2}\ {\IfNoValueF{##1}{##1}}{\IfNoValueF{##2}{(##2)}} }
    }%
}

\sgMeta{x}
\sgMeta{y}
\sgMeta{z}
\sgMeta{C}
\sgMeta{D}
\sgMeta{N}
\sgMeta{M}
\sgMeta{ff}{f}
\sgMeta{pf}{p}
\sgMeta{P}
\sgMeta{el}{\overline{e}}
\sgMeta{X}
\sgMeta{Y}
\sgMeta{Z}
\sgMeta{t}
\sgMeta{T}
\sgMeta{s}
\sgMeta{S}
\sgMeta{m}
\sgMeta{n}
\sgMeta{u}
\sgMeta{U}
\sgMeta{v}
\sgMeta{V}
\sgMeta{A}
\sgMeta{B}
\sgMeta{J}{\text{J}}
\sgMeta{Nat}{\text{Nat}}
\sgMeta{Int}{\text{Int}}
\sgMeta{Bool}{\text{Bool}}
\sgMeta{Zero}{0}
\sgMeta{NatElim}{\text{NatElim}}
\sgMeta{VecElim}{\text{EqElim}}
\sgMeta{EqElim}{\text{VecElim}}
\sgMeta{Refl}{\text{Refl}}
\sgMeta{subst}{\text{subst}}

\sgMeta{G}{\Gamma}
\sgMeta{ep}{\varepsilon}
\sgMeta{times}{\times}
\sgMeta{lambda}{\lambda}
\sgMeta{Unit}{\mathbf{1}}
\sgMeta{unit}{()}

\sgMeta{spine}{\seq{e}}

\mathlig{||}{\bnfalt}
\mathlig{::=}{\bnfdef}
\mathlig{+::=}{\bnfadd}
\mathlig{-::=}{\bnfsub}
\mathlig{|>}{\vartriangleright}
\mathlig{<|}{\vartriangleleft}
\mathlig{<=}{\Leftarrow}
\mathlig{=>}{\Rightarrow}
\mathlig{>->}{\rightarrowtail}
\mathlig{|->}{\mapsto}
\mathlig{|=>}{\Mapsto}
\mathlig{==}{\equiv}
\mathlig{=-=}{\cong}
\mathlig{**}{\times}
\mathlig{<<}{\langle}
\mathlig{>>}{\rangle}

\newcommand{\set}[1]{{\{ #1 \}}}

\newcommand{\mynote}[3][red]
    {{\color{#1} \fbox{\bfseries\sffamily\scriptsize#2}
    {\small$\blacktriangleright$\textsf{\emph{#3}}$\blacktriangleleft$}}~}

\newcommand{\je}[1]{\mynote{JE}{#1}}
\usepackage{mathtools}

\newcommand{\gradualstyle}[1]{{\ensuremath{\color{RoyalBlue}\mathrm{{#1}}}}}
\newcommand{\staticstyle}[1]{{\ensuremath{\color{BrickRed}\mathsf{{#1}}}}}

\newcommand{\staticdesc}{\staticstyle{\textsf{red sans-serif font}}\xspace}
\newcommand{\gradualdesc}{\gradualstyle{\text{blue serif font}}\xspace}

\newcommand{\g}[1]{\gradualstyle{#1}}
\newcommand{\s}[1]{\staticstyle{#1}}
\newcommand{\ix}[1]{\color{black}{\mathit{#1}}}

\NewDocumentCommand{\sType}{m}{{\staticstyle{\mathbf{Type}_{#1}}}}
\NewDocumentCommand{\gType}{m}{{\gradualstyle{\mathbf{Type}_{#1}}}}

\newcommand{\stepstostar}{\longrightarrow^{*}}
\newcommand{\stepstoplus}{\longrightarrow^{+}}
\newcommand{\etasteps}{\longrightarrow^{*}_{\eta}}
\newcommand{\stepsto}{\longrightarrow}

\newcommand{\bnfalt}{\mathbf{\,\,\mid\,\,}}
\newcommand{\bnfdef}{\mathbf{\ \Coloneqq\ }}
\newcommand{\defbnf}{\bnfdef} %
\newcommand{\bnfadd}{{\mathbf{\ +\!\!\Coloneqq\ }}}
\newcommand{\bnfsub}{{\bf - : : =}}

\makeatletter
\newcommand{\mathboxed}[1]{\text{\fboxsep=.2em\fbox{\m@th$\displaystyle#1$}}}
\makeatother

\newcommand{\ie}{i.e.\ }
\newcommand{\eg}{e.g\ }

\newcommand{\errsym}{\mho}
\newcommand\qm{\gradualstyle{\normalfont{\textbf{?}}}}

\newcommand\err{\gradualstyle{\errsym}}

\newcommand{\Type}{\mathbf{Type}}

\newcommand{\rrule}{\textsc}

\newcommand{\seq}[1]{\overline{#1}}

\newcommand{\ala}{\`a la }

\newcommand{\bN}{\mymathbb{N}}

\newcommand{\bOne}{\mymathbb{1}}
\newcommand{\bB}{\mymathbb{B}}

\definecolor{lightgray}{gray}{0.90}
\newcommand{\Gbox}[1]{\colorbox{lightgray}{$#1$}}

\newcommand{\MLTT}{Martin-L\"{o}f Type Theory}

\newmdenv[
usetwoside=false,
topline=false,
bottomline=false,
rightline=false,
leftmargin=0.2in,
linewidth=0.75pt,
skipabove=\topsep,
skipbelow=\topsep,
nobreak=false
]{leftrule}

\newcommand{\case}[2]{
  \noindent $\blacktriangleright$ \textbf{Case} \text{#1} \textbf{:}
  {
    
    \begin{leftrule}
      
      #2
    \end{leftrule}
  }
  \noindent \ignorespaces
}

\newcommand{\lcase}[2]{
  \noindent $\blacktriangleright$ \textbf{Case} \text{#1} \textbf{:} #2
}

\newcommand{\mcase}[2]{\case{$#1$}{#2}}

\newcommand{\rlcase}[2]{\lcase{\rrule{#1}}{#2}}
\newcommand{\mlcase}[2]{\lcase{$#1$}{#2}}

\newcommand{\rcase}[2]{\case{\rrule{#1}}{#2}}

\newsavebox{\saveboxedarray}
\newenvironment{inferbox}[0]
{\begin{minipage}{\textwidth}\mprset{center}\begin{mathpar}}
    {\end{mathpar}\end{minipage}}

 \newenvironment{boxedarray}[1]
 {\begin{lrbox}{\saveboxedarray}\begin{math}\begin{array}{#1}}
                                              {\end{array}\end{math}\end{lrbox}\fbox{\usebox{\saveboxedarray}}}
                                        \reservestyle{\command}{\textsf}

\newcommand{\myparagrapht}[1]{\noindent{\bf #1}}
\newcommand{\myparagraph}[1]{\vspace{0.5em}\myparagrapht{#1.}}

\usepackage{tikz-cd}

\usepackage{letltxmacro}

\LetLtxMacro{\oldfigure}{\figure}
\LetLtxMacro{\oldendfigure}{\endfigure}

\LetLtxMacro{\oldcaption}{\caption}

\renewenvironment{figure}
{\oldfigure}
{\vspace{-2ex}\oldendfigure}

\renewcommand{\caption}[1]{\vspace{-0.25ex}\vspace{-\baselineskip}\oldcaption{#1}}

\usepackage{proof-at-the-end}

\pgfkeys{/prAtEnd/apxproof/.style={
restate,no link to proof,proof at the end,no link to theorem
  }
}

\pgfkeys{/prAtEnd/apxlem/.style={
all end
  }
}

\pgfkeys{/prAtEnd/global custom defaults/.style={
}
}

\include{proofSetting}

\usepackage{iftex}
\ifPDFTeX
\usepackage[utf8]{inputenc}
\usepackage[inline]{enumitem}
\DeclareUnicodeCharacter{2029}{}
\fi

\usepackage{underscore}

\usepackage[nomain,
            order=word,
            hyperfirst=false,
            acronym,
            shortcuts,
            nonumberlist]{glossaries}

\NewDocumentCommand{\gSucc}{g}{\gradualstyle{\gtt{S}\IfValueT{#1}{\ #1}}}

\NewDocumentCommand{\gNil}{g}{\gradualstyle{{Nil}\IfValueT{#1}{\ #1}}}
\NewDocumentCommand{\sNil}{g}{\staticstyle{{Nil}\IfValueT{#1}{\ #1}}}

\NewDocumentCommand{\gCons}{gggg}{\gradualstyle{\gtt{Cons}\IfValueT{#1}{\ #1}\ \IfValueT{#2}{\ #2}\ \IfValueT{#3}{\ #3}\ \IfValueT{#4}{\ #4}}}

\NewDocumentCommand{\gVec}{gg}{\gradualstyle{{Vec}\IfValueT{#1}{\ #1}\ \IfValueT{#2}{\ #2}}}
\NewDocumentCommand{\sVec}{gg}{\staticstyle{{Vec}\IfValueT{#1}{\ #1}\ \IfValueT{#2}{\ #2}}}

\NewDocumentCommand{\conv}{gg}{\staticstyle{\stt{conv}\IfValueT{#1}{\ #1}\IfValueT{#2}{\ #2}}}

\usepackage{lstautogobble}
\usepackage{listings}

\lstdefinelanguage{Agda}
  {morekeywords={let,in,as,data,record,import,infix,infixl,infixr,module,open,renaming,using,where,\_},
   morekeywords=[2]{Set,Set1,Set2,Type},
  literate=*
     {?}{$\mathrm{\qm}$}1
     {->}{$\mathrm{\to}$}2,
   otherkeywords={=,:,(,),\{,\},:=,;},
   sensitive=true,
   morecomment=[n]{\{-}{-\}},
   morecomment=[l]{--},
   morestring=[b]{"}
  }[keywords,comments,strings]

\lstnewenvironment{gradualCode}{\lstset{language=Agda,
    basicstyle=\rmfamily\color{RoyalBlue},
    columns=flexible, mathescape=true}}{}

\lstnewenvironment{agdaCode}{\lstset{language=Agda,
    basicstyle=\rmfamily\color{Black},
    columns=flexible, mathescape=true}}{}

\lstnewenvironment{staticCode}{\lstset{language=Agda,
    basicstyle=\sffamily\color{BrickRed},
        literate=*
     {?}{$\mathrm{\qm}$}1
     {Type}{$\mathcal{U}$ }1
     {->}{$\mathrm{\to}$}2,
     columns=flexible, mathescape=true}}{}

\newcommand{\Code}{\staticstyle{\mathbf{Code}}}
\newcommand{\El}{\staticstyle{\mathbf{El}}}

\newcommand{\T}[1]{ {\color{black}\mathcal{T}\llbracket} \g{ #1 } {\color{black}\rrbracket}}

\newcommand{\V}[1]{{\color{black}\mathcal{V}\llbracket} \s{ #1 } {\color{black}\rrbracket} }

\usepackage{sansmath}

\newcommand{\lang}{{\sf{GrInd}}\xspace}

\newcommand{\surflang}{{\sf{GrIf}}\xspace}
\newcommand{\surflangDesc}{\underline{Gr}adual language with \underline{If}-else branching\xspace}
\newcommand{\surfelang}{{\sf{GrEq}}\xspace}
\newcommand{\surfilang}{\lang}

\newcommand{\slang}{{\sf{StIf}}\xspace}
\newcommand{\slangDesc}{\underline{St}atic language with \underline{If}-else branching\xspace}
\newcommand{\silang}{{\sf{StInd}}\xspace}
\newcommand{\selang}{{\sf{StEq}}\xspace}

\newcommand{\clang}{{\sf{CastIf}}\xspace}
\newcommand{\clangDesc}{\underline{Cast} calculus with \underline{If}-else branching\xspace}
\newcommand{\cilang}{{\sf{CastInd}}\xspace}
\newcommand{\celang}{{\sf{CastEq}}\xspace}

\newcommand{\tlang}{{\sf{StIR}}\xspace}

\newcommand{\cast}[2]{\g{\langle #2 <= #1 \rangle}}
\newcommand{\castnog}[2]{\g{\langle} #2 \g{<=} #1 \g{\rangle}}
\newcommand{\castenv}[2]{{\langle} #2 {<=} #1 {\rangle}}
\newcommand{\rmeet}[3]{\g{#1 \sqcap_{#3} #2 }}

\newcommand{\grefl}[3]{\g{refl_{#1 |- #2 =-= #3}}}

\newcommand{\elabsto}{\rightarrowtriangle}
\newcommand{\echeck}[3]{#1 \elabsto #3 <= #2}
\newcommand{\esynth}[3]{#1 \elabsto #3 => #2}
\newcommand{\etype}[3]{#1 \elabsto #3 : \gType{}_{=>\g{#2}} }
\newcommand{\redsto}{\leadsto}

\newcommand{\squbc}{\sqsubseteq^{C}}
\newcommand{\squbG}{\sqsubseteq^{\Gamma}}
\newcommand{\squbstar}{\sqsubseteq^{C*}}
\newcommand{\sqeqc}{\sqsupseteq\sqsubseteq^{C}}
\newcommand{\sqeqstar}{\sqsupseteq\sqsubseteq^{C*}}
\newcommand{\squbs}{\sqsubseteq^{S}}

\newcommand{\qml}{\qm_{\gType{\ell}}}
\newcommand{\qmat}[1]{\g{\qm_{@ #1}}}
\newcommand{\attagl}[2]{\g{\langle#1\rangle_{\g\ell}#2}}

\crefformat{section}{\S#2#1#3}
\crefformat{subsection}{\S#2#1#3}
\crefformat{subsubsection}{\S#2#1#3}
\crefrangeformat{section}{\S\S#3#1#4 to~#5#2#6}
\crefmultiformat{section}{\S\S#2#1#3}{ and~#2#1#3}{, #2#1#3}{ and~#2#1#3}

\newcommand{\tagOf}{\mathit{tagOf}}
\newcommand{\typeTagOf}{\mathit{typeTagOf}}
\newcommand{\levelOne}{{\mathcal{L}^1}}

\newcommand\psynth{{=>}^{*}}

\newcommand{\pc}{\g{=-=}}

\acmYear{2021}

\ifdef{\reviewmode}{
  \settopmatter{printfolios=true,printccs=false,printacmref=false}
  \setcopyright{none}
  \renewcommand\footnotetextcopyrightpermission[1]{}
  \pagestyle{plain}
  \raggedbottom
}{
\setcopyright{acmcopyright}
\copyrightyear{2021}
\acmDOI{10.1145/1122445.1122456}
}

\AtBeginDocument{
  \providecommand\BibTeX{{
    \normalfont B\kern-0.5em{\scshape i\kern-0.25em b}\kern-0.8em\TeX}}}

\acmJournal{PACMPL}
\acmVolume{37}
\acmNumber{4}
\acmArticle{111}
\acmMonth{7}

\begin{document}

\title{Approximate Normalization and Eager Equality Checking for Gradual Inductive Families}

  \author{Joseph Eremondi}

  \affiliation{
    \department{Department of Computer Science}              
    \institution{University of British Columbia}            
    \country{Canada}                    
  }
  \email{{jeremond@cs.ubc.ca}}

  \author{Ronald Garcia}
  \affiliation{
    \institution{University of British Columbia}            
    \country{Canada}                    
  }
  \email{rxg@cs.ubc.ca}

    \author{\'{E}ric Tanter}
    \affiliation{
      \department{Computer Science Department (DCC)}              
      \institution{University of Chile}            
      \country{Chile}                    
    }
    \email{etanter@dcc.uchile.cl}

\renewcommand{\shortauthors}{Eremondi, Garcia, and Tanter}

\begin{abstract}

Harnessing the power of dependently typed languages can be difficult.
Programmers must manually construct proofs
to produce well-typed programs,
which is not an easy task.
In particular,
migrating
code to these languages is challenging.
\textit{Gradual typing} can make dependently-typed languages easier
to use by mixing static and dynamic checking in a principled way.
With gradual types, programmers can incrementally migrate code to a dependently typed language.

However, adding gradual types to dependent types creates a new challenge:
mixing decidable type-checking and incremental migration in a full-featured language is a precarious balance.
Programmers expect type-checking to terminate, but dependent type-checkers
evaluate terms at compile time,
which is problematic because gradual types can introduce non-termination
into an otherwise terminating language.
Steps taken to mitigate this non-termination must not jeopardize the smooth transitions
between dynamic and static.

We present a gradual dependently-typed language that supports inductive type families,
has decidable type-checking, and provably supports smooth migration between static and dynamic,
as codified by the refined criteria for gradual typing proposed by Siek~et~al.~(2015).
Like Eremondi~et~al.~(2019), we use approximate normalization for terminating compile-time evaluation.
Unlike Eremondi~et~al., our normalization does not require comparison of variables, allowing us to show termination with a syntactic model
that accommodates inductive types.
Moreover, we design a novel a technique for tracking constraints on type indices,
so that dynamic constraint violations signal run-time errors eagerly.
To facilitate these checks, we define an
algebraic notion of gradual precision, axiomatizing certain semantic
properties of gradual terms.

\end{abstract}

\ifdef{\reviewmode}{

}{

}

\maketitle

\section{Introduction}
\label{sec:intro}
Dependently typed languages are an expressive tool for
specifying and verifying program properties.

Terms and types may depend on each other, allowing programs, specifications, and proofs of correctness
to be expressed using the same language.

Programmers have the power of a
higher-order logic available to formulate and prove program invariants.

However, when a programmer uses dependent types, they have replaced one problem with two problems:
they must write a correct algorithm and also prove that it is
correct.
Well-typed programs in dependently-typed
languages are guaranteed to meet their specifications, but producing those well typed programs is challenging.
Migrating programs from non-dependently typed languages is particularly tricky.
In practice, dependent types are currently used by a few experts, who often have close ties to the developers of the tools they are  using~\citep{qedAtLarge}.

In this paper, we contribute to the recent line of work integrating \textit{gradual types} with dependent types,
with the aim of making it easier to learn
and migrate code to dependently-typed languages such as Idris~\citep{idrisPaper}.
Gradual types
facilitate the migration of code between dynamic and static type disciplines.
Programmers may assign any expression the
\textit{unknown type} $\qm$.
For gradual type-checking, type equality is replaced with type \textit{consistency}~\citep{ANDERSON200353,gradualTypeInitial}:
equality up to occurrences of $\qm$, allowing terms with precise and imprecise
types to be used together.

This mixing is principled:
gradual languages provide \textit{safety},
ensuring that type inconsistencies at run-time cause an error to be raised,
but  no other unsafe behavior is possible.
Fully gradual languages satisfy the \textit{gradual guarantees}~\citep{refinedCriteria}, which ensure that programs can be smoothly
migrated between the static and dynamic paradigms, while knowing that any errors
indicate a fundamental inconsistency in the program's type structure.

\myparagraph{Example: Migrating Quicksort}

To demonstrate the utility of gradual types for moving code to a dependently typed language, we describe the example due to \citet{Eremondi:2019:ANG:3352468.3341692}:
migrating quicksort.
Consider the classic list type and a standard quicksort implementation.
\\
\begin{minipage}{0.33\textwidth}
  \scriptsize
\begin{flalign*}
  &\s{\textbf{data}\ List : (X : \sType{}) -> \sType{}\ }
  \s{\textbf{where}}\\
  &\quad  \s{Nil : List\ X } \\
  & \quad \s{  Cons : X -> List\ X -> List\ X}
\end{flalign*}
\end{minipage}
\hfill\vline\hfill
\begin{minipage}{0.58\textwidth}
  \scriptsize
\begin{flalign*}
  &\s{sort : List\ Float -> List\ Float}\\
  &\s{sort\ Nil = Nil}\\
  &\s{sort\ (Cons\ h\ t) =}
  \s{(sort\ (filter\ (<\ h)\ t)) }
    \s{\ \doubleplus (Cons\ h\ (sort\ (filter\ (>\ h)\ t))) }
\end{flalign*}
\end{minipage}

In place of lists, dependent types allow programmer to use
\textit{size-indexed vectors}:

\begin{minipage}{0.35\textwidth}
\scriptsize
\begin{flalign*}
  &\s{\textbf{data}\ Vec : (X : \sType{}) -> (n : \bN) -> \sType{}\ \textbf{where}}\\
  &\qquad  \s{Nil : Vec\ X\ 0 } \\ 
  &\qquad \s{  Cons : X -> Vec\ X\ n -> Vec\ X\ (1 + n)}
\end{flalign*}
\end{minipage}
\hfill\vline\hfill
\begin{minipage}{0.55\textwidth}
  \scriptsize
\begin{flalign*}
  &\s{sort : (n : \bN) -> Vec\ Float\ n -> Vec\ Float\ n} \\
  &\s{sort\ 0\ Nil = Nil} \\
  &\s{sort\ (1 + n') (Cons\ h\ t) =} \\
   &\qquad \s{(sort\ {\color{black}\fbox{???}}\ (filter\ (<\ h)\ t)) }
   \s{\ \doubleplus (Cons\ h\ (sort\ {\color{black}\fbox{???}}\ (filter\ (>\ h)\ t))) }
 \end{flalign*}
\end{minipage}

Vectors are an \textit{indexed inductive type family}:
for each element type $\s{X}$ and number $\s{n}$, $\s{Vec\ X\ n}$
is a distinct type, inhabited only by vectors with $n$ elements.
They have the same constructors as lists,
but the constructors' return types specify a value for the index $\s{n}$
that reflects the length of the result.
This lets programmers specify types that rule out erroneous inputs,
such as $\s{head : Vec\ X\ (1 + n) -> X}$, 
which only accepts non-empty vectors.

Although lists and vectors have the same constructors, we encounter two problems
when adapting $\s{sort}$ to vectors.
The first problem is length indexing. The dependent $\s{sort}$ needs an argument
for the vector's length. What length should we give to the recursive calls?
We do not statically know how many elements are smaller
than $\s{h}$.
Also, the result of $\s{sort}$ should have length $\s{n} = $\s{1 + n'},
but it has length $\s{n_{1} + 1 + n_{2}}$, where $\s{n_{1}}$ and $\s{n_{2}}$
are the lengths of $\s{filter\ (<\ h)\ t}$
and $\s{filter\ (>\ h)\ t}$.
Even the type of $\s{filter}$ is non-obvious:
it must return type $\s{(n : \bN) \times Vec\ \bN\ n}$
\ie a dependent pair with a length and a vector of that length.

The second problem is that the recursive calls
do not take structurally smaller arguments, so termination is not obvious
to the type-checker.
While the programmer could use a function for well-founded induction on
$\s{\bN}$, they must prove
that $\s{n_{1}}$ and $\s{n_{2}}$ are smaller than $\s{n}$.

Now, development is blocked until the above problems are solved.
Both are solvable: the programmer can write a proof that
$\s{n \equiv n_{1} + 1 + n_{2}}$
and that $\s{n_{1} < n \equiv true}$ and $\s{n_{2} < n \equiv true}$.
However, they cannot run or test their $\s{sort}$ function
until those proofs are provided!

\textbf{Gradual Types to the Rescue:}

Gradual dependent types allow the programmer to run and test their sort function
before writing the missing proofs, by letting
$\qm$ stand in for any term.
We write gradual terms using \gradualdesc,
to distinguish them from the \staticdesc of static terms.
Instead of returning a dependent pair,
we write a gradual $\g{filter}$, so that
$\g{filter\ (<\ h)\ \gt}$ and $\g{filter (>\ h)\ \gt}$ check at type
$\g{Vec\ Float\ \qm}$.
We then use $\qm$ as the length argument for the recursive calls to $\g{sort}$,
so concatenating gives length $\g{\qm + 1 + \qm}$, which is \textit{consistent with} $\g{n}$.
We use $\qm$ to represent imprecision in the type index,
but preserve the knowledge that $\g{sort}$ produces a $\g{Vec}$.

Likewise, gradual types help with the termination problem by allowing $\qm$ as a \textit{proof}.
The programmer can show termination by using a
$\g{wfInd}$ library function for well-founded induction with the type
$\g{(P : \bN \!->\! \gType{} ) ->
   ((n : \bN) \!->\! (rec : (m : \bN) \!->\! m \!<\! n \!==\! true \!->\! P\ m) \!->\! P\ n )
  \!->\! (n : \bN) -> P\ n} $.\\
Each call to $\g{rec}$ still requires a proof that $\g{m < n \equiv true}$,
but with gradual dependent types, the programmer can use $\qm$ as a placeholder for that proof.

The resulting gradual code is:

{\small
\begin{flalign*}
  \scriptsize
  &\g{sort : (n : \bN) -> Vec\ Float\ n -> Vec\ Float\ n} \\
  &\g{sort = wfInd\ (\lambda n -> Vec\ Float\ n -> Vec\ Float\ n )\ sort' \textbf{ where }}  \\
  &\quad \g{sort'\ 0\ rec\ Nil = Nil } \\
  &\quad \g{sort'\ (1 + n')\ rec\ (Cons\ h\ t) = }
    \g{(rec\ \qm\ \qm\ (filter\ (<\ h)\ t)) }
   \g{\ \doubleplus (Cons\ h\ (rec\ \qm\ \qm\ (filter\ (>\ h)\ t))) }
 \end{flalign*}
 }

Gradual types are useful not only because of what they allow statically, but what they prevent dynamically,
and gradual dependent types are no exception.
By keeping run-time information about types and indices,
we ensure that unsafe operations are never performed.
The keen-eyed reader will notice that \textbf{our implementation of} $\g{sort}$ \textbf{is flawed}:  the list is partitioned
using $\g{<}$ and $\g{>}$, so duplicate occurrences of $\g{h}$ are removed.
When $\g{sort}$ runs on a list containing duplicates, \textit{a run-time error is raised},
notifying the programmer that the actual length does not match the one specified in the type.
A major contribution of this work is the ability to raise such errors when a list
is constructed, while still satisfying important metatheoretic properties about the language.

\myparagraph{Related Work}
The flexibility of gradual typing
means that some terms do not terminate~\citep{gradualTypeInitial}.
Since types depend on terms, dependent type-checking evaluates some terms at compile-time,
meaning that
non-termination must be managed in order for type-checking to remain decidable.
We must also account for the possibility of dynamic type errors \textit{during type-checking}.
        \lang builds on two existing languages with different approaches to this challenge: GDTL~\citep{Eremondi:2019:ANG:3352468.3341692} and GCIC~\citep{bertrand:gcic}.

  \textbf{GDTL} is a foundational (but minimal) calculus
for gradual dependent types.
Like GDTL, \lang gives dynamic semantics to programs with $\qm$ standing in for arbitrary types or terms.
Both have an exact but possibly-diverging run-time semantics, but use \textit{approximate normalization} to provide decidable
normalization during type-checking by producing $\qm$ when termination is not guaranteed.
Unlike GDTL,
we show that all features of \lang except non-positive types can be simulated in a strongly-normalizing
language. Thus, we approximate in fewer places,
allowing more types to be checked statically.

        While GDTL provides dynamic checking of dependent properties like our $\g{sort}$ example,
        the proof of decidable type-checking does not extend to inductive types.
        GDTL's approximate normalization was based on hereditary substitution,
        but extending this to inductive types is an open problem,
        even without gradual types.

        Instead we use a syntactic model to show termination for \lang,
        which is possible because
        \lang defers the comparison
        of bound variables until they have concrete values. We can then prove termination
        by simulating approximate normalization in a strongly-normalizing language,
        which was not possible for GDTL.

  \textbf{GCIC} provides a gradual version of the
Calculus of Inductive Constructions (CIC). It
integrates gradual types with inductive families in three languages,
giving a choice from any two of decidable type-checking, fully embedding CIC,
        and graduality~\citep{10.1145/3236768}, a strengthened version of the gradual guarantees.
\lang achieves decidable type-checking, fully embeds CIC, and satisfies the gradual guarantees.
        As with GCIC, we define \lang's semantics using a gradual cast calculus,
        and prove decidable type-checking using simulation by CIC.
        However, GCIC's version of gradual propositional equality is inhabited for every pair of terms,
        regardless of whether they are consistent, and safety is ensured using casts when eliminating an equality proof. So for $\g{sort}$, no error is raised when wrong-sized output is given, only when that output is accessed in an unsafe way.
        Our version of gradual propositional equality keeps information through the run of a program,
        allowing more eager reporting of when gradual values do not match their types.

\myparagraph{Our Contribution to Gradual Dependent Types}

This work contributes to the theoretical foundations of gradual dependent types, helping fulfill the goals
and address the challenges above.
We present \lang, a gradual dependently-typed language with inductive families.
Accompanying \lang are
proofs of important metatheoretic properties: decidable type-checking, and
the criteria for principled gradual typing, namely
safety and the gradual guarantees~\citep{refinedCriteria}.

Our work has applied and theoretical contributions.
\textbf{Our applied contribution} is a gradual dependently typed language with
run-time checking of equality constraints.
With static dependent types, the propositional equality $\s{t_{1} ==_{T} t_{2}}$ type denotes
proofs that $\s{t_{1}}$ and $\s{t_{2}}$ are equal.
In \lang, the programmer can prove $\g{t_{1} \pc_{T} t_{2}}$ for any consistent $\g{t_{1}}$ and $\g{t_{2}}$.
We call this \textit{propositional consistency},
since it relaxes equality to accommodate imprecision.
\textbf{Our insight} is to implement propositional consistency between two terms by tracking
\textit{evidence} that the two terms are consistent, akin to the AGT~\citep{agt}
or threesomes~\citep{Siek:2009:TWB:1570506.1570511} approach.
This evidence is represented by a third term at least as precise as both terms.
New evidence can be composed with
existing evidence.
We propose that, for variables, this composition should be deferred until
they have concrete values, since this allows equalities to be checked lazily as the program executes,
as well as allowing us to prove important properties of \lang.

\textbf{Our theoretical contribution} is a proof that \lang satisfies both the gradual guarantees
and decidable type-checking, which have never previously been achieved for a dependently-typed language with inductive types.
\textbf{Our insight} is that these metatheoretic properties can be proven in a language
with propositional consistency by carefully
defining what it means for one term to be more precise than another.
We establish decidable type-checking via the termination of approximate normalization
using a \textit{syntactic model} in the style of \citet{10.1145/3018610.3018620}. That is, we provide a
type-preserving translation to a strongly normalizing language, and
establish a simulation between the dynamic semantics. This allows the semantics of \lang to be understood through
the well-known Calculus of Inductive Constructions  (CIC)~\citep{paulinmohring:hal-01094195}, and provides a path
towards implementation.
To establish that the gradual guarantees are compatible with evolving evidence of propositional consistency, we define \textit{algebraic precision}.
Algebraic precision extends the usual syntactic precision relation
with laws about how evidence of consistency is composed in \lang, and with laws relating casts between different types.
We show that normalizing \lang terms preserves algebraic precision, and that reducing
algebraic precision of a type preserves its consistency with other types.

\begin{figure}

\begin{tikzcd}
  \textbf{\parbox{5em}{\underline{St}atic\\ language}}                                                                                              &  & \textbf{\parbox{5em}{\underline{Gr}adual\\
      language}}                                                                     &  & \textbf{\parbox{5em}{\underline{Cast}\\ calculus}}                                                                             &  & \textbf{\parbox{5em}{\underline{St}atic\\ target}} \\[-20pt]
\slang \arrow[rr, "\text{Embedded in}"] \arrow[dd, "\text{Add equality} "', dotted]                                  &  & \surflang \arrow[rr, "\text{Elaborates to}"] \arrow[dd, "\text{Add equality} "', dotted]     &  & \clang \arrow[rrdd, "\text{Translates to}"] \arrow[dd, "{\text{Add equality, composition} }"', dotted] &  &                        \\[-14pt]
                                                                                                                   &  &                                                                                            &  &                                                                                                 &  &                        \\
\selang \arrow[rr, "\text{Embedded in}"] \arrow[dd, "\text{Add inductives}"', dotted]                              &  & \surfelang \arrow[rr, "\text{Elaborates to}"] \arrow[dd, "\text{Add inductives}"', dotted] &  & \celang \arrow[rr, "\text{Translates to}"] \arrow[dd, "\text{Add inductives}"', dotted]         &  & \text{\small\parbox{5em}{\tlang \\ (\slang with \\ induction-recursion)}}                 \\[-14pt]
                                                                                                                   &  &                                                                                            &  &                                                                                                 &  &                        \\[-14pt]
\text{\parbox{5em}{\silang\\ \small (Non-Cumulative,\\ Predicative CIC) }} \arrow[rr, "\text{Embedded in}"]  &  & \surfilang \arrow[rr, "\text{Elaborates to}"]                                              &  & \cilang \arrow[rruu, "\text{Translates to}"]                                                    &  &
\end{tikzcd}
\vspace{-2em}

\caption{Relationships between calculi in this paper}
\label{fig:languages}
\end{figure}
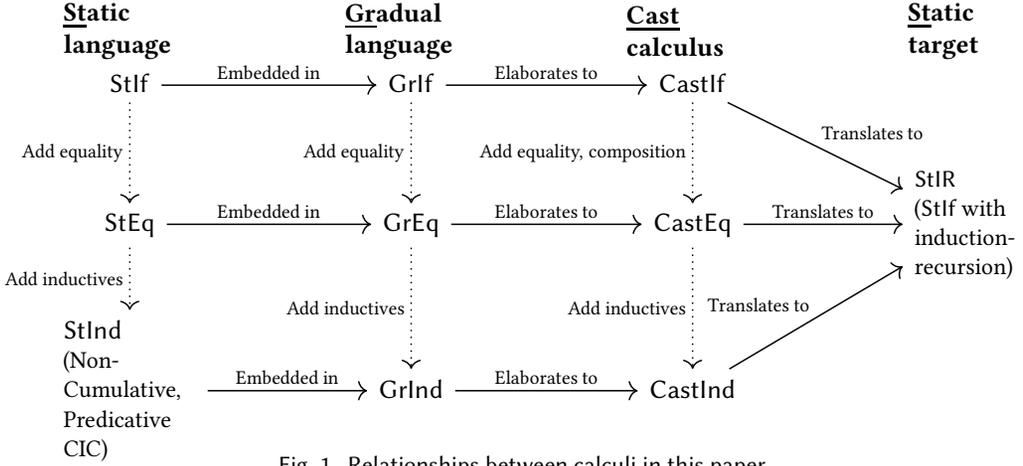

\myparagraph{Road map and Calculi}
To mitigate the complexity of our development,
we present several languages,
progressing from a small language to one with inductive types.
\cref{fig:languages} shows
the relationship between these languages.

First we gradualize \slang (\cref{sec:static}), a subset of CIC with dependent functions,
a universe hierarchy, booleans, and conditionals. We add $\qm$ to obtain \surflang, our first gradual
surface language (\cref{sec:surface-syntax}), which replaces type equality with type consistency.
We define the dynamic semantics of \surflang terms by \textit{elaboration} to \clang,
which extends \surflang with explicit casts (\cref{sec:cast,sec:elab}).
We then develop \selang, \surfelang, and \celang,
which introduce propositional equality to \slang, \surflang and \clang respectively
(\cref{sec:equality}).
For the static and gradual languages, we need only the usual $\g{refl}$ constructor,
but for \celang we must keep evidence of propositional equality
between two terms.
To compose evidence, we also add an explicit composition operator
in \celang.

We use propositional consistency to add support for inductive types (\cref{sec:inductives}):
\selang, \surfelang and \celang become \silang, \surfilang, and \cilang.

In addition to defining these languages,
we develop their metatheory.
We first show the results for \surflang and \clang,
but they are easily extended to
\surfelang/\celang and \lang/\cilang.
We present a precision relation in \cref{sec:precision} and
prove the gradual guarantees.

Likewise, we show the termination of approximate normalization by a syntactic model (\cref{sec:translation}),
where we translate the cast calculi to \tlang,
which is \silang (i.e. predicative, non-cumulative CIC) extended with inductive-recursive types~\citep{DYBJER20031}.
Operationally, the reader may think of \tlang as being Agda~\citep{agdaPaper}. A mechanization of
the translation in Agda is ongoing.

\section{\slang: The Static Starting Language}
\label{sec:static}

We begin with \slang, a \slangDesc, from which we progress to \lang.
\slang is a small subset of CIC~\citep{paulinmohring:hal-01094195}
with dependent functions,
booleans with non-dependent conditionals,
and a hierarchy of type universes. For simplicity, we omit cumulativity and subtyping.
We use \slang to introduce the structure of a gradual surface language (\cref{sec:surface-syntax})
and cast calculus (\cref{sec:cast}).

\subsection{Syntax and Typing}

\slang includes universes, variables, type ascriptions, function types,
functions, applications, booleans, and if-else branching.
The dependent function type $\s{(x : T_{1}) -> T_{2}}$ is like a normal function type,
but the variable $\sx$ is bound, and may occur in $\s{T_{2}}$
to refer to the function argument value.
Normal and neutral terms are a mutually-recursively defined
subset of terms: normal terms cannot be further evaluated,
and neutral terms are those normal terms that are immediately blocked because they eliminate a variable.
\begin{displaymath}
  \small
  \begin{array}{rcll}
    \textsc{Term} \ni\s{t},\sT & \defbnf & \sType{\ell} \bnfalt \sx \bnfalt \s{t :: T}
    \bnfalt \s{(x : T_1) -> T_2} \bnfalt \s{\lambda x \ldotp t} \bnfalt \s{t_1\ t_2}
    \bnfalt \s{\bB} \bnfalt \s{true} \bnfalt \s{false} \bnfalt \s{if\ t_1\ t_2\ t_3}\\
\textsc{Normal} \ni \sv, \sV & \defbnf
& \sN
                                       \bnfalt \sType{\ell}
                                       \bnfalt \s{v :: V}
\bnfalt \s{(\sx : \s{V_1}) -> \s{V_2}}
                                         \bnfalt \s{\lambda \sx \ldotp \sv}
                                       \bnfalt \s{\bB} \bnfalt \s{true} \bnfalt \s{false}

\\
\textsc{Neutral} \ni \sN     & \defbnf
& \sx
                                         \bnfalt \sN\ \sv
                                         \bnfalt \s{if\ N\ v_2\ v_3}
  \end{array}
\end{displaymath}

\begin{figure}
  
  \begin{boxedarray}{@{}l@{}}
    \boxed{\Gamma |- \st => \sT\quad \textit{(Synthesis)}}\qquad
  \boxed{\Gamma |- \st <= \sV\quad \textit{(Checking)}} \\

    \begin{inferbox}
      \inferrule[SVar]{\Gamma(\sx) = \sT}{\Gamma |- \sx => \sT}

      \inferrule[SType]{ }{\Gamma |- \sType{\ell} => \sType{1+\ell}}

      \inferrule[SBool]{ }{\Gamma |- \s{\bB} => \sType{0}}

      \inferrule[STF]{
        \s{b} \in \set{\s{true}, \s{false}}
      }{\Gamma |- \s{b} => \s{\bB}}

      \inferrule[SAscr]{
        \Gamma |- \sT => \sType{\ell}\\\\
        \Gamma |- \sT \etasteps \sV : \sType{\ell}\\\\
        \Gamma |- \st <= \sV}
      {\Gamma |- \s{t :: T} => \sT}

      \inferrule[SApp]{
        \Gamma |- \s{t_0} => \sT \\
        \Gamma |- \sT => \sType{\ell}\\\\
        \Gamma |- \sT \! \etasteps \! \s{(x: V_1)\! ->\! V_2} : \sType{\ell} \\\\
      \Gamma |- \s{t_1} <= \s{V_1}
    }{\Gamma |- \s{t_0\ t_1} => [\s{t_1} / \sx]\s{V_2}}

    \inferrule[SConv]{
      \Gamma : \sV => \sType{\ell}\\
      \Gamma |- \st => \s{T} \\\\
      \Gamma |- \s{T} <= \sType{\ell} \\\\ 
      \Gamma |- \s{T} \!\etasteps\! \s{V'}\! : \sType{\ell} \\
      \s{V} =_\alpha \s{V'}
    }{\Gamma |- \st <= \sV}

      \inferrule[SPi]{
        \Gamma |- \s{T_1} => \sType{\ell_1} \\\\
        (\sx : \s{T_1})\Gamma |- \s{T_2} => \sType{\ell_2} }
      {\Gamma |- \s{(x : T_1) -> T_2} => \s\Type{}_{\max(\s{\ell_1},{\s{\ell_2}})}
      }

    \inferrule[SIf]{
      \Gamma |- \s{t_1} <= \s{\bB}\\\\
      \Gamma |- \s{t_2} \!<=\! \s{V}\\\!\!\!\!\!\!
      \Gamma |- \s{t_3} \!<=\! \s{V}
    }{ \Gamma |- \s{if\ t_1\ t_2\ t_3} <= \sV}

    \inferrule[SLam]{
      (\sx : \s{V_1})\Gamma |- \st <= \s{V_2}}
    {\Gamma |- \s{\lambda x \ldotp t} <= \s{(x : V_1) \!->\! V_2}}

  \end{inferbox}
\\\\\small
\boxed{\s{t_1} \leadsto \s{t_2} \ \textit{(Reduction)}} \
\boxed{\Gamma |- \s{t} \stepsto_\eta \s{v} : \sT \textit{ ($\eta$-expansion, non-structural rules) } } \
\boxed{\Gamma |- \s{t} \etasteps \s{v} : \sT \textit{ (Normalization) }} \\
  \begin{inferbox}
    \s{((\lambda x \ldotp t_2) :: (x : T_1) -> T_2) \ t_1}
    \redsto_\beta   \s{(}[\s{t_1} / \sx]\s{t_2}\s{)::}[\s{t_1} / \sx]\s{T_2}

    \s{if\ true\ t_1\ t_2}  \redsto_{\mathsf{true}}  \s{t_1}

    \s{if\ false\ t_1\ t_2}   \redsto_\mathsf{false}   \s{t_2}

    \s{t :: T} \redsto_{::}  \st
      \textit{ when } \st \in \set{\s{true}, \s{false}, \s{\bB}, \sType{\ell} }

      \inferrule[EtaNe]{
    (\sx : \sV_1)\Gamma |- \sN\ \sx \stepsto_\eta \sv : \s{V_2}
  }
  {\Gamma |- \sN \stepsto_\eta \s{\lambda (x : V_1) \ldotp \sv} : \s{(x : V_1) -> V_2}}

  \inferrule{\st \stepstostar \s{v'} \\
    \Gamma |- \s{v'} \stepsto_\eta \sv : \sV
  }
  {\Gamma |- \s{t} \etasteps \sv : \sV}
  \end{inferbox}
\end{boxedarray}
  \caption{\slang: Static and Dynamic Semantics }
  \label{fig:static-types-reductions}
\end{figure}

The type system  (\cref{fig:static-types-reductions}) follow a
 bidirectional style~\citep{localTypeInference}:
synthesis treats the type as outputs,
while checking takes as input a type and checks a term against it.
We are very precise about where we normalize: before checking against a type (\rrule{SApp, SAscr}),
and at the synthesis-checking boundary (\rrule{Conv}).

The bidirectional style simplifies the process of adding
gradual types to \slang (\cref{sec:surface-syntax}),
since we can check for consistency at the synthesis-checking boundary.
Additionally, it allows top-level type annotations
to provide information for deeper terms, so the can write
fewer annotation.

The rules \rrule{SApp} and \rrule{SConv}
enable types to depend on terms.
\rrule{SApp} allows the type of an application to depend on the argument's \textit{value}: to synthesize
an application's type, we synthesize a type for the function, normalize it
to a function type, and check the argument against the domain.
The return type is the codomain, but with the bound parameter replaced
by the concrete argument.
\rrule{SConv} accounts for computation in types: a term checks against a type
if it synthesizes a type that normalizes to (is convertible with) the checked type,
modulo $\alpha$-equivalence.

The remaining rules are standard. Variables synthesize types from the context (\rrule{SVar}).
The type $\s{\bB}$ synthesizes $\sType{0}$ (\rrule{SBool}), while functions synthesize
the greater universe of the domain and codomain universes (\rrule{SPi}). Each universe synthesizes
a universe one level up (\rrule{SType}), while booleans synthesize $\s{\bB}$ (\rrule{STF}).
If-expressions check against a type if the scrutinee is boolean and both branches check against
that type (\rrule{SIf}), and functions check against a function type if their bodies
check against the return type in a context extended with the argument's type.

\subsection{Semantics}
\label{subsec:static-semantics}

We describe the semantics of \slang as the compatible closure of a set of reductions.
Each reduction corresponds to an elimination form in our language
or the removal of a redundant annotation. The core reduction rules are defined
as a relation $\_\!\!\!\redsto\!\!\!\_ \subseteq \textsc{Term} \times \textsc{Term}$, which we specify in \cref{fig:static-types-reductions}.
The reductions are standard,

including $\beta$ reductions, if-branching,
and removal of redundant ascriptions.

\slang enjoys confluence and strong normalization,
so we do not need
evaluation contexts. Instead, we define a contextual
step relation ${\_\!\!\!\stepsto\!\!\!\_ \subseteq \textsc{Term} \times \textsc{Term}}$
which applies a single reduction to any sub-term.
The multi-step relation ${\_\!\!\!\stepstostar\!\!\!\_ \subseteq \textsc{Term} \times \textsc{Term}}$
is the reflexive-transitive closure of $\stepsto$.
Since normal forms do not reduce, ${\s{T} \stepstostar \s{V}}$ denotes the maximally reduced
form of $\sT$.

As a final consideration,
we wish $\s{f : V_1 -> V_2}$ to be equivalent to $\s{\lambda x \ldotp f\ x }$
when compared in e.g. \rrule{Cong}.
We \textit{$\eta$-expand} normal forms before comparing them syntactically.

\Cref{fig:static-types-reductions} includes the standard type-directed $\eta$-expansion rules: a neutral term (possibly a variable)
with a function type is expanded into a $\lambda$-abstraction.
Structural rules (omitted) mirror the structure of the
typing rules.
Note that $\eta$-expansion only takes normal forms,
which simplifies the reasoning about confluence.

The $\eta$-rules provide the final piece we need to fully define
our normalization relation ${\Gamma |- \s{t} \etasteps \s{v} : \sT}$:
a term is contextually reduced until it is in normal form, then $\eta$-expanded (\cref{fig:static-types-reductions}).

\section{\surflang: The Surface Language}
\label{sec:surface-syntax}

With our static starting point defined, we now turn to \surflang, a \surflangDesc.
In this section, we give the syntax of \surflang, along with some typing intuitions.
The idea behind \surflang is that a term checks against any type that is consistent with the type it synthesizes.
Because these conversions are implicit, we elaborate them into a calculus with explicit casts (\cref{sec:cast}),
for which a dynamic semantics is easily defined.

We denote \surflang terms with the metavariables $\gs$ and $\gS$.
Compared to \slang, \surflang differs only by the introduction of $\qmat{\ell}$, that is,
the unknown term ascribed with \textit{the level of} its type. Using bidirectional
typing means that the programmer need not ascribe $\qm$ with its precise type.
We conjecture that level annotations can be inferred in practice, and write $\qm$
when the level is not relevant.
{\small
\begin{displaymath}
  \textsc{Term} \subseteq \textsc{GTerm} \ni \g{s}, \g{S} \bnfadd \g{\qmat{\ell}}
\end{displaymath}
}

The key feature of \surflang is that, whenever \slang compares types
for equality, \surflang instead compares them for consistency, which allows the programmer to use
$\qm$ to express type imprecision.
Because a we need to explicitly track type information to ensure safety, we cannot define a semantics for \surflang directly.
However, the \slang typing rules depend on the semantics (\eg \rrule{SConv}),
so we cannot gradualize them without a dynamic semantics.

Instead, we defer the definition of \surflang's type system until \cref{sec:elab}, but present here admissible rules conveying
the essence of \surflang typing (\cref{fig:surface-type-lemmas}).
In place  of equality, we use $=-=^{*}$, \textit{syntactic consistency modulo conversion}, whose definition we defer until \cref{sec:elab}. Roughly, $\g{T_{1} =-=^{*} \g{T_{2}}}$ means that $\g{T_{1}}$ and $ \g{T_{2}}$
normalize to values that are equal up to occurrences of $\qm$.
These type rules can be viewed as programmer-friendly explanation of \surflang typing.
We can also view it as a list of desiderata for \surflang's full type system.
Indeed, the full type system for \surflang admits each rule as a lemma.

\begin{figure}
  \begin{boxedarray}{@{}l@{}}
    \boxed{\Gamma |- \gs => \gS\quad \Gamma |- \gs <= \gS \textit{ (Checking and Synthesis)}}\
  \rrule{GVar}, \rrule{GType}, \rrule{GBool}, \rrule{GTF}, \rrule{GIf:}
  \text{ analogous to \slang }
    \\
\begin{inferbox}
  \mprset {fraction ={\cdot\cdots\cdot}}
    \inferrule[GLam]{
      \g{S} =-=^{*}  \g{(x : S_1) \!->\! S_2}\\
      (\gx : \g{S_1})\cdot |- \gs <= \g{S_2}}
    {\cdot |- \g{\lambda x \ldotp s} <= \gS}

      \inferrule[GApp]{
        \cdot |- \g{s_0} => \gS \\
        \gS =-=^{*} \g{(x: S_1)\! ->\! S_2} \\
      \cdot |- \g{s_1} <= \g{S_1}
    }{\cdot |- \g{s_0\ s_1} => [\g{s_1} / \gx]\g{S_2}}

    \inferrule[GAscr\ (\rrule{GPi} \textup{\textit{follows same pattern}})]{
        \cdot |- \gS => \g{S_2}\\
        \g{S_2} =-=^{*} \gType{\ell}\\
        \cdot |- \gs <= \gS}
      {\cdot |- \g{s :: S} => \gS}

    \inferrule[GConv]{
      \cdot |- \gs => \g{S'} \\
      \g{S} =-=^{*} \g{S'} 
    }{\cdot |- \gs <= \gS}

  \inferrule[GUnk]{\cdot |- \gS <= \gType{\ell}}
  {\cdot |- \qmat{\ell} <= \g{S} }

\end{inferbox}
\end{boxedarray}
\caption{\surflang: Admissible Type Judgment Properties}
\label{fig:surface-type-lemmas}
\end{figure}

The \rrule{GUnk} rule allows $\qmat{\ell}$ to check against any type, so long as
that type has the correct level.
The main rule is \rrule{GConv}, where the equality check of \rrule{SConv} is replaced
by a consistency check.
In particular, this ensures that any term that synthesizes a type checks against $\qmat{\ell}$.
Similarly, \rrule{GLam} allows functions to check against arrow types or $\qmat{\ell}$,
which is also consistent with an arrow type.

In \rrule{GAscr}, instead of requiring that the type synthesize $\gType{\ell}$,
we require that it synthesize a type consistent with $\gType{\ell}$.
We omit rule \rrule{GPi}, but it uses the same relaxation as \rrule{SPi}.
\rrule{GApp} allows applications whose functions synthesize a type consistent
with a function type, and use that function type to determine the synthesized result.

\section{\clang: A Gradual Cast Calculus}
\label{sec:cast}

Much of the subtlety of gradual dependent types lies in defining the run-time semantics.
Because types depend on terms and type-checking normalizes terms, the dynamic semantics
deeply affect typing.
Here we define \clang, a \clangDesc.
\clang features type imprecision while introducing explicit casts.
Given these casts, we define a static and  dynamic semantics
for \clang. The dynamic semantics for the surface language \surflang is defined by
elaboration to \clang.

\subsection{Syntax}

We denote \clang terms using the metavariables $\g{t}$ and $\g{T}$.
The syntax for \clang is mostly the same as \slang, with changes highlighted in \Gbox{\text{grey}}.

\begin{displaymath}
  \small
  \begin{array}{rcll}
     \textsc{CTerm} \ni \g{t},\gT & \defbnf & \Gbox{\qm_{\gT}} \bnfalt \Gbox{\err_{\gT}} \bnfalt \Gbox{\attagl{h}{t}}
                                            \bnfalt \Gbox{\cast{T_1}{T_2}\gt}
                                            \bnfalt \g{(\lambda x \Gbox{: \gT} \ldotp t)}
                                            \bnfalt \g{if_{\Gbox{\gT}}\ t_1\ t_2\ t_3}
    \\ &&
          \bnfalt \g{(x : T_1) -> T_2}
           \bnfalt \g{t_1\ t_2}
          \bnfalt \g{\bB} \bnfalt \g{true} \bnfalt \g{false}
    \bnfalt \gType{\ell} \bnfalt \gx
    \\
    \textsc{TypeTag} \ni \Gbox{\g{h}} & \defbnf & \Gbox{\g{\Pi} \bnfalt \g{\bB} \bnfalt \gType{}} \qquad
    \textsc{Tag} \ni \Gbox{\g{H}} \defbnf \Gbox{\g{h} \bnfalt \g{\lambda} \bnfalt \g{true} \bnfalt \g{false}}
  \end{array}
\end{displaymath}
Conditionals and functions have type annotations: some reductions produce casts
that require dynamic type information.
$\qm_{\gT}$ is the \textit{unknown} term of type $\gT$,
and $\err_{{\gT}}$, the \textit{error} term of type $\gT$.
These are the least and most precise terms of type $\gT$ respectively.
Unlike with \surflang, we explicitly ascribe $\qm$ with its type, though its universe level
can be synthesized from the ascribed type.
Casts from $\g{T_{1}}$ to $\g{T_{2}}$ are written $\cast{T_1}{T_2}\gt$.
\textit{Tagged} terms $\attagl{h}{t}$
construct elements of $\qml$.
If the type of $\gt$ has tag $\g{h}$,
then we inject $\gt$ into the unknown type $\qml$ by tagging it.

The metafunction $\typeTagOf$ gives the tag its type must have if it is well typed (\cref{fig:term-tags}).
We also define $\tagOf$ to get a term's constructor, which is useful in \cref{sec:equality}.

\begin{figure}
  \small

\begin{minipage}{0.3\textwidth}
  \begin{flalign*}
    & \mathrlap{\tagOf : \textsc{CTerm} \rightharpoonup \textsc{Tag} } \\
    & \tagOf(\g{true}) = \g{true}\\
    & \tagOf(\g{false}) = \g{false}\\
    & \tagOf(\g{(\lambda x : T \ldotp v)}) = \g{\lambda}\\
    & \tagOf(\g{(x : V_1)->V_2}) = \g{\Pi}\\
    & \tagOf(\g{\bB}) = \g{\bB}\\
    & \tagOf(\gType{\ell}) = \gType{}
  \end{flalign*}
\end{minipage}
\begin{minipage}{0.3\textwidth}
  \begin{flalign*}
    & \mathrlap{\typeTagOf : \textsc{CTerm} \rightharpoonup \textsc{TypeTag} } \\
    & \typeTagOf(\g{true}) = \g{\bB}\\
    & \typeTagOf(\g{false}) = \g{\bB}\\
    & \typeTagOf(\g{(\lambda x : T \ldotp v)}) = \g{\Pi}\\
    & \typeTagOf(\g{(x : V_1)->V_2}) = \gType{}\\
    & \typeTagOf(\g{\bB}) = \gType{}\\
    & \typeTagOf(\gType{\ell}) = \gType{}
  \end{flalign*}
\end{minipage}
  \begin{minipage}{0.3\textwidth}
    \begin{flalign*}
    & {\levelOne_{\ell} : \textsc{Tag} \to \textsc{CTerm} } \\
    & \levelOne_\ell(\g{\Pi}) = \g{(x : \qml)-> \qml}\\
    & \levelOne_\ell(\g{\bB}) = \g{\bB}\\
    & \levelOne_\ell(\gType{}) = \gType{\ell}\\
    & \mathrlap{\levelOne_{\ell\not\Pi} : \textsc{Tag} \rightharpoonup \textsc{CTerm} } \\
    & \levelOne_{\ell\not\Pi}(\g{\Pi}) \  \textit{ undefined }\\
    & \levelOne_{\ell\not\Pi}(\g{h}) = \levelOne_\ell(\g{h}) \textit{ o/w }
  \end{flalign*}
\end{minipage}
  \caption{Tags and Unknown Types for Terms}
    \label{fig:term-tags}
  \end{figure}

Normal and neutral terms are similar to those of \slang.
Terms $\qm_{\gT}$  and $\err_{\gT}$  are normal when $\gT$ is, and wrapping a normal term
in $\attagl{h}{}$ produces a normal term.
Casts cannot reduce
when either the term or the types are neutral.
There are a few exceptions to this:
  casting $\err_{\gV}$ reduces to $\err_{\gT}$ for target type $\g{T}$,
  and casting to or from $\err_{\gType{\ell}}$ reduces to $\err$.
  Likewise,
  a neutral term is cast between function types by $\eta$-expanding.
  Casting a neutral term to $\qml$ reduces by wrapping
        in $\attagl{h}{}$ when the source type is not neutral, or produces the original term when the source type is $\qml$.
All reduction rules are given in \cref{subsec:clang-semantics}.
\begin{flalign*}
    &\gv, \gV  \defbnf
                              \Gbox{\qm_{\gV}} \bnfalt \Gbox{\err_{\gV}} \bnfalt \Gbox{\attagl{h}{v}}
    \bnfalt \gN
      \bnfalt \gType{\ell}
      \bnfalt \g{(\lambda \gx : T \ldotp \gv)}
      \bnfalt \g{(x : V_{1})->V_{2}}
      \bnfalt \g{\bB} \bnfalt \g{true} \bnfalt \g{false}
    \\
    & \gN      \defbnf
                                              \gx
      \bnfalt \gN\ \gv
      \bnfalt \g{if_{\gV}\ N\ v_2\ v_3}
                                             \bnfalt \Gbox{\cast{\!\!N}{V\!\!}\g{v_{\not\err}}}
                                             \bnfalt \Gbox{\cast{\!\!V}{N\!\!}\g{v_{\not\err}}}
                                             \bnfalt \Gbox{\cast{\!\!V_{\not\err\not\qm\not\Pi}}{V_{\not\err\not\qm\not\Pi}\!\!}\g{N}}
                                             \bnfalt \Gbox{\!\!\cast{\qml}{V_{\not\err\not\qm}\!\!}\g{N}}
    \\
  &\textit{where }\quad \mathrlap{
    \g{v_{\not\err}}, \g{V_{\not\err}} \textit{ is } \gv \textit{ without } \g{\err_{\gV}},\quad
    \g{V_{\not\err\not\qm}} \textit{ is } \g{V_{\not\err}} \textit{ without }  \g{\qm_{V}},\quad
    \g{V_{\not\err\not\qm\not\Pi}} \textit{ is } \g{V_{\not\err\not\qm}} \textit{ without } \g{(x : V_{1}) \!->\! V_{2}}
    }
\end{flalign*}

\subsection{Typing}

\begin{figure}
  \begin{boxedarray}{@{}l@{}}
    \boxed{\Gamma |- \gt => \gT\quad \Gamma |- \gt <= \gV \textit{(Synthesis and Checking)} }
  \  \rrule{CVar}, \rrule{CType}, \rrule{CBool}, \rrule{CTF},
  \rrule{CPi}, \rrule{CApp:} \text{ same as \slang }
\\
\begin{inferbox}
  {\inferrule[CConv]
  {\Gamma |- \gt => \g{T'}\\
    \Gamma |- \g{T'} => \gType{\ell}\\
    \Gamma |- \g{T} <= \gType{\ell}\\\\
    \Gamma |- \g{T} \etasteps \g{V} : \gType{\ell}\\
    \Gamma |- \g{T'} \etasteps \g{V} : \gType{\ell}
  }
  {\Gamma |- \gt <= \gT}}

  \inferrule[CIf]{
    \Gamma |- \g{t_1} <= \g{\bB}\\
    \Gamma |- \gT => \gType{\ell}\\\\
    \Gamma |- \g{t_2} <= \g{T}\\
    \Gamma |- \g{t_3} <= \g{T}
  }{ \Gamma |- \g{if_{\gT}\ t_1\ t_2\ t_3} => \gT}
  \vspace{-0.75em}

  {\inferrule[CCast]
    {
      \Gamma |- \g{T_1} => \gType{\ell} \\
      \Gamma |- \g{t} <= \g{T_1}\\
      \Gamma |- \g{T_2} <= \gType{\ell}
    }
    {\Gamma |- \cast{T_1}{T_2}\gt => \g{T_2}}}

  \inferrule[CLam]{
    (\gx : \g{T_1})\Gamma |- \gt => \g{T_2}\\
  }
  {\Gamma |- \g{(\lambda x : T_1 \ldotp t)} => \g{(x : T_1) \!->\! T_2}}

  {\inferrule[CTag]
    { \Gamma |- \g{t} <= \levelOne_{\ell\not\Pi}(\g{h})}
    {\Gamma |- \attagl{\g{h}}{t} => \qml}}

  {\inferrule[CTagFun]
    {\Gamma |- \gv <= \qml}
    {\Gamma |- \attagl{\Pi}{(\lambda x : \qml \ldotp v)} => \qml}}

  {\inferrule[CUnk]
    {\Gamma |- \g{T} => \gType{\ell}}
    {\Gamma |- \qm_{\gT} => \gT}}

  {\inferrule[CErr]
    {\Gamma |- \g{T} => \gType{\ell}}
    {\Gamma |- \err_{\gT} => \gT}}
  \end{inferbox}
\end{boxedarray}
  \caption{\clang: Typing }
  \label{fig:clang-types}
\end{figure}

Typing for \clang, given in \cref{fig:clang-types}, is similar to \slang.
However, the addition of type annotations means that all forms can synthesize,
and since conversion is the only checking rule, we check against non-normal types.
All casts are explicit, so we need not reason about consistency,
although casts between inconsistent types produce $\err$.
\rrule{CLam} and \rrule{CIf} are synthesis versions of the \slang rules,
and \rrule{CConv} ensures synthesized and input types have the same normal form.
In \rrule{CUnk} and \rrule{CErr}, $\qm$ and $\err$ synthesize their ascriptions.
In \rrule{CCast},  $\cast{T_{1}}{T_{2}}\gt$ synthesizes $\g{T_2}$,
provided $\gt$ checks against $\g{T_1}$.

Finally, there are rules that synthesize $\qml$.
\rrule{Tag} injects a value of non-function type $\qml$ if it is tagged accordingly.
The term is typed against a \textit{level-1 unknown type}, written
$\levelOne$, which is the least precise type with the given tag (\cref{fig:term-tags}).
The name \textit{level-1 type} refers to the depth of knowledge we have about the type:
$\qm$ is a level-0 imprecise type (zero knowledge), $\g{\qm -> \qm}$ is level-1 imprecise type, etc.
When casting to $\qm$, we always cast through the corresponding $\levelOne$ type first.
$\levelOne$ is trivial for $\g{\bB}$ and $\gType{\ell}$, but is
relevant with the introduction of inductive types.
\rrule{TagFun} allows a function to be injected into $\qml$,
provided that the function does not refer to its argument
This assumption is critical for ensuring that functions are properly approximated
before being cast to type $\qml$.

\subsection{Semantics}
\label{subsec:clang-semantics}

\myparagraph{Reductions} \clang's reductions (\cref{fig:clang-reductions})
come in three groups. First, we have rules inherited from \slang:
\rrule{True}, \rrule{False} and \rrule{$\beta$}.
Second, we have the unknown- and error-producing rules \rrule{$\beta$?, $\beta\errsym$, If?} and \rrule{If$\errsym$},
which are versions of the \slang rules for $\qm$ and $\err$.
Unlike eager exceptions,
an expression containing $\err$ might not reduce to $\err$, such as when it is
passed to a function ignoring its argument.

The remaining rules are for casts, labeled according to their source and destination types.
\rrule{To?} does casts from non-functions to $\qm$ by casting through the level-1 type and affixing a tag.
For functions, \rrule{$?\Pi$} injects a function into $\qml$. To ensure termination,
we create a $\lambda$-abstraction that ignores its argument and produces the least precise value in the image of the original function.
Since this upper bound has type $[\qm_\g{T_1} / \gx]\g{T_2}$, we cast that result into $\qml$.
As we show in \cref{sec:precision}, all functions in \clang are monotone,
so applying $\qm_{\g{T_1}}$ yields the least precise value in the function's image.
The remaining rules handle trivial casts from a type to itself, as well as casts
that produce $\err$, either from $\err$ in the source, destination, or cast term,
or from casting when tags do not match.
Finally, \rrule{$\Pi\Pi$} casts between function types by producing a new function
that casts its argument to the source domain, applies the given function,
then casts to the target codomain. Because function types are dependent,
we cast the argument when computing the target return type.

\begin{figure}
  \begin{boxedarray}{@{}l@{}}
  \boxed{\g{t_1} \leadsto \g{t_2} \quad \textit{(Reductions)}}\qquad\text{\rrule{$\beta$,True,False} as in \slang}\\
  \begin{inferbox}
    
    \g{\qm_{\g{(x : T_1)->T_2}}\ \gt}   \redsto_{\beta?}  \qm_{[\g{t} / \gx]\g{T_2}}

    \g{\err_{\g{(x : T_1)->T_2}}\ \gt}   \redsto_{\beta\errsym}  \err_{[\g{t} / \gx]\g{T_2}}

    \g{if_{\gT}\ \qm_{\g{\bB}}\ t_1\ t_2}   \redsto_\textsc{If?}   \qm_{\gT}

    \g{if_{\gT}\ \err_{\g{\bB}}\ t_1\ t_2}   \redsto_{\textsc{If}\errsym}   \err_{\gT}

    \inferrule{}
    {\cast{\qml}{\qml}\gt \leadsto_{\rrule{UnkUnk}} \gt }

    \inferrule{\typeTagOf(\gt) = \gType{}}
    {\cast{\gType{\ell}}{\gType{\ell}}\gt \leadsto_{\textsc{Type}} \gt }

    \inferrule{\tagOf(\g{T_1}) \neq \tagOf(\g{T_2})}
    {\cast{T_1}{T_2}\gt \redsto_{\textsc{tag}\errsym} \err_{\g{T_2}}}

    \inferrule{\g{h} = \tagOf(\gT)}
    {\cast{\qml}{T}\attagl{h}{t} \redsto_{\rrule{From?}}
      \cast{\ix{\levelOne_{\ell}(\g{h})}}{\gT} \gt }

     \inferrule{\g{h} \neq \tagOf(\gT)}
    {\cast{\qml}{T}\attagl{h}{t} \redsto_{?\errsym} \err_{\gT}}

    \inferrule{\g{h} = \tagOf(\gT)}
    {\cast{T}{\qml}\gt \redsto_{\mathsf{To?}}
      \attagl{h}{(\cast{\gT}{\ix{\levelOne_{\ell\not\Pi}(\g{h})}} \gt)} }

    \inferrule{\typeTagOf(\gt) = \g{\bB}}
    {\cast{\bB}{\bB}\gt \leadsto_{\bB\bB} \gt }

    \inferrule{}
    {\cast{\err_{\gType{\ell}}}{\gT}\gt \redsto_{\textsc{From}\errsym} \err_{\gT}}

    \inferrule{}
    {\cast{T_1}{T_2}\err_{\g{T_1}} \redsto_{\rrule{Cast}\errsym} \err_{\g{T_2}} }

    \inferrule{}
    {\cast{\gT}{\err_{\gType{\ell}}}\gt \redsto_{\textsc{To}\errsym} \err_{\err_{\gType{\ell}}}}

    \inferrule{}
    {\cast{(y:T_1)->T_2}{\qml}\gt \redsto_{?\Pi}
      \attagl{\Pi}{\g{(\lambda x : \qml \ldotp \cast{{\ix{[\qm_{\g{T_1}} / \gy]\g{T_2}}}}{\qml}(\gt\ \qm_{T_1})) }} }

    \inferrule{}
    {\cast{(y : T_1)\!->\!T_2}{(x: T'_1)\!->\! T'_2}\gt \leadsto_{\Pi\Pi} \g{(\lambda x : T'_1 \ldotp} \cast{{\ix{[\cast{T'_1}{T_1}\gx / \gy]\g{T_2}}}}{T'_2}\g{(t\ (\cast{T'_1}{T_1}x)))}}
  \end{inferbox}
\end{boxedarray}
  \caption{\clang: Reductions}
  \label{fig:clang-reductions}
\end{figure}

\myparagraph{$\eta$-Expansion} $\eta$-expansion is largely the same as in
\slang, but must account for $\qm$ and $\err:$
{\small
\begin{displaymath}
  \inferrule
  {(\gx : \g{V_1}) |- \g{V_2} \stepsto_{\eta} \g{V'}}
  {\Gamma |- \qm_{\g{(x : V_1)\!->\!V_2}} \stepsto_{\eta} \g{(\lambda x : V_{1} \ldotp \qm_{V'})} : \g{(x : V_1) \!->\! V_2} }
\qquad
  \inferrule{ }
  {\Gamma |- \err_{\g{(x : V_1)\!->\!V_2}} \stepsto_{\eta} \g{(\lambda x : V_{1} \ldotp \err_{V'})} : \g{(x : V_1) \!->\! V_2}}
\end{displaymath}
These rules allow normalization to reflect the behavioral equivalence
of $\qm_{\g{(x : V_{1}) -> V_{2}}}$ to $\g{\lambda x : V_{1} \ldotp \qm_{\g{V_{2}}}}$ and $\err_{\g{(x : V_{1}) -> V_{2}}}$ to $\g{\lambda x : V \ldotp \err_{{\g{V_{2}}}}}$ for function types.
}
\subsection{Metatheory: Confluence and Safety}

Standard techniques yield the required
properties of \clang, namely confluence, progress and preservation.
By treating $\err$ as a value in \clang,
we show that terms are never ``stuck'', but some reach error
in a controlled manner.

\clang is confluent, which is needed to show type preservation:

\begin{theoremEnd}[apxproof]{lemma}[Confluence]
  If $\g{t} \stepstostar \g{t_1}$ and $\g{t} \stepstostar \g{t_2}$,
  then $\g{t_1} \stepstostar \g{t_3}$ and $\g{t_2} \stepstostar \g{t_3}$
  for some $\g{t_3}$.
\end{theoremEnd}
\begin{proofEnd}
  We use the parallel reduction strategy of \citep{TAKAHASHI1995120}.
  The parallel reduction $\Rrightarrow$ may contract zero or more redexes in a given term,
  but none of the \textit{resulting} redexes.
  Substitution preserves $\Rrightarrow$, since replacing a variable with a term
  never eliminates a redex, only introduces one.

  For each $\g{t}$, there is a term
  $\g{t*}$ which is the \textit{maximal} parallel reduction of $\g{t}$, by reducing
  all sub-redices of $\gt$ one step.
  $\g{t*}$ is maximal in that if $\g{t} \Rrightarrow \g{t'}$ then $\g{t'}\Rrightarrow \g{t*}$.
  The maximality of $\g{t*}$ gives a single-step diamond property for $\Rrightarrow$, and because $\leadsto \ \subseteq\ \Rrightarrow \ \subseteq\  \stepstostar$, we obtain confluence of $\stepstostar$.

  Proof of maximality: We perform induction on $\g{t_{start}}$, our goal is to show that, for every
  $\g{t_{mid}}$ such that $\g{t_{start}} \Rrightarrow \g{t_{mid}}$, then $\g{t_{mid}}\Rrightarrow \g{t_{start}*}$.

  \mlcase{\g{\bB}, \gType{\ell}, \g{true}, \g{false}}{Then $\g{t_{mid}} = \g{t_{start}} = \g{t_{start}*}$, so we can take 0 steps.}

  \mcase{\g{\lambda (x : T \ldotp t)}}{
    Then $\g{t_{start}*} = \g{(\lambda x : T* \ldotp t*)}$.
    $\g{t_{mid}}$ must have the form $\g{\lambda (x : T' \ldotp t')}$, where $\g{T} \Rrightarrow \g{T'}$
      and $\g{t} \Rrightarrow \g{t'}$, so by our hypothesis, $\g{T'}\Rrightarrow\g{T*}$ and $\g{t'}\Rrightarrow\g{t*}$,
      yielding our result.}

  \case{any other $\gt$ such that $\gt \not\leadsto \g{t'}$ for any $\g{t'}$}
  {
    same logic as lambda: only contextual steps are possible, so stepping each immediate sub-term maximally (by the inductive hypothesis) gives the maximal result. }

  \mcase{\g{(\lambda x : T \ldotp t_1)\ t_2}}{
    Then $\g{t_{start}*} = [\g{t_{2}*}/\gx]\g{t_{1}}$.
    We have two possibilities for $\g{t_{mid}}$.

    \textbf{First case:} $\g{t_{mid}} = \g{(\lambda x : T' \ldotp t'_1)\ t'_2}$ where
    $\g{T} \Rrightarrow \g{T'}$, $\g{t_{1}} \Rrightarrow \g{t'_{1}}$ and $\g{t_{2}} \Rrightarrow \g{t'_{2}}$.
    Then $\g{(\lambda x : T' \ldotp t'_1)\ t'_2} \Rrightarrow [\g{t'_{2}}/\gx]\g{t'_{1}} $.
    Since $\Rrightarrow$ is preserved under substitution,
    and since $\g{t_{1}*}$ and $\g{t_{2}*}$ are maximal by our hypothesis,
    we have
    $\g{(\lambda x : T' \ldotp t'_1)\ t'_2} \Rrightarrow [\g{t_{2}*}/\gx]\g{t_{1}*} $.

    \textbf{Second case:} $\g{t_{mid}} = [\g{t'_{2}}/\gx]\g{t'_{1}}$ where
    $\g{T} \Rrightarrow \g{T'}$, $\g{t_{1}} \Rrightarrow \g{t'_{1}}$ and $\g{t_{2}} \Rrightarrow \g{t'_{2}}$.
    Then the result follows directly from our inductive hypothesis.
}

  \case{Remaining cases

    }
    {
      Same idea as above: since each remaining form is a redex.
      Each of the given redex rules has an elimination form as the top-level syntactic construct, and all sub-terms are constrained only by applying introduction forms.
  These introduction forms only ever step contextually, so any steps within the sub-terms preserve
  the contractability of the top-level term.
  The right hand side of each redex is composed entirely from sub-terms
  of the LHS, or from irreducible normal forms (like the level-1 type of a given tag).

      We always have two cases: either $\g{t_{mid}}$ contracts the top-level redex,
      or it does not.

  \textbf{First Case:}  $\g{t_{mid}}$ is $\g{t_{start}}$ with some sub-terms
  possibly stepped. Since the remaining reduction rules are all abstract in these sub-terms, $\g{t_{mid}}$
  can always be reduced by the same rule as $\g{t_{start}}$. The result then follows
  by the maximality

  \textbf{Second Case:} $\g{t_{start}} \stepsto \g{t_{next}}$ by some rule,
  and $\g{t_{mid}}$ is $\g{t_{next}}$ with some of the redexes that occur in both $\g{t_{start}}$
  and $\g{t_{next}}$ contracted.
  The right-hand sides of reduction rules are all formed through some combination of
  substitutions and applying syntax constructors, so we obtain
  $\g{t_{mid}} \Rrightarrow \g{t_{start}*}$ by contracting all the remaining redexes that also
  occur in $\g{t_{start}}$.

  The proof for \rrule{$?\Pi$} and \rrule{$\Pi\Pi$} also takes into account the fact that part of $\g{t_{start}}$
  is duplicated in the right-hand side of the redex.
}

\end{proofEnd}

Confluence allows us to show canonicity by a direct induction over
typing derivations.

\begin{theoremEnd}[apxproof]{lemma}[Canonical Forms]
  \label{lem:canonical-forms}
Consider $\gv$ where $\Gamma |- \gv <= \gT$ or $\Gamma |- \gv => \gT$.
  \begin{enumerate}
  \item If $\gT \stepstostar \g{(x : T_1) -> T_2} $,
    then $\gv$ is either neutral,
  $\err_\g{(x : T_1)->T_2}$, $\g{(\lambda x : T_{1} \ldotp v')}$ for some $\g{v'}$,
  or $\textrm{\qm}_\g{(x : T_1)->T_2}$;
\item If $\gT \stepstostar \g{\bB}$, then $\gv$ is neutral, $\qm_{\g{\bB}}$,
    $\err_{\g{\bB}}$, $\g{true}$ or $\g{false}$;
  \item If and $\gT \stepstostar \gType{\ell}$, then $\gv$ is neutral, or one of $\qml$, $\err_{\gType{\ell}}$, $\g{\bB}$,
    $\gType{\ell-1}$ (if $\g\ell > 0$), or $\g{(x : V_1) -> V_2}$.
  \end{enumerate}
\end{theoremEnd}
\begin{proofEnd}
  We begin by noting that arrow types only reduce to other arrow types,
  and $\g{\bB}$ and $\g{\gType{\ell}}$ do not further reduce.

  By induction on the typing derivation.

  \rlcase{CVar}{ is neutral.}

  \rlcase{CType, CBool}{trivial.}

  \rlcase{CTF}{is boolean literal.}

  \rlcase{CPi}{Then $\gV = \g{(x : V_{1}) -> V_{2}} $}

  \rcase{CLam}{ Then $\gv = \g{\lambda x : V \ldotp v'}$ and $\g{T} = \g{(x : V) -> T'}$
    for some $\g{T'}$, yielding our result.
 }

  \rlcase{CApp, CIf, CCast}{ if normal, must be neutral.}

  \rlcase{CUnk}{Then $\gv$ is $\qm_{\gV}$ where $\g{T} = \gV$. By our premise,
    since $\g{V}$ does not reduce, it is either $\g{(x : V_1)->V_2}$, $\g{\bB}$ or $\gType{\ell}$,
  so in each case, we have the desired form for $\g{\qm_{V}}$.}

  \rlcase{CErr}{Same reasoning as for $\qm$.}

  \rcase{CConv}{
    Consider the proof for (1), the proof is similar in the remaining cases.
    Then $\Gamma |- \gv => \g{T'}$ where both $\g{T}$ and $\g{T'}$ normalize to the same type.
    If $\g{T} \stepstostar \g{(x : T_{1}) -> T_{2}}$, we know it has a normal form by the premise of
    \rrule{CConv}, so we must have $\g{T}\stepstostar \g{(x : V_{1})->V_{2}}$.
    So then $\g{T'} \stepstostar \g{(x : V_{1}) -> V_{2}}$, meaning we can apply our inductive
    hypothesis to on the derivation of $\Gamma |- \gv => \g{T'}$ to obtain our result.
  }

  \rrule{CTag, CTagFun}{ Impossible, since $\qm_{\gV}$ cannot steo to any of the desired types. }

\end{proofEnd}

\begin{theoremEnd}[apxproof]{corollary}[Progress]
  \label{cor:progress}
  If $\Gamma |- \g{t} : \g{T}$, then $\g{t}$ is a normal form, or $\g{t} \stepsto \g{t'}$
  for some $\g{t'}$.
\end{theoremEnd}
\begin{proofEnd}
  By induction on the typing derivation.
  If $\g{t}$ is a normal form, we are done. Similarly if $\g{t}$ only steps contextually,
  the result follows from our induction hypothesis, since each sub-term has a typing derivation.

  In the remaining cases, we show that $\g{t}$ is a redex. If it is an application, then we can apply
  one of \rrule{$\beta$}, \rrule{$\beta?$} or \rrule{$\beta\errsym$} by \cref{lem:canonical-forms} (1).
  If it is an $\g{if}$ expression, we can apply one of $\rrule{true}$, \rrule{false},
  \rrule{If?} or \rrule{If$\errsym$} by \cref{lem:canonical-forms} (2).
  Otherwise, it must be a cast $\cast{T_{1}}{T_{2}}\g{t''}$.
  If $\g{T_{1}}$ or $\g{T_{2}}$ is not a value, then we step contextually by our hypothesis.
  Assume henceforth that they are values.

  \case{$\g{T_{1}}$ and $\g{T_2}$ both have tags}{
    Then either the tags are the same, or they are distinct.
    If distinct, then we step by \rrule{Tag$\errsym$}.
    Otherwise, by \cref{lem:canonical-forms} (3) the tag must be $\g{\bB}, \g{\Pi}$ or $\gType{}$,
    meaning we can step by one of \rrule{$\bB\bB$}, \rrule{$\Pi\Pi$} or $\rrule{TypeType}$ respectively.
  }

  \case{$\g{T_{1}}$ or $\g{T_{2}}$ or $\g{t''}$ is neutral}{
    We have defined such a cast to be neutral any time a step is not possible.
  }

  \case{One of $\g{T_{1}}$, $\g{T_{2}}$ or $\g{t''}$ is $\err$:}{
    Then we step by one of \rrule{To$\errsym$}, \rrule{From$\errsym$}
    or \rrule{Cast$\errsym$}.
  }

  \case{$\g{T_{1}}$ and $\g{T_{2}}$ are both $\qm$}{
    For the cast to be well typed, they must both be $\qml$ for some $\g\ell$.
    Then we step by \rrule{UnkUnk}.
  }

  \case{$\g{T_{1}}$ is $\qml$}{
    If no previous case has applied, $\g{T_{2}}$ cannot be neutral or $\err$ or $\qm$,
    so by \cref{lem:canonical-forms} (3) it must have a defined tag, so we can
    step by \rrule{From?} or \rrule{Tag$\errsym$}.
  }

  \case{$\g{T_{2}}$ is $\qml$}{
    If no previous case has applied, $\g{T_{1}}$ cannot be neutral or $\err$ or $\qm$,
    so by \cref{lem:canonica-forms} (3) it must have a defined tag, so we can
    step by \rrule{From?} or \rrule{Tag$\errsym$}.
  }

  All other forms are impossible, by our assumption that $\g{T_{1}}$ and $\g{T_{2}}$
  are values, and \cref{lem:canonical-forms} (3).

\end{proofEnd}

\begin{theoremEnd}[apxlem]{lemma}[Weakening]
    \label{lem:weak}
    If $\Gamma_1 \Gamma_2 |- \gt : \gV$ and $\gx$ is free in $\Gamma_2$,
    then $\Gamma_1 (x : \g{V'}) \Gamma_2 |- \gt : \gV$.
  \end{theoremEnd}
  \begin{proofEnd}
    By induction on typing derivation.

    \rlcase{Var}{inserting new variable does not change lookup.}

    \rlcase{Type, Bool, TrueFalse}{immediate, since no reference to context}

    \lcase{All remaining rules}{follows from IH, and fact that unused variables do not affect $\eta$-expansion.}
  \end{proofEnd}

  The final piece of the safety puzzle is showing that
  each reduction step preserves types.

\begin{theoremEnd}[apxlem]{lemma}[Substitution Preserves Conversion]
  \label{lem:subst-conv}
  If $\g{T} \stepstostar \g{T''}$
  then $[\gt / \gx]\g{T} \stepstostar [\gt / \gx]\g{T'}$
\end{theoremEnd}
\begin{proofEnd}
    We know that $\g{T} \Rrightarrow^{*} \g{T'}$ since $\stepsto\ \subseteq\ \Rrightarrow $.
    Then $[\g{t'} / \g{x}]\g{T} \Rrightarrow^{*} [\g{t'} / \g{x}]\g{T'}$,
    since substitution preserves $\Rrightarrow$.
    But $\Rrightarrow^{*} \subseteq \stepstostar$ since each parallel reduction can be decomposed
    into sequential individual reductions, giving us our result.
\end{proofEnd}

  \begin{theoremEnd}[apxlem]{lemma}[Substitution Preserves Typing]
    \label{lem:subst-typing}
  If $\Gamma_1 (\g{x'} : \g{T'})\Gamma_2 |- \gt : \gT$ and $\Gamma_2 |- \g{t'} : \g{T'}$, then $([\g{t'} / \g{x'}]\Gamma_1)\Gamma_2 |- [\g{t'} / \g{x'}]\gt : [\g{t'} / \g{x'}]\gT$.
\end{theoremEnd}
\begin{proofEnd}
  By induction on the typing derivation of $\gt$.

  \rcase{Var}{
    If $\gx = \g{x'}$ same variable, then follows from \cref{lem:weak}.
  If they are not equal and $\gx$ is earlier in the context than $\g{x'}$,
  then $\g{x'}$ cannot occur in the type of $\g{x}$ and the result holds.
  If $\g{x}$ is later in the context, then new context has substituted type,
  giving us our result.
}

  \rlcase{Type, Bool, TrueFalse, Err}{ immediate.}

  \rcase{Lam , Pi}{Follows from IH, premise that substituted type normalizes,
  and fact that lambdas/arrow-types only ever reduce contextually.}

  \rlcase{App}{ From IH and distinct variables assumption.}

  \rlcase{If, Unk, Err, Tag, TagFun}{Follows from IH.}

  \rcase{Conv}{
    Suppose that $\Gamma_1 (\g{x'} : \g{T'})\Gamma_2 |- \gt : \gT$ by rule \rrule{Conv},
    so $\Gamma_1 (\g{x'} : \g{T'})\Gamma_2 |- \gt : \g{T'}$ where $\g{T} \stepstostar \g{T''}$
    and $\g{T'} \stepstostar \g{T''} $.
    Then $[\g{t'} / \g{x'}]\g{T} \stepstostar [\g{t'} / \g{x'}]\g{T''}$
    and $[\g{t'} / \g{x'}]\g{T'} \stepstostar [\g{t'} / \g{x'}]\g{T''}$
    by \cref{lem:subst-conf}.
    Thus, we can apply \rrule{Cong} with IH to obtain our result.
    \je{TODO: how does eta work with this?}
  }
\end{proofEnd}

\begin{theoremEnd}[apxproof]{lemma}[Preservation]
  \label{lem:preservation}
  If $\Gamma |- \g{t_{start}} : \g{T}$ and $\g{t_{start}} \stepsto \g{t_{end}}$ then $\Gamma |- \g{t_{end}} : \gT$.
\end{theoremEnd}
\begin{proofEnd}
  By induction on the typing derivation, inversion on the step.

  \case{\g{t_{end}} = \err_{\gT}}{\je{TODO: show is *same* type}}

  \lcase{$\g{t_{start}}$ typed by \rrule{Conv}, any reduction rule}{We can build the resulting derivation from the IH, plus \cref{lem:subst-conv}. \je{TODO: right lemma? why?}}

  \lcase{$\g{t_{start}} \stepsto \g{t_{end}}$ contextually and $\g{t_{start}}$ not typed using \rrule{App} or \rrule{Conv}}{Follows from the IH.}

  \lcase{$\g{t_{start}} \stepsto \g{t_{end}}$ contextually and $\g{t_{start}}$ typed with \rrule{App}}{We construct the resulting typing derivation
  from IH, finally applying \rrule{Conv} to account for any steps taken in the argument.}

\case{$\g{t_{start}} \leadsto_\beta \g{t_{end}}$, $\g{t_{start}}$ typed with \rrule{App}}
{suppose $\Gamma |- \g{(\lambda x : T'_{1} \ldotp t_2)\ t_1} : [\g{t_1} / \gx]\g{T_2}$,
  where $\Gamma |- \g{(\lambda x : T'_{1} \ldotp t_2)} : \g{(x : T_1)->T_2}$,
  and $\Gamma |- \g{t_1} : \g{T_1}$.
  Then we know that $\g{(\lambda x : T'_{1} \ldotp t_2)}$ must be typed by some sequence of \rrule{Conv},
  followed by \rrule{Lam}. So $\Gamma |- \g{(\lambda x : T'_{1} \ldotp t_2)} : \g{(x : T'_1) -> T'_2}$ where
  $\g{(x : T'_1) -> T'_2} =_\leadsto \g{(x : T_1) -> T_2}$.
    Confluence gives us the injectivity of the arrow constructor, so we know that
    $\g{T'_1} =_\leadsto \g{T_1}$ and $\g{T'_2} =_\leadsto \g{T_2}$.
    We can then apply \cref{lem:subst-typing} and \rrule{Conv} to obtain our result.
  }

  \lcase{$\g{t_{start}} \leadsto_{\beta?} \g{t_{end}}$, $\g{t_{start}}$ typed with \rrule{App} }{
  Same reasoning as above (e.g. using Pi-injectivity), with with \rrule{Unk} in place of \rrule{Lam}}.

\lcase{$\g{t_{start}} \leadsto_{\textsc{True}} \g{t_{end}}$ or $\g{t_{start}} \leadsto_{\textsc{False}} \g{t_{end}}$, $\g{t_{start}}$ typed with \rrule{If} }
  { We simply use the sub-derivation for the appropriate branch.
  }

  \lcase{$\g{t_{start}} \leadsto_{\textsc{If?}} \g{t_{end}}$, $\g{t_{start}}$ typed with \rrule{If} }
  { We can type directly with \rrule{Unk}.
  }

  \case{$\g{t_{start}} \leadsto_{\textsc{from?}} \g{t_{end}}$, $\g{t_{start}}$ typed with \rrule{Cast}}
  {We have $\cast{\qml}{T}\attagl{h}{t} \redsto_{?\textsc{tag}}
    \cast{\ix{\mathsf{\levelOne_\ell}(\g{h})}}{\gT} \gt$,
    and we need to show that $\Gamma |- \g{t} : \levelOne_\ell(\g{h})$.
    We know that $\attagl{h}{t}$ must be typed with some sequence of
    \rrule{Conv}, ending with \rrule{UnkType}, which gives us a typing for
    $\attagl{h}{t}$ at type $\qml$. The sub-derivation for $\g{t}$
    then gives  $\Gamma |- \g{t} : \levelOne_\ell(\g{h})$, so we can use \rrule{Cast}
    to construct the complete derivation.
    }

  \case{$\g{t_{start}} \leadsto_{\textsc{To?}} \g{t_{end}}$, $\g{t_{start}}$ typed with \rrule{Cast}}
  {We can build the typing derivation directly using \rrule{UnkType} and \rrule{Cast}. }

  \case{$\g{t_{start}} \leadsto_{\textsc{?}\Pi} \g{t_{end}}$, $\g{t_{start}}$ typed with \rrule{Cast}}
  {
    We have $\cast{(y:T_1)->T_2}{\qml}\gt \redsto_{?\Pi}
    \attagl{\Pi}{\g{(\lambda x : \qml \ldotp \cast{{\ix{[\qm_{\g{T_1}} / \gy]\g{T_2}}}}{\qml}(\gt\ \qm_{T_1})) }}$.
    By our premise, $\Gamma |- \gt : \g{(y:T_1)->T_2} $, so
    \rrule{App} gives $\Gamma |- \gt\ \qm_{\g{T_1}} : [\qm_{\g{T_1}} / \gy]\g{T_2}$.
    Applying \rrule{Cast} gives
    $\Gamma |- \g{\cast{{\ix{[\qm_{\g{T_1}} / \gy]\g{T_2}}}}{\qml}(\gt\ \qm_{T_1})} : \qml$.
    Finally, \rrule{TagFun} gives our results.
  }

  \case{$\g{t_{start}} \leadsto_{\Pi\Pi} \g{t_{end}}$, $\g{t_{start}}$ typed with \rrule{Cast}}
  {\je{TODO: fix for lambda ascr}
    We have:\\ $\cast{(y : T_1)->T_2}{(y: T'_1)-> T'_2}\gt $
    $\leadsto_{\Pi\Pi} \g{(\lambda x : T'_{1} \ldotp} \cast{{\ix{[\cast{T'_1}{T_1}\gx / \gy]\g{T_2}}}}{T'_2}\g{(t\ (\cast{T'_1}{T_1}x)) )_{\g{(y : T'_{1})->T'_{2}}}}$.
    By our premise, we have $\Gamma |- \gt : \g{(y : T_1)->T_2} $, so by \cref{lem:weak},
  we have $(\gx : \g{T'_1})\Gamma |- \gt : \g{(y : T_1)->T_2}$.
  \rrule{Cast} gives us $(\gx : \g{T'_1})\Gamma |- \cast{T'_1}{T_1}\gx : \g{T_1}$,
  so \rrule{App} gives $(\gx : \g{T'_1})\Gamma |- \g{t\ (\cast{T'_1}{T_1}\gx)} : [\cast{T'_1}{T_1}\gx/\gy] \g{T_2}$. Then, \rrule{Cast} gives \\$(\gx : \g{T'_1})\Gamma |- \g{\cast{T'_2}{{\ix{[\cast{T'_1}{T_1}\gx / \gy]\g{T_2}}}}(t\ (\cast{T'_1}{T_1}\gx))} : \g{T'_2} $.
  Finally, applying \rrule{Lam} yields our result.
  }

  \lcase{Steps by $\bB\bB$, \textsc{Type}, or $??$, $\g{t_{start}}$ typed with \rrule{Cast}}{We can produce the sub-derivation for the term being cast.}
\end{proofEnd}

These together allow us to show type safety for \clang:

\begin{theorem}[Type Safety]
  If $\Gamma |- \gt : \gT$, then either $\gt$ diverges, or $\gt \stepstostar \gv$
  where $\Gamma |- \gv : \gT$.
\end{theorem}

\begin{theoremEnd}[apxlem]{lemma}[Arrow Typing Inversion]
  \label{lem:pi-inversion}
  If $\Gamma |- \g{(x : T_1)->T_2} : \g{T_3}$,
  then $\Gamma |- T_1 : \gType{\ell_1}$ and
  $(x : T_1)\Gamma |- \g{T_2} : \gType{\ell_2}$,
  where $\g{T_3} $ is convertible with $\gType{}_{\max(\g{\ell_1},\g{\ell_2})}$.
\end{theoremEnd}
\begin{proofEnd}
  By induction on the typing derivation.

  \rlcase{Conv}{Follows from our hypothesis, along with the contextual stepping rules for arrow types.}

  \rlcase{Pi}{Immediate.}

\end{proofEnd}

Type safety is not the only important property of a gradual language.
We prove the gradual guarantees
and decidability of type-checking in \cref{sec:precision,sec:translation} respectively.

\section{Surface Typing and Elaboration}
\label{sec:elab}

The trickiness of typing \surflang is that we want to consider types equivalent whenever they have consistent normal forms. \surflang does not have to specify its dynamic semantics: this was precisely
why we defined \clang. Following \citet{bertrand:gcic}, we express the typing of \surflang \textit{terms}
in terms
of \clang \textit{types}. To type applications of dependent functions, we \textit{elaborate}
\surflang terms into \clang, which are substituted into the appropriate types.

\begin{figure}
  \begin{boxedarray}{@{}l@{}}
    \boxed{\Gamma |- \esynth{\gs}{\gT}{\gt}\quad \Gamma |- \echeck{\gs}{\gT}{\gt}\quad\Gamma |- \etype{\gS}{\ell}{\gT} \textit{(Synthesis, Checking and Level Synthesis)  }}\\
    \begin{inferbox}

      \rrule{EVar,EType,EBool,ETF,ELam,EIf}\text{: straightforward adaptations from \slang }

      \inferrule[EAscr\ (EPi \textup{\textit{ similar}})]{
        \Gamma |- \etype{\gS}{\ell}{\gT}\\\\
        \Gamma |- \gT \etasteps \gV : \gType{\ell}\\\\
        \Gamma |- \echeck{\gs}{\gV}{\gt}}
      {\Gamma |- \esynth{\g{s :: S}}{\gT}{\gt}}
    {
      
    \inferrule[EConsistent]{
      \Gamma |- \esynth{\gs}{\g{T}}{\gt}\\\!\!\!\!
      \Gamma |- \g{T} => \gType{\ell}\\\\
      \Gamma |- \g{T} \etasteps \g{V'} : \gType{\ell}\\\\
      \Gamma |- \gV : \gType{\ell}\\\!\!\!\!
      {\g{V} =-= \g{V'}}
    }{\Gamma |- \echeck{ \gs}{\gV}{\cast{V'}{V}\gt} \!}
    }

    \inferrule[EConv]{
      \Gamma |- \esynth{{\gs}}{\gT}{\gt}\\\\
      \Gamma |- \gt => \gType{\ell}\\\\
      \Gamma |- \gT \etasteps \gV : \gType{\ell}\quad
      \gV =-= \gV \\
    }{\Gamma |- \echeck{{\gs}}{\gV}{\gt}}

    {
      \inferrule[ELamUnk]{
        (\gx : \qml)\Gamma |- \echeck{ \gs}{\qml}{ \gt}
      }
      {\Gamma |- \echeck{\g{\lambda x \ldotp s }}{\qml}{ \cast{\qml\!->\!\qml}{\qml}\g{(\lambda x : \qml \ldotp t)}}}
    }

    {
      \inferrule[EUnk]{\Gamma |- \gV => \gType{\ell}}
      {\Gamma |- \echeck{\qmat{\ell}}{\gV}{\qm_{\gV}}}
    }

    {
    \inferrule[EAppUnk]{
      \Gamma |- \esynth{\!\g{s_0}\!}{\!\g{T_0}\!}{\! \g{t_0}\!} \quad
      \g{T_0} \!\stepstostar\! \qml \quad
      \Gamma |- \echeck{\!\g{s_1}\!}{\!\qml\!}{\! \g{t_1}}
    }{\Gamma |- \esynth{\g{s_0\ s_1}}{\qml}{ \g{(\cast{\qml}{\qml\!\!->\!\!\qml}t_0)\ t_1}}}
    }

    \inferrule[ELevel]{
      \Gamma |- \esynth{\gS}{\g{T'}}{ \gT}\\
      \gT \!\stepstostar\! \gType{\ell}
      }
      {\Gamma |- \etype{\gS}{\ell}{\gT}}

      \inferrule[EApp]{
        \Gamma |- \esynth{\g{s_0}}{\g{T_0}}{ \g{t_0}} \\\!\!\!\!
        \Gamma |- \g{T_0} => \gType{\ell} \\\\
        \Gamma |- \g{T_0} \!\etasteps\! \g{(x : V_1) -> V_2}\\
        \Gamma |- \echeck{\g{s_1}}{\g{V_1}}{\g{t_1}}
    }{\Gamma |- \esynth{\g{s_0\ s_1}}{[\g{t_1} / \gx]\g{V_2}}{\g{t_0\ t_1}}}

    {
      \inferrule[EUnkLevel]{
        \Gamma |- \esynth{\gS}{ \g{T'}}{ \gT}\\
      \gT' \stepstostar \qm_{\gType{1+\ell}}}
        {\Gamma |- \etype{\gS}{\ell}{ \cast{\qm_{\gType{1+\ell}}}{\gType{\ell}}\gT}}
    }
  \end{inferbox}
\end{boxedarray}
  \caption{\surflang: Elaboration}
  \label{fig:elab}
\end{figure}

We combine the typing and elaboration of \surflang terms.
Once again, we use bidirectional typing.
 $\Gamma |- \echeck{\gs}{\gV}{\gt}$ means that
surface term $\gs$ elaborates to \clang term $\gt$ when checked against normalized \clang type $\gV$,
while $\Gamma |- \esynth{\gs}{\gT}{\gt}$ expresses that $\gs$ elaborates to $\gt$ while
synthesizing (possibly non-normal) type $\gT$.

In several places, we must check that a given \surflang term is
a type.
However, in \surflang, such a term might synthesize $\qm_{\gType{1+\ell}}$,
which is consistent with, but not equal to, $\gType{\ell}$.
We write
 $\Gamma |-  \etype{\gs}{\ell}{\gt}$ to
check that a term's type is consistent with $\gType{\ell}$ for some $\g\ell$,
with elaboration $\gt$, where the level $\g{\ell}$ is synthesized.

The elaboration rules are given in \cref{fig:elab}.
We omit the direct adaptations from \slang,
including \rrule{ELam} and \rrule{EIf} which use the type against which they check
to generate the elaboration's ascriptions.

In \rrule{EUnk}, $\qmat{\ell}$ checks against any type with the correct level.
        The level ascription is necessary to avoid Girard's paradox~\citep{Coquand86}.
\rrule{EConv} checks a term against the type it synthesizes, and
\rrule{EConsistent}
checks a term against a type whose normal form is consistent with its synthesized type.
We insert the appropriate cast to give the elaborated term the correct type.
\rrule{EConv} is redundant compared with \rrule{EConsistent}, but can be seen as an optimization
since it generates no casts.
\rrule{EApp} works like \rrule{SApp}, but substitutes the elaboration
of the argument to obtain the correct return type.
The rules \rrule{EAppUnk} and \rrule{ELamUnk} account for the fact that functions may have type $\qml$,
since it is consistent with function types. They insert the appropriate casts to $\g{\qm -> \qm}$
in their elaborations. Likewise, \rrule{EAscr} checks that the ascription's type is
consistent with $\gType{\ell}$ for some $\g\ell$.

With elaboration defined in terms of \clang types, we define
typing for \surflang terms and types:
{\small
\begin{displaymath}
  \inferrule{\cdot |- \etype{\gS}{\ell}{\gT} \\
    \cdot |- \echeck{\gs}{\gT}{\gt}
  }
{\cdot |- \gs : \gS}
\end{displaymath}
}
A \surflang term is well typed if its type elaborates at some level,
and the term elaborates and checks against the elaborated type.
The rules of \cref{fig:surface-type-lemmas} hold as lemmas for this relation,
with the minor modification that types are elaborated before they are added to the type environment.

The key property of elaboration is that it produces well typed
\clang expressions.

\begin{theoremEnd}[apxlem]{lemma}[Conversion Inversion]
  \label{lem:conv-inv}
  If $\Gamma |- \gt <= \gT$, then $\Gamma |- \gt => \g{T'}$ such that $\Gamma |- \gT \etasteps \gV$
  and $\Gamma |- \g{T'} \etasteps \gV$.
\end{theoremEnd}
\begin{proofEnd}
  \rrule{EConv} is the only checking rule for \clang, so the result follows from inversion.
  \end{proofEnd}

\begin{theoremEnd}[apxproof]{lemma}[Elaboration Preserves Typing]
  \label{lem:elab-typing}
  If $\Gamma |- \echeck{ \gs}{\gT}{ \gt}$ or $\Gamma |- \esynth{\gs}{\gT}{ \gt}$,
  then $\Gamma |- \gt <= \gT$.
  Likewise, if $\Gamma |- \etype{\gS}{\ell}{\gT}$,
  then $\Gamma |- \gT <= \gType{\ell}$.
\end{theoremEnd}
\begin{proofEnd}
  By induction on the typing derivation.

  \rlcase{EVar, EType, EBool, ETF, EUnk}{Immediate.}

  \rlcase{EPi, EIf, ELam, ELevel}{Immediately follows from the inductive hypothesis.}

  \rlcase{EAscr}{From IH we have $\Gamma |- \gt <= \gV$
    so by \cref{lem:conv-inv} we have $\Gamma |- \gt => \g{T'}$ where $\g{T'} \stepstostar \gV $.
    Since we know $\g{T} \etasteps \g{V}$, we can apply \rrule{CConv}
    to obtain our result.
  }

  \rcase{EApp}{From IH, we have that $\Gamma |- \g{t_0} <= \g{T_0}$,
    so by \cref{lem:conv-inv} and \cref{lem:confluence} we have $\Gamma |- \g{t_{0} => \g{T'_{0}}}$
    where $\g{\Gamma |- \g{T'_{0}} } \etasteps \g{(x : V_{1}) -> V_{2}}$.
    By our IH, we have $\Gamma |- \g{t_{1}} <= \g{V_{1}}$,
    giving us everything we need to use \rrule{CApp} to produce our result.
  }

  \rcase{EAppUnk}{IH gives that $\Gamma |- \g{t_0} => \g{T_0}$, \cref{lem:conv-inv}
    gives $\Gamma |- \g{t_0} => \g{T'_{0}}$ where $\Gamma |- \g{T'_{0}} \etasteps \qml $.
    By \rrule{CConv} we have $\Gamma |- \g{t_{0}} <= qml$.
    Then \rrule{CCast} gives that $\Gamma |- \cast{\qml}{\qml -> \qml}\g{t_{0}} => \g{\qml -> \qml}$.
    Finally, IH gives that $\Gamma |- \g{t_1} <= \qml$, so we can use \rrule{CApp}
    to obtain our result.
  }

  \rcase{ELamUnk}{
    IH gives $(\gx : \qml)\Gamma |- \gt <= \qml$, so \rrule{CLam} gives $\Gamma |- \g{(\lambda x : \qml \ldotp t)} : \g{\qml -> \qml}$. Finally, \rrule{CCast} gives us our result.
  }

  \rcase{EUnkLevel}{
    IH gives $\Gamma |- \gt <= \gT$, so applying \cref{lem:conv-inv} and \rrule{CConv} along with our premise gives
    $\Gamma |- \gt <= \qm_{\gType{1+\ell}}$.
    Both $\gType{\ell}$ and $\qm_{\gType{1+\ell}}$ have type $\gType{1+ell}$, so \rrule{CCast}
    gives $\Gamma |- \cast{\qm_{\gType{1+\ell}}}{\gType{\ell}}\gt <= \gType{\ell}$, which is our goal.
  }

  \rcase{EConsistent}{Our IH gives us $\Gamma |- \gt <= \g{T}$, so \cref{lem:conv-inv}
    gives $\Gamma |- \gt => \g{T'}$ where $\Gamma |- \g{T'} \etasteps \g{V'}$.
    So by \rrule{CConv} we have $\Gamma |- \gt <= \g{V'}$
    Then by \rrule{CCast},
    $\Gamma |- \cast{V'}{V}\gt <- \g{V}$, giving us our result.
  }

  \rlcase{EConv}{Same as above, but no need to apply \rrule{CCast}}
\end{proofEnd}

We also want \slang to conservatively extend \surflang.
There is a trivial embedding
of \slang into \surflang, since each syntactic construct of \slang
appears in \surflang.

Likewise, there is a direct

embedding of
well-typed \slang terms into
\clang.
We write $\lceil \st \rceil$ and $\lfloor  \st \rfloor$
to denote the embeddings of $\st $ from \slang into \clang and \surflang respectively.

\begin{theoremEnd}[apxproof]{lemma}[Conservative Extension of \slang]
  $\cdot |- \st <= \sV $ iff $ \cdot |- \lceil \st \rceil : \lfloor  \sV \rfloor$. Moreover, if $\cdot |- \st <= \sV $
  and $\st \etasteps \sv : \sV$ then $\lfloor \st \rfloor \etasteps \lfloor \sv \rfloor : \lfloor \sV \rfloor$
 \end{theoremEnd}
\begin{proofEnd}
  Let:
  \begin{itemize}
\item $\gs = \lceil \st \rceil$
\item $\gS = \lceil \sT \rceil$
\item $\gT = \lfloor \sT \rfloor$ be the embedding of $\sT$ into \clang;
\item $\Gamma'$ be $\Gamma$ with $\lfloor \_ \rfloor$ applied to each contained type.
\end{itemize}

  We simultaneously prove that  $\Gamma |- \st => \sT $ iff $\Gamma' |- \gs => \gT$,
  and that $\Gamma |- \sT : \sType{}_{=>\s\ell}$ iff
  $\Gamma |- \gS : \gType{}_{=>\g\ell}$.

  For proving that $\st$ well typed implies $\gs$ well typed, we note that each rule of \slang
  has a corresponding rule in \clang that is identical. Moreover, none of these rules
  insert casts, so each reduction of the $\st$ is directly simulated by the elaboration
  of $\gs$, modulo the removal of type ascriptions (which \clang does not use).
  In particular, the approximations of \clang only occur when casting to $\qml$, so they
  never arise for fully static terms.

  For proving the opposite direction, the reasoning is similar: if $\qm$ does not occur in $\gs$,
  then each rule used directly corresponds to a static rule. In particular, if $\qm$ does not occur in $\gT$,
  then there are no cases in which \rrule{EConsistent} is used that \rrule{EConv}
  cannot instead be used. This means that if $\gs$ is fully static, then its elaboration
  contains no casts, so we  simulate the reductions of its elaboration by the reductions
  of $\st$.

  For the semantics, there is a direct simulation, since each reduction step in \slang
  is avaliable in clang.

  \je{More detail?}
\end{proofEnd}

In \cref{sec:translation} we prove strong normalization for \clang,
from which we show decidability of \surflang typing.

\begin{theoremEnd}[apxlem]{definition}[Well-formed Context]
  We say that a context is well formed if it is $\cdot$,
  or if it is $(\gx : \gT)\Gamma$, where $\Gamma |- \gT : \gType{\ell}$ for some $\g\ell$,
  and $\Gamma$ is well formed.
\end{theoremEnd}

\begin{theoremEnd}[apxlem]{lemma}[Well-formed Synthesis]
  \label{lem:elab-synth-wf}
  If $\Gamma |- \esynth{\gs}{\gT}{\gt}$ and $\Gamma$ is well-formed, then $\Gamma |- \gT : \gType{\ell}$
  for some $\ell$. Moreover, this $\gT$ is unique.
\end{theoremEnd}
\begin{proofEnd}
  The uniqueness follows from the syntax-directedness of synthesis.

  We show typing by induction on the synthesis derivation.

  \rlcase{EVar}{From the well-formedness of $\Gamma$.}

  \rlcase{EType, EBool, ETF, EPi, EAppUnk}{trivial}.

  \rlcase{EAscr}{ Follows from \cref{lem:elab-typing}.}

  \rlcase{EApp}{Follows from our IH on the typing of $\g{s_0}$,
  inversion on the typing of $\g{(x : T_1) -> T_2}$,
  applying \cref{lem:elab-typing} to $\g{t_1}$, and
  \cref{lem:subst}.}

\end{proofEnd}

\begin{theoremEnd}[apxlem]{lemma}[Well-formed Premises]
  \label{lem:wf-premise}
  Given a derivation $\mathcal{D} :: \Gamma |- \echeck{\gs}{\gT}{\gt}$,
  where $\Gamma$ is well-formed (see \cref{lem:elab-synth-wf}),
  and $\Gamma |- \gT : \gType{\ell}$, then
  the environments for the premises of $\mathcal{D}$ are well-formed,
  and
  each type in $\mathcal{D}$ has type $\gType{\ell}$ for some $\g\ell$ in its environment.
\end{theoremEnd}
\begin{proofEnd}
  By cases on the derivation.

  \rlcase{EVar}{Follows from the well-formedness of $\Gamma$}.

  \rlcase{EType, EBool, ETrue, EUnk, ELamUnk}{ trivial.}

  \rlcase{EAscr}{Environments trivial. Typing of $\gT$ follows from \cref{lem:elab-typing} and \cref{lem:preservation}.}

  \rlcase{EPi}{Well-formedness of $(\gx : \g{S_1})\Gamma$ follows from premise and \cref{lem:elab-typing}}.

  \rcase{EApp}{$\Gamma |- \g{T_0} : \gType{\ell}$ by \cref{lem:elab-synth-wf},
    and \cref{lem:preservation} gives that $\Gamma |- \g{(x : T_1) : T_2} : \gType{\ell}$.
    Inversion on this derivation gives the well-typedness of $\g{T_1}$.
    \je{TODO: what type, since might be smaller!}
  }

  \rlcase{EAppUnk}{Same as above, except that $\qml$ is trivially well-typed.}

  \rlcase{EConsistent, EConv}{Follows from premises, \cref{lem:preservation} and \cref{lem:elab-synth-wf}.}

  \rlcase{EIf}{Follows from our assumption that $\gT$ is well typed.}

  \rcase{ELam}{\cref{lem:pi-inversion}
    gives that $\g{T_1}$ is well typed with level at most $\g\ell$, so
    extending $\Gamma$ with $\g{T_1}$ preserves well-formedness. Likewise, the same lemma
    gives the typing of $\g{T_2}$ under $(\gx : \g{T_1})\Gamma$.
  }

  \rlcase{ELamUnk}{Follows from the well-typedness of $\qml$.}
\end{proofEnd}

\begin{theoremEnd}[apxproof]{lemma}[Decidable Type Checking]
  Given $\gs$ and $\gS$, it is decidable whether $\cdot |- \gs : \gS$.
\end{theoremEnd}
\begin{proofEnd}
  We instead prove the stronger statement, that the following are decidable:
  \begin{itemize}
  \item Given $\Gamma, \gs, \gT$, where $\Gamma$ is well formed (see \cref{lem:elab-synth-wf})
    and $\Gamma |- \gT : \gType{\ell}$, does there exist $\gt$ such that
    $\Gamma |- \echeck{\gs}{\gT}{ \gt}$;
  \item Given $\gs$ and well-formed $\Gamma$, are there $\gt, \gT$ such that
    $\Gamma |- \esynth{\gs}{\gT}{ \gt}$;
  \item Given $\gs$ and well-formed  $\Gamma$, is there some $\g\ell$ such that $\Gamma |- \g{s} : \gType{}_{=>\g\ell} $.
  \end{itemize}
  Given these, we decide whether $\gs : \gS$ by deciding whether $\cdot |- \gS : \gType{}_{=>\g\ell}$
  for any $\g\ell$, and if it does, deciding whether $\Gamma |- \echeck{ \gs}{\gT}{ \gt}$ for any $\gt$.

  The key facts are that:
  \begin{itemize}
  \item The rules are syntax directed (with a few one-step lookaheads), so at most one rule
    can apply to a given term for synthesis, or term/type combination for checking. If we decide that any premise of this rule cannot hold, then the judgment.
          (The exceptions are the pair \rrule{EConsistent} and \rrule{EConv}, and the pair \rrule{EApp} and \rrule{EAppUnk}. For \rrule{EConsistent} and \rrule{EConv} we check if the synthesized type's normal form $\g{V'}$
          is equal to the $\gV$ against which we are checking. For \rrule{EApp} and \rrule{EAppUnk}
          we check if the function synthesizes $\qml$ or an arrow type.)
  \item Any side-conditions involving $\stepstostar$ or $\etasteps$ are decidable, by our
    assumption of strong normalization.
  \item The proofs of progress and preservation, along with \cref{lem:elab-typing},
    \cref{lem:elab-synth-wf} and \cref{lem:wf-premise} are all constructive.
  \end{itemize}

We proceed by cases on which judgment were are deciding along with induction on the shape of $\gs$.

  \mlcase{(=>, \gx)}{Must use \rrule{EVar}, just look in $\Gamma$}

  \mlcase{(=>, \gType{\ell}, \g{\bB}, \g{true}, \g{false})}{Only one rule in each case, and it synthesizes under any context.}

  \mcase{(=>, \gs :: \gS)}{
    \cref{lem:elab-typing} gives us a typing derivation for $\gT$,
    and \cref{lem:preservation} lets us transform this into a derivation for $\gV$
  which allows us to apply our hypothesis and decide $\Gamma |- \echeck{ \gs}{\gV}{ \gt}$. }

  \mlcase{(=>, \g{(x : S_1)-> S_2})}{Must use \rrule{EPi}, can use IH to decide premises.}

  \mlcase{(=>, \g{s_0\ s_1})}{Must use \rrule{EApp} or \rrule{EAppUnk} .
    Our IH lets us decide whether $\g{s_0}$ synthesizes a type, and we use strong normalization
    to decide if it converts to an arrow or $\qml$. In the former case, \cref{lem:elab-synth-wf}, along with preservation
    and \cref{lem:pi-inversion} give us that $\g{V_1}$ is well-typed, allowing us to use IH
    to decide whether $\g{s_1}$ checks against $\g{V_1}$.
    In the latter case, we only need  check if $\g{s_1}$ checks against $\qml$,
    which we can do by our hypothesis.
  }

  \mlcase{<= \gT, \g{if\ s_1\ s_2\ s_3}}{Must use \rrule{EIf}, check each premise by IH. Well-typing of $\gT$ follows from premise.}

  \mlcase{<= \g{(x : T_1) -> T_2}, \g{\lambda x \ldotp s}}{Must use \rrule{ELam}, just use IH to check premise.
  \Cref{lem:pi-inversion} gives the well-formedness/typing of the type and environment.}

  \mlcase{<= \qml, \g{\lambda x \ldotp s}}{Must use \rrule{ELamUnk}, just use IH to check premise.}

  \mlcase{<= \gT, \qmat{\ell}}{Must use \rrule{EUnk}. We simply then examine the derivation of typing for $\gT$ (which we have by our assumption) to compare the levels.
  \je{TODO: level uniqueness?}}

  \lcase{$<= \gT$, all other cases}{Must use \rrule{EConsistent} or \rrule{EConv}. To determine which one, we synthesize $\g{T'}$ for $\gt$ by IH, normalize it and and compare with $\gV$ for $\alpha$-equivalence.
    By our premise, we already have the typing derivation for $\gV$, and we can obtain the
    derivation for $\g{V'}$ from \cref{lem:elab-synth-wf} and \cref{lem:progress}, which allows us to determine what level
    they type against. Then we decide whether the normal form types are equal or consistent, since if not,
    we cannot possibly apply either rule.
  }

  \mlcase{: \gType{}_{=>\g\ell}, \gS}{We simply synthesize the type for $\gS$, normalize, and check if it is $\gType{\ell}$ or $\qm_{\gType{1+\ell}}$}.

\end{proofEnd}

As a final consideration, we describe exact run-time semantics for \surflang programs.
Approximate normalization is well-suited to normalizing types, but the actual execution of
programs should be exact. Thankfully, exact execution is obtained with simple modifications
to \clang: \rrule{CLamUnk} lets tagged functions refer to their bodies, \rrule{$?\Pi$} is removed, and \rrule{To?}
is relaxed to work on function types. The key to decidable type-checking is that
elaboration (\cref{fig:elab}) and \clang typing (\cref{fig:clang-types}) still use
approximate normalization to normalize types.

\section{Precision, Consistency, and the Gradual Guarantees}
\label{sec:precision}

The elaboration from \surflang to \clang refers to a consistency relation $=-=$
that extended equality to account for imprecision.
However, defining such a relation raises several challenges.
There is a delicate balance between ensuring that consistency is decidable,
proving the gradual guarantees, and keeping \clang simple to prove type
checking terminates.
Our choices must be flexible enough to accommodate the evidence of equality from \cref{sec:equality}.
In this section, we define a precision relation on \clang terms,
and use it to establish the relationship between the syntactic precision
of \surflang terms and the consistency of their \clang types.
We begin by explaining the delicate balance between the above criteria,
and summarizing our strategy for achieving that balance. We then define
precision and consistency relations for \clang (\cref{subsec:precision,subsec:consistency}) and
establish the relationship between them (\cref{subsec:sgg}),
which yields the gradual guarantees as a corollary.

\subsection{Balancing Decidability and the Gradual Guarantees}

In most gradually-typed languages, consistency is defined
as equality up to occurrences of $\qm$. However, because of the type-term overlap in \surflang
and the desire to consider the consistency modulo normalization,
elaboration to \clang uses \clang types, for which we have a dynamic semantics.
But \clang has casts, which our
consistency relation must take into account.

Moreover, in order for the static gradual guarantee (SGG) to hold,
there must be an intimate relationship between consistency and precision.
For non-dependent gradual types, type precision can be defined in terms of consistency:
$\g{T} \sqsubseteq \g{T'} \triangleq \forall \g{T''} \ldotp \g{T'' =-= T'} \Rightarrow \g{T''} =-= \g{T}$.
The SGG follows from showing that syntactically precision-related terms
have precision-related types.
However, dependent types break this approach: types may contain terms, so we need
a precision relation for arbitrary terms, that a priori preserves consistency as precision is lost.
Rules like \rrule{EConsistent} normalize terms, so
precision-related terms must relate to precision-related values \ie
the dynamic gradual guarantee (DGG) must hold.
\rrule{EApp} uses elaborated terms in the synthesized types, so precision-related \surflang terms
must have related elaborations.

However, the syntactic approach does not scale to gradual dependent types, particularly
when we need to treat terms as evidence of equality, and compose that evidence.
Consider $\rmeet{(1+x)}{(1+y)}{T}$, the composition of $\g{1 + x}$ and $\g{1 + y}$.
For this to be valid evidence that $\g{1+x}$ and $\g{1+y}$ are consistent, it should be more precise than either,
but this does not hold with syntactic precision. Precision must reflect the \textit{behavioral} property that the composition is more precise for \textit{any} $\g{x}$ and $\g{y}$.
Similarly, we need some behavioral properties of casts to show that elaboration
is monotone in the elaborated term.
However, precision cannot include \textit{all} behavioral properties: if precision relates all observationally equivalent terms,
then all observationally equivalent terms should be consistent, which would either make consistency undecidable or
too coarse to exclude distinct static terms.

We design a new precision relation to compare
        cast-calculus terms, called \textit{algebraic precision} because, in addition
        to the usual syntactic rules, it has axioms describing the behavior of casts
        and compositions.
        We show that the elaborations of syntactic precision-related terms are related by algebraic precision.
        These in turn evaluate to precision-related terms \ie the dynamic gradual guarantee.
        Reducing a term's algebraic precision keeps it consistent
        with everything the original term was consistent with, which lets us prove the static
        gradual guarantee.

Moreover, algebraic precision provides a principled
        framework for reasoning about the precision of casts. The return type of a dependent function application depends on the value of its argument, so terms related by syntactic precision may not have the same type.
        With algebraic precision such terms are compared by casting one to the type of the other:
        the algebraic laws establish a Galois connection between these casts,
        so precision holds regardless of  which term we cast.

\subsection{Precision}
\label{subsec:precision}

\begin{figure}
	\begin{boxedarray}{@{}l@{}}
		\boxed{\Gamma |- \g{t_1} \squbc \g{t_2} \textit{ (Algebraic Precision: Generating and Structural Rules)}}
		\\
		\begin{inferbox}
		  \inferrule[PrecGenUnk]{
			\Gamma |- \gt \psynth \g{V}\\\\
			\Gamma |- \gT \etasteps \g{V}
		  }{\Gamma |- \gt \squbc \qm_{\gT}}

		  \inferrule[PrecGenErr]{
			\Gamma |- \gT \etasteps \g{V}\\\\
			\Gamma |- \gt \psynth \g{V}
		  }{\Gamma |- \err_{\gT} \squbc \gt}

			\inferrule[PrecCongApp]{
				\Gamma |- \g{t_1} \psynth  \g{(x : V_1) -> V_2} \\
				\Gamma |- \g{t'_1} \psynth \g{(x : V_1) -> V_2}\\\\
				\ix{[\g{t_2}/ \gx]\g{V_2}} \stepstostar \g{T}\\
				\ix{[\g{t'_2}/ \gx]\g{V_2}} \stepstostar \g{T'}\\\\
				\Gamma |- \g{t_1} \squbc \g{t'_1} \\
				\Gamma |- \g{t_2} \squbc \g{t'_2}
			}
			{\Gamma |-
			  \cast{\g{T}}{\g{T'}}\g{(t_1 \ t_2)}
			  \squbc \g{t'_1 \ t'_2}}

			\inferrule[PrecCongPi]{
				\Gamma |- \g{T_1} \squbc \g{T'_1} \\
				(\gx : \g{T'_1})\Gamma |- [\cast{T'_1}{T_1}\gx / \gx]\g{T_2} \squbc \g{T'_2}
			}
			{\Gamma |- \g{(x : T_1) -> T_2} \squbc \g{(x : T'_1) -> T'_2}}

			\inferrule[PrecCongCast]{
				\Gamma |- \g{t} \squbc \g{t'}
			}
			{\Gamma |- \cast{T_1}{T_2}\gt \squbc \cast{T_1}{T_2}\g{t'}}

			\inferrule[PrecCongLam]{
				(\gx : \gT)\Gamma |- \g{t_1} \squbc \g{t_2}
			}{ \Gamma |- \g{\lambda x \!:\!\! T \ldotp t_1} \squbc \g{\lambda x \!:\!\! T \ldotp t_2}}

			\inferrule[PrecCongTag]{\Gamma |- \g{t_1} \squbc \g{t_2}}{\Gamma |- \attagl{h}{t_1} \squbc \attagl{h}{t_2}}

			\inferrule[PrecCongIf]{
				\Gamma |- \g{t_1} \squbc \g{t'_1}\quad
			    \Gamma |- \g{t_2} \squbc \g{t'_2}\quad
				\Gamma |- \g{t_3} \squbc \g{t'_3}
			}
			{\Gamma |- \g{if_{\gT}\ t_1\ t_2\ t_3} \squbc \g{if_{\gT}\ t'_1\ t'_2\ t'_3}}
		\end{inferbox}
	  \end{boxedarray}
	\caption{Algebraic Precision for \clang: Generating and Structural Rules}
	\label{fig:ax-precision-cong}
\end{figure}

\begin{figure}
	\begin{boxedarray}{@{}l@{}}
		\boxed{\Gamma |- \g{t_1} \squbc \g{t_2} \textit{ (Algebraic Precision, axiomatic rules)}}\
		\boxed{\Gamma |- \g{t_1} \squbstar \g{t_2} \textit{ (Precision Modulo Normalization)}}
		\\
		\begin{inferbox}
			\inferrule[PrecAxRefl]{ }{
				\gt \squbc \gt}

			\inferrule[PrecAxEta]{ }
			{\Gamma |- \cast{(x : T_1)->T_2}{(x : T'_1)->T'_2}\gt
			  \squbc \g{\lambda x \ldotp \cast{T_2}{T'_2}(\gt\ \cast{T'_1}{T_1} \gx)}}

			\inferrule[PrecAxCastBot]{ }
			{\cast{T_1}{T_2} \err_\g{T_1} \squbc \err_{\g{T_2}}}

			\inferrule[PrecAxTrans]{
				\Gamma |- \g{t_1} \squbc \g{t_2}\\
				\Gamma |- \g{t_2} \squbc \g{t_3}
			}{
				\Gamma |- \g{t_1} \squbc \g{t_3}
			}

			\inferrule[PrecAxUpDown]{
				\Gamma |- \g{T_1} \squbstar \g{T_2}\\
			}
			{ \Gamma |- \gt \squbc \cast{T_2}{T_1}\cast{T_1}{T_2}\g{t} }

			\inferrule[PrecAxDownUp]{
				\Gamma |- \g{T_1} \squbstar \g{T_2}\\
			}
			{ \Gamma |- \cast{T_1}{T_2}\cast{T_2}{T_1}\g{t} \squbc \gt }

			\inferrule[PrecAxIntermed]{\Gamma |- \g{T_1} \squbstar \g{T_2}}
			{\Gamma |- \cast{T_1}{T'}\cast{T}{T_1}\gt \squbc \cast{T_2}{T'}\cast{T}{T_2}\gt}

			\inferrule[PrecAxComposeUp]{\Gamma |- \g{T_1} \squbstar \g{T_2} \squbstar \g{T_3}}
			{\Gamma |- \cast{T_2}{T_3}\cast{T_1}{T_2}\gt \sqeqc \cast{T_1}{T_3}\gt}

		  \inferrule[PrecNorm]{\g{T_1} \stepstostar \g{V_1} \\
			\g{T_2} \stepstostar \g{V_2} \\
			\g{V_1} \squbc \g{V_2}
		  }
		  {\Gamma |- \g{T_1} \squbstar \g{T_2}}
		\end{inferbox}
	  \end{boxedarray}
	\caption{Algebraic Precision for \clang: Axioms and Normalization}
	\label{fig:ax-precision-ax}
\end{figure}

\Cref{fig:ax-precision-cong,fig:ax-precision-ax} define algebraic precision
by generating rules, structural rules, and axioms.
We use $\psynth$ as a shorthand for synthesizing then normalizing a type.
$\qm_{\gT}$ and $\err_{\gT}$
are the least and greatest terms with type $\gT$ (\rrule{PrecGenUnk}, \rrule{PrecGenErr}).
\rrule{PrecCongLam}, \rrule{PrecCongTag} and \rrule{PrecCongIf} are straightforward structural rules.
\rrule{PrecCongCast},
allows precision related terms \textit{for casts of the same source and destination type}.
Casts with different sources are compared with the axiomatic rules below.
In \rrule{PrecCongApp},  reducing the precision of the argument may reduce the precision
of the application's type, so we compare the result of application after
casting to the less precise type.
\rrule{PrecCongPi} performs a similar cast, since the bound variables may have different types.

The axiomatic rules provide the essential properties for precision.
\rrule{PrecAxUpDown} and \rrule{PrecAxDownUp} establish a Galois connection
for casts between precision-related types.
Casting to a more precise type,
then back, produces the original term, but (possibly) more precise,
since new errors may have been introduced, or ascribed type information made more precise.
Conversely, casting to a less precise type may lose precision due to an approximation.
In both cases, types must be normalized before comparing for precision.

We also have rules establishing the well-behavedness
of precision and casts.
Rules \rrule{PrecAxRefl} and \rrule{PrecAxTrans} provide reflexivity and transitivity,
which would otherwise be violated by the addition of axioms.

\rrule{PrecAxCompose} asserts that casting through an intermediate type is equivalent
to casting directly, so long as all casts are upcasts.
Similarly, \rrule{PrecAxIntermed} asserts that casting through an intermediate type
is monotone with respect to that intermediate type,
regardless of the source and destination.

\rrule{PrecAxCastBot} asserts that $\err$ is inescapable:
we cannot take $\err$ at one type and get a non-error term by casting it to a different type,
so $\err$ actually behaves like a dynamic error.
Finally, \rrule{PrecAxEta} relates a cast between function types to the $\eta$-expanded
version where the argument and result are cast separately.

\Cref{fig:ax-precision-admitted} gives some rules admitted by axiomatic precision,
which help to show why our axioms ensure the well-behavedness of casts.
For example, The rules \rrule{PrecAdAdjR} and \rrule{PrecAdAdjL}
establish that, when comparing terms of different types, it does not matter
whether we cast the more or less precise term.
\rrule{PrecAdTrivialCast} shows that casting from a type to itself produces an
equi-precise result, and \rrule{PrecAdIsoCast} shows that casts between equi-precise
types form an isomorphism (up to equi-precision).

\begin{figure}
	\begin{boxedarray}{@{}l@{}}
		\boxed{\g{t_1} \squbc \g{t_2} \textit{ (Algebraic Precision: Admissible Rules)}}
		\\
		\begin{inferbox}
  \mprset {fraction ={\cdot\cdots\cdot}}
			\inferrule[PrecAdAdjR]{
				\Gamma |- \g{T_1} \squbstar \g{T_2}\\
				\Gamma |- \g{t_1} \squbc \cast{T_2}{T_1}\g{t_2}
			}
			{ \Gamma |-\cast{T_1}{T_2}\g{t_1} \squbc \g{t_2} }

			\inferrule[PrecAdAdjL]{
				\Gamma |- \g{T_1} \squbstar \g{T_2}\\
				\Gamma |- \cast{T_1}{T_2}\g{t_1} \squbc \g{t_2}
			}
			{
				\Gamma |- \g{t_1} \squbc \cast{T_2}{T_1}\g{t_2}}

			\inferrule[PrecAdComposeDown]{\Gamma |- \g{T_1} \squbstar \g{T_2} \squbstar \g{T_3}}
			{\Gamma |- \cast{T_2}{T_1}\cast{T_3}{T_2}\gt \sqeqc \cast{T_3}{T_1}\gt}

			\inferrule[PrecAdIsoCast]{
				\Gamma |- \g{T_1} \sqeqstar \g{T_2} \\
			}
			{
			\Gamma |- 	\gt \sqeqc \cast{T_2}{T_1}\cast{T_1}{T_2}\gt
			  }

			\inferrule[PrecAdTrivialCast]{ }{
				\Gamma |- \gt \sqeqc \cast{T}{T}\gt
			}

			  \inferrule[PrecAdSimpleApp]{
				\Gamma |- \g{t_1} \psynth \g{V_1->V_2}\\
				\Gamma |- \g{t'_1} \psynth \g{V_1 -> V_2}\\
				\Gamma |- \g{t_1} \squbc \g{t'_1}\\
				\Gamma |- \g{t_2} \squbc \g{t'_2}
			  }
			  {\Gamma |- \g{t_1\ t_2} \squbc \g{t'_1\ t'_2} }
		\end{inferbox}
	\end{boxedarray}
	\caption{Some Admissible Rules Algebraic Precision}
	\label{fig:ax-precision-admitted}
\end{figure}

\begin{theoremEnd}[apxlem]{lemma}[Admissible Rules for Algebraic Precision]
	The rules in \cref{fig:ax-precision-admitted} are admissible given
	the rules in \cref{fig:ax-precision-cong,fig:ax-precision-ax}.
\end{theoremEnd}
\begin{proofEnd} We show the derivation tree for each rule.

	\rcase{PrecAdAdjR}{
		Our goal is $\cast{T_1}{T_2}\g{t_1} \squbc \g{t_2}$.
		Then $\cast{T_1}{T_2}\g{t_1} \squbc \cast{T_1}{T_2}\cast{T_2}{T_1}\g{t_2}$ by \rrule{PrecCongCast}.
		We get our goal by \rrule{PrecAxTrans} and \rrule{PrecAxDownUp}.
	}

	\rcase{PrecAdAdjL}{
		Our goal is $\g{t_1} \squbc \cast{T_2}{T_1}\g{t_2}$.
		By \rrule{PrecCongCast} we have $\cast{T_2}{T_1}\cast{T_1}{T_2}\g{t_1} \squbc \cast{T_2}{T_1}\g{t_2}$.
		Then, by \rrule{PrecAxTrans} and \rrule{PrecAxUpDown} we have our goal.
	}

	\rcase{PrecAdTopDown}{
		By \rrule{PrecAxTrans}, \rrule{PrecGenUnk} and our premise we have $\cast{T_1}{T_2}\qm_{T_1} \squbc \g{t}$, so by \rrule{PrecAdAdjL} we have our goal.
	}

	\rcase{PrecAdComposeDown}{
		We show the $\squbc$ case, noting that the $\sqsupseteq$ case is dual.
		Goal is $\cast{T_2}{T_1}\cast{T_3}{T_2}\gt \sqeqc \cast{T_3}{T_1}\gt$.
		By \rrule{PrecAdAdjL}, it suffices to prove $\cast{T_1}{T_3}\cast{T_2}{T_1}\cast{T_3}{T_2}\gt \squbc \gt$. We show this goal using three steps that we connect with two applications of \rrule{PrecAxTrans}.

		\begin{enumerate}
			\item By \rrule{PrecAxUpCompose} we have\\
			      $\cast{T_1}{T_3}\cast{T_2}{T_1}\cast{T_3}{T_2}\gt \squbc \cast{T_2}{T_3}\cast{T_1}{T_2}\cast{T_2}{T_1}\cast{T_3}{T_2}\gt$.
			\item By \rrule{PrecAxDownUp} we have\\
			      $\cast{T_1}{T_2}\cast{T_2}{T_1}\cast{T_3}{T_2}\gt \squbc \cast{T_3}{T_2}\gt$,
			      so by \rrule{PrecCongCast} we have
			      $\cast{T_2}{T_3} \cast{T_1}{T_2}\cast{T_2}{T_1}\cast{T_3}{T_2}\gt \squbc \cast{T_2}{T_3} \cast{T_3}{T_2}\gt $.
			\item By \rrule{PrecAxDownUp} we have
			      $\cast{T_2}{T_3}\cast{T_3}{T_2} \gt \squbc \gt$.
		\end{enumerate}
	}

	\rcase{PrecAdTrivialCast}{
		We show the $\squbc$ case, noting that the $\sqsupseteq$ case is dual.
		Our goal is $\gt \squbc \cast{T}{T}\gt$.
		By \rrule{PrecAxUpDown} we have $\gt \squbc \cast{T}{T}\cast{T}{T}\gt$,
		and by \rrule{PrecAxUpCompose} we have $\cast{T}{T}\cast{T}{T}\gt \squbc \cast{T}{T}\gt$,
		so \rrule{PrecAxTrans} gives us our goal.
	}

	\rcase{PrecAdIsoCast}{
		Our goal is $\gt \sqeqc \cast{T_2}{T_1}\cast{T_1}{T_2} \gt$.
		By \rrule{PrecAdTrivialCast} we get $\gt \sqeqc \cast{T_1}{T_1} \gt$,
		and by \rrule{PrecAxUpCompose} combined the equi-precision of
		$\g{T_1}$ and $\g{T_2}$, we get
		$\cast{T_1}{T_1} \gt \sqeqc \cast{T_2}{T_1}\cast{T_1}{T_2}\gt $.
		Our result then follows from \rrule{PrecAxTrans}.
	  }

	  \rlcase{PrecAdSimpleApp}{ We have $\cast{\ix{[\g{t_{1}}/\gx]\g{V_{2}}}}{\ix{[\g{t_{1}}/\gx]\g{V_{2}}}}\g{(t_{1}\ t_{2})}$
		by \rrule{PrecCongApp}. Then applying \rrule{PrecAdTrivialCast} and \rrule{PrecAxTrans}
		yields our result.
	  }

\end{proofEnd}

\subsection{Consistency}
\label{subsec:consistency}

\Cref{fig:consistency}  defines \textit{consistency}, specifying how we compare
terms in the \rrule{EConsistent} elaboration rule.
The rules come in three varieties.
First, we have structural rules for each syntactic construct (except casts).
Then, we have the \rrule{CstUnk} rules:
because we have an explicit error term $\err$, we want it to be inconsistent with \textit{all types},
including itself. So we cannot allow all types to be consistent with $\qm$, but must check that each
part is. These rules directly mirror the structural rules.

$\err$ is not consistent with any terms, even itself.
Finally, we have \rrule{CstCastL, CstCastR, CstTagL} and \rrule{CstTagR}, which ignore
casts when comparing for precision.

\begin{figure}
	\begin{boxedarray}{@{}l@{}}
	  \boxed{\g{v_{1}} =-= \g{v_{2}} \textit{ (Value Consistency)}}
	  \qquad\rrule{CstCastR, CstTagR} \textit{ defined symmetrically }
		\\
		\begin{inferbox}
			\inferrule[CstUnkBase]{\gv \textit{is one of } \qm_\g{V'}, \gx, \\\\\gType{\ell}, \g{\bB}, \g{true}, \g{false} }{\qm_{\gV} =-= \gv}

			\inferrule[CstUnkFlip]{\qm_{\gV} =-= \gv}{\gv =-= \qm_{\gV}}

			\inferrule[CstUnkPi]{
				\qm_{\gV} =-= \g{V_1}\\
				\qm_{\gV} =-= \g{V_2}
			}
			{\qm_{\gV} =-= \g{(x : V_1) \!\!->\!\! V_2}}

			\inferrule[CstUnkLam]{\qm_{\g{V_1}} =-= \g{v}}
			{
				\qm_{\g{V_1}} =-= \g{\lambda x : \gV \ldotp v}
			  }

			\inferrule[CstUnkApp]{
				\qm_{\gV} =-= \g{N}\\\\
				\qm_{\gV} =-= \g{v}
			}
			{\qm_{\gV} =-= \g{N\ v}}

			\inferrule[CstUnkIf]{
				\qm_{\g{\bB}} =-= \g{N}\\\\
				\qm_{\gV} =-= \g{v_1}\\
				\qm_{\gV} =-= \g{v_2}
			}
			{\qm_{\gV} =-= \g{if_V\ N\ v_1\ v_2}}

			\inferrule[CstBase]{ \gv \textit{is one of }\qm_\g{V'}, \gx, \\\\\gType{\ell}, \g{\bB}, \g{true}, \g{false} }{\gv =-= \gv}

			\inferrule[CstIf]{
				\g{v_1} =-= \g{v'_1}\\\\
				\g{v_2} =-= \g{v'_2}\\
				\g{v_3} =-= \g{v'_3}
			}
			{\g{if_{V}\ v_1\ v_2\ v_3} =-= \g{if_{V'}\ v'_1\ v'_2\ v'_3}}

			\inferrule[CstApp]{\g{N} =-= \g{N'}\\\\
				\g{v} =-= \g{v'}
			}
			{\g{N\ v} =-= \g{N'\ v'}}

			\inferrule[CstPi]{
				\g{V_1} =-= \g{V'_1}\\
				\g{V_2} =-= \g{V'_2}
			}
			{\g{(x : V_1) \!\!->\!\! V_2} =-= \g{(x : V'_1) \!\!->\!\! V'_2}}

			\inferrule[CstLam]{\g{v_1} =-= \g{v_2}}
			{
				\g{\lambda x : V_1 \ldotp v_1 } =-= \g{\lambda x : V_2 \ldotp v_2 }
			}

			\inferrule[CstCastL]{\g{v_1} =-= \g{v_2}}{\g{v_1} =-= \cast{V_1}{V_2}\g{v_2}}

			\inferrule[CstTagL]{\g{v_1} =-= \g{v_2}}{\g{v_1} =-= \attagl{h}{v_2}}
		\end{inferbox}
	\end{boxedarray}
	\caption{Consistency for \clang}
	\label{fig:consistency}
\end{figure}

\begin{theoremEnd}[apxlem]{lemma}[Symmetry and Weak Reflexivity]
	\label{lem:cst-sym-refl}
	\begin{itemize}
		The following properties hold for consistency:
		\item If $\g{v_1} =-= \g{v_2}$ then $\g{v_2} =-= \g{v_1}$;
		\item If $\gv =-= \gv$ then $\qm_{\gV} =-= \gv$ ;
		\item If $\g{v_1} =-= \g{v_2}$ then $\g{v_1} =-= \g{v_1}$.
	\end{itemize}
\end{theoremEnd}
\begin{proofEnd}
  (1) by induction on the derivation of $=-=$. For the \rrule{CstUnk} rules we smply apply
  \rrule{CstUnkFlip}. For \rrule{CstUnkFlip} our result is immediate from our premise.
  ALl other cases follow directly from the IH.

  (2) By induction on the derivation of $=-=$.

	\case{Any \rrule{CstUnk} rule}{ $\g{v}$ must be some $\qm_{\g{V'}}$, so we apply \rrule{CstUnkBase} to get our result.}

	\case{ \rrule{CstCastL}}{
		Then $\g{v}$ is $\g{\cast{V_{1}}{V_{2}}\g{v'}}$ for some $\g{v'}$.
		Then follows from IH applied to premise.
	}

	\case{\rrule{CstCastL}}{
		Then $\g{v}$ is $\g{\cast{V_{1}}{V_{2}}\g{v'}}$ for some $\g{v'}$.
		Then IH gives us $\g{v'} =-= \qm_{\gV}$, so we apply \rrule{CstCastR}
		to get our result.
	}

	\rlcase{CstTagL, CstTagR}{Same idea as above.}

	\case{Remaninig Cases}{Follow by combining IH with the matching \rrule{CstUnk} rule}

  (3) By induction on the derivation of $=-=$.

	\case{ \rrule{CstUnkFlip}}{
		Apply IH on our premise}

	\case{Any other \rrule{CstUnk}}{ we know $\g{v_{1}}$
		must be some $\qm_{\gV}$ so our result follows from \rrule{CstUnkBase}.}

	\rlcase{CstCastL,CstTagL}{
		Follows from IH applied to premise.
	}

	\rcase{CstCastR,CstTagR}{Apply IH to premise, then construct result with \rrule{CstCastL}}

	\case{All other cases:}{Follows after applying IH to premises and building derivation with same rule.}
\end{proofEnd}

Our consistency relation is big enough and small enough:
we exclude all non-equal static types, but reduce
precision without compromising existing consistencies that hold.

\begin{theoremEnd}[apxproof]{lemma}[Consistency of Static Terms Implies $\alpha$-Equivalence]
	If $\lfloor \s{v} \rfloor =-= \lfloor \s{v'} \rfloor $ then $\s{v} =_{\alpha} \s{v'}$.
\end{theoremEnd}
\begin{proofEnd}
  By induction on the derivation of $=-=$. The \rrule{CstUnk} rules are impossible,
  along with \rrule{CstCastL,CstCastR,CstTagL} and \rrule{CstTagR}, since $\qm$
  and casts are never produced by our embedding.
  The remaining cases all follow immediately from our hypothesis, along with the
  congruence of $\alpha$-equivalence under the various syntactic constructors of \clang.
\end{proofEnd}

\begin{theoremEnd}[apxlem]{lemma}[Reducing Precision Preserves Self Consistency]
  \label{lem:self-cst-prec}
  If $\Gamma |- \g{v_{1}} \squbc \g{v_{2}}$ and $\g{v_{1}} =-= \g{v_{1}}$ then
  $\g{v_{2}} =-= \g{v_{2}}$
\end{theoremEnd}
\begin{proofEnd}
  By induction on derivation of $\squbc$.

  \case{Any rule, $=-=$ form with \rrule{CstUnkFlip}}{
	Then $\g{v_{1}} = \g{\qm_{V}}$. Cases on $\squbc$ derivation.
	If \rrule{PrecAxRefl}, trivial. If \rrule{PrecAxTrans}, follows from two applications of IH.
	If \rrule{PrecAxUpDown}, then apply \rrule{CstCastR} twice.
	Assume henceforth that $=-=$ is not initially build with \rrule{CstUnkFlip}.
  }

  \rlcase{PrecGenUnk}{Follows from \rrule{CstUnkBase}}

  \rlcase{PrecGenErr}{Impossible since $\err$ is not self-consistent}

  \rcase{PrecCongApp}{Then $=-=$ derived with \rrule{CstCastL} or \rrule{CstCastR}.
	In either case, get $\g{t_{1}\ t_{2}}$ self-consistent as a premise. Inversion gives
	$\g{t_{1}}$ and $\g{t_{2}}$ each self-consistent, so we can apply IH to get $\g{t'_{1}}$
	and $\g{t'_{2}}$ self-consistent. Result follows from \rrule{CstApp}.
  }

  \rcase{PrecCongPi}{Inversion gives $\g{T_{1}}$ and $\g{T_{2}} $ both self-consistent.
	Then $[\cast{T'_{1}}{T_{2}}\gx/\gx]\g{T_{2}}$ is self-consistent, since it differs from $\g{T_{2}}$
	only in casts, which consistency ignores. Then we can apply our IH and \rrule{CstPi} for our result.
	}

	\lcase{Remaining \rrule{PrecCong*}}{Follows from IH plus application of same rule.}

	\rlcase{PrecAxRefl}{Trivial}

	\rlcase{PrecAxEta}{Inversion gives $\g{t}$ self-consistent, so adding casts preserves self-consistency.}

	\rlcase{PrecAxCastBot}{Impossible}

	\rlcase{PrecAxTrans}{Apply IH twice.}

	\rcase{PrecAxEta}{Inversion gives $\g{t}$ self-consistent, so adding casts preserves self-consistency.}

	\rlcase{PrecAxUpDown, PrecAxDownUp, PrecAxIntermed, PrecAxComposeUp}{Trivial, adding/removing casts does not affect self-consistency.}
  \end{proofEnd}

  \begin{theoremEnd}[apxlem]{lemma}[Cast/Tag Removal]
	\label{lem:cast-tag-removal}
	If $\tagOf(\gv) = \g{h}$, and $\gv =-= \g{v'}$,
	then either $\g{v} = \qm_{\g{V}}$ for some $\gV$,
	or $\g{v'}$ is some sequence of casts/tags applied
	to a term $\g{v''}$ where $\tagOf(\g{v''})$.
  \end{theoremEnd}
  \begin{proofEnd}
	By induction on $=-=$ derivation. Trivial for \rrule{CstFlip} and \rrule{CstBase}.
	Follows from IH for \rrule{CstCastL} and \rrule{CstTagL}.
	Immediate for \rrule{CstPi} and \rrule{CstLam}. All other cases don't have defined tag on LHS.
  \end{proofEnd}

\begin{theoremEnd}[apxlem]{lemma}[Consistency is Upward Closed on the Left]
	\label{lem:cst-closed-left}
	If $\g{v_{L1}} =-= \g{v_R}$ and $\Gamma |- \g{v_{L1}} \squbc \g{v_{L2}}$
	then $\g{v_{L2}} =-= \g{v_R}$.
\end{theoremEnd}
\begin{proofEnd}
  We perform induction on the derivation of $\g{v_{L1}} \squbc \g{v_{L2}}$.

  \rlcase{Any rule, $=-=$ with \rrule{CstUnkFlip}}{
	Then $\g{v_{R}} = \qm_{\gV}$. The result follows from \cref{lem:self-cst-prec}.
  }

	\rlcase{PrecGenUnk}{Then $\g{v_{L2}}$ is $\qm_{\gT}$.
		We obtain our result from \cref{lem:cst-sym-refl},
		since $\g{v_{R}}$ must be self-consistent.
	}

	\rcase{PrecCongCast}{
	  If $=-=$ formed with \rrule{CstCastL},
		then $\g{t_{L1}}  = \cast{V_{1}}{V_{2}} \gv \squbc \cast{V_{1}}{V_{2}}\g{v'} = \g{V_{L2}} $.
		Our premise gives $\g{v} \squbc \g{v'}$. We perform inversion on the $=-=$ derivation.
		For \rrule{CstCastL}, we have
		$\gv \squbc \g{v'}$ and $\gv =-= \g{v_R}$, so we get our result from IH and \rrule{CstCastL}.
		The proof is symmetric for \rrule{CstCastR}.

	}

	\rlcase{PrecCongTag}{Same reasoning as for \rrule{PrecCongCast}}

	\rlcase{PrecCongPi}{
	  \Cref{lem:cast-tag-removal} and inversion givs $\g{v_{R}}$ is some sequence of casts/tags
	  applied to $\g{(x : V_{1}) -> V_{2}}$
	  with $\g{T_{1}} =-= \g{V_{1}}$ and $\g{T_{2}} =-= \g{V_{2}}$.
	  Then we know $\g{T_{1}}\squbc \g{T'_{1}}$ and $[\cast{T'_{1}}{T_{1}}\gx/\gx]\g{T_{2}} \squbc \g{T'_{2}}$.
	  Since casts do not affect consistency, we know that $[\cast{T'_{1}}{T_{1}}\gx/\gx]\g{T_{2}} =-= \g{V_{2}}$,
	  giving us what we need to apply IH. Our result then comes by applying \rrule{CstPi}
	  with \rrule{CstTagL} and \rrule{CstCastL} the proper number of times.
	}

	\rlcase{PrecCongApp}{
	  Similar to \cref{lem:cast-tag-removal}, the derivation of $\cast{T}{T'}\g{(t_{1}\ t_{2})}$
	  must contain one use of \rrule{CstCastR}, some number of \rrule{CstCastL} and \rrule{CstTagL},
	  then \rrule{CstApp} for $\g{t_{1}\ t_{2}} =-= \g{v_{1}\ v_{2}}$ for some $\g{v_{1}}, \g{v_{2}}$.
	  IH gives us $\g{t'_{1}} =-= \g{v_{1}}$ and $\g{v'_{2}} =-= \g{v'2}$, so our result comes
	  from \rrule{CstApp} and the proper number of \rrule{CstCastL} and \rrule{CstTagL}.
	}

	\lcase{Remaining \rrule{PrecCong*}}{Follows from inversion on the derivation of $=-=$ and the IH.}

	\rlcase{PrecAxRefl}{Trivial, since normal forms cannot be further stepped.}

	\rcase{PrecAxTrans}{
		Then we have $\g{t_1} \squbc \g{t_2}$ and $\g{t_2} \squbc \g{t_3}$, where $\g{t_1} =-= \g{v_R}$.
		Our goal is $\g{t_3} =-= \g{v_R}$.
		By IH, we have $\g{t_2 =-= \g{v_R}}$. We can then apply our IH again
		to get $\g{t_3 =-= \g{v_R}}$.
		Each use of the IH uses a structurally smaller part of the $\squbc$ derivation,
		so this is well-founded.
	}

	\rlcase{PrecAxUpDown}{We apply \rrule{CstCastL} twice}.

	\rcase{PrecAxDownUp}{
		First suppose that $=-=$ is derived with two uses of \rrule{CstCastL}.
		Then we have $\gt =-= \g{v_R}$, which is our goal.

		If instead $=-=$ is derived with \rrule{CstUnkFlip},
		then the following two rules must be \rrule{CstCastL}, so the same reasoning applies.

		The final case is where $=-=$ uses \rrule{CstCastR} or \rrule{CstTagR}.
		Then we can repeatedly apply inversion to obtain $\cast{T_1}{T_2}\cast{T_2}{T_1}\gt =-= \g{v'_R}$,
		where $\g{v'_R}$ is $\g{v_R}$ with the top-level casts and tagging removed.
		This must in turn be derived with two uses of \rrule{CstCastL}, giving us
		$\gt =-= \g{v'_R}$. Repeated uses
		of \rrule{CstCastR} and \rrule{CstTagR} then yield our goal.
	}

	\rcase{PrecAxComposeUp, PrecAxIntermed, PrecAxDepApp}{
		Similar reasoning to above: casts don't affect consistency, so the appropriate amount
		of peeling away casts and re-adding them yields the result.
	}

	\rlcase{PrecGenErr,PrecAxCastBot}{
		Impossible, since $\err$ and $\cast{T_1}{T_2}\err$ are not consistent with any terms.
	  }

	\rlcase{PrecAxEta}{Impossible, since casting between function types is not a normal form.}

\end{proofEnd}

\begin{theoremEnd}[apxproof]{lemma}[Reducing $\squbc$ Preserves $=-=$]
	\label{lem:consistency-upward-closed}
	If $\g{v_1} =-= \g{v_2}$, $\Gamma |- \g{v_1} \squbc \g{v'_1}$ and $\g{v_2} \squbc \g{v'_2}$
	then $\g{v'_1} =-= \g{v'_1}$.
\end{theoremEnd}
\begin{proofEnd}
	By \cref{lem:cst-closed-left} we have $\g{v'_{1}} =-= \g{v_{2}}$.
	By \cref{lem:cst-sym-refl}, we have $\g{v_{2}} =-= \g{v'_{1}}$.
	Applying \cref{lem:cst-closed-left} again yields our result.
\end{proofEnd}

\subsection{Showing the Normalization Guarantee}
\label{subsec:sgg}

We have axiomatized precision, but we still must show that those axioms are meaningful.
We establish the validity of our axioms by showing that the
precision properties established syntactically reflect the precision properties of the values
to which they normalize. The proposition takes the form of the dynamic gradual guarantee,
so we call it the \textit{Normalization Gradual Guarantee}.

\begin{theoremEnd}[apxlem]{lemma}[Substitution Monotone in Inserted Term]
	\label{lem:subst-mono-subterm}
	If $\Gamma |- \g{t_1} \squbc \g{t_2}$,
	then $\Gamma |- [\g{t_1} / \gx]\gt \squbc [\g{t_2} / \gx]\gt$.
\end{theoremEnd}
\begin{proofEnd}
	By induction on $\g{t}$.
	If $\gt$ is $\gx$, holds by our premise.
	If $\gt$ is $\gy \neq \gx$, holds by \rrule{PrecAxRefl}.
	Otherwise, we can apply a congruence rule plus IH.
\end{proofEnd}

\begin{theoremEnd}[apxlem]{lemma}[Substituting Monotone in Term Being Substituted In]
	\label{lem:subst-mono-interm}
	$\g{t_{low}} \squbc \g{t_{high}}$,
	then $[\g{t_{sub}} / \gx]\g{t_{low}} \squbc [\g{t_{sub}} / \gx]\g{t_{high}}$.
\end{theoremEnd}
\begin{proofEnd}
  By induction on the derivation of  $\g{t_{low}} \squbc \g{t_{high}}$.

  \rcase{PrecAxRefl}{the result trivially holds since both sides are the same}

  \rcase{PrecCongPi, PrecCongApp}{Straightforward from IH,
	and that  equality of normal forms and synthesis are preserved under substitution.
	In both cases we use our bound variable assumption,
	which allows us to commute the substitutions in the premise and the substitution from the lemma.
  }

  \lcase{All other cases}{Straightforward: substitution does not change the structure of any
	of the conclusions or premises, so given that both equality of normal forms and synthesis are
	preserved under substitution, each case we apply IH to the premises and re-build the derivation
  with the results and whatever rule was initially used.}
\end{proofEnd}

\begin{theoremEnd}[apxlem]{lemma}[Substitution is Monotone]
	If
	$\g{t_{1}} \squbc \g{t_{2}}$
	$\g{t'_{1}} \squbc \g{t'_{2}}$
	then $[\g{t'_{1}} / \gx]\g{t_{1}} \squbc [\g{t'_{2}} / \gx]\g{t_{2}}$.
\end{theoremEnd}
\begin{proofEnd}
	Follows from \cref{lem:subst-mono-subterm},
	\cref{lem:subst-mono-interm},
	and \rrule{PrecAxTrans}.
\end{proofEnd}

\begin{theoremEnd}[apxlem]{lemma}{Lambda Cast Chains}
  (1) If $\cast{T_{1}}{T_{2}}\ldots\cast{T_{n-1}}{T_{n}}\g{(\lambda x : T_{ascr} \ldotp t)} \squbc \g{t'}$,
  and that term checks against $\g{T*}$ under $\Gamma$,
  then $\g{t'} $ has the form $ \cast{T'_{1}}{T'_{2}} \ldots \cast{T'_{m-1}}{T_{m}}\g{t''}$,
  where $\g{t''}$ is either $\g{(\lambda x : T'_{ascr} \ldotp t''')}$
  
  or $\g{t''}$ is $\qm_{\gT}$ for some $\gT$.

  (2) For $\g{b} \in \set{\g{true}, \g{false}}$,
  if $\cast{T_{1}}{T_{2}}\ldots\cast{T_{n-1}}{T_{n}}\g{b} \squbc \g{t'}$,
  and that term checks against $\g{T*}$ under $\Gamma$,
  then $\g{t'} $ has the form $ \cast{T'_{1}}{T'_{2}} \ldots \cast{T'_{m-1}}{T_{m}}\g{t''}$,
  where $\g{t''}$ is either $\g{b}$, $\g{(\lambda x : T'_{ascr} \ldotp t''')}$,
  
  or  $\qm_{\gT}$ for some $\gT$.

  (3)
  If $\cast{T_{1}}{T_{2}}\ldots\cast{T_{n-1}}{T_{n}}\g{b} \squbc \g{\qm_{\gT}}$,
  and that term checks against $\g{T*}$ under $\Gamma$,
  then $\g{t'} $ has the form $ \cast{T'_{1}}{T'_{2}} \ldots \cast{T'_{m-1}}{T_{m}}\g{t''}$,
  where $\g{t''}$ is either $\g{\qm{\g{T'}}}$ or $\g{(\lambda x : T'_{ascr} \ldotp t''')}$

\end{theoremEnd}
\begin{proofEnd}
  Proof of (1): by induction on the derivation of $\squbc$.

  \rlcase{PrecGenUnk}{Trivial, since is $\qm$}

  \rlcase{PrecCongCast}{Follows from IH}.

  \rlcase{PrecCongLam}{Trivial}

  \rlcase{PrecAxEta}{Trivial, RHS is a lambda. Impossible to be well-typed }

  \rlcase{PrecAxRefl}{Trivial}

  \rlcase{PrecAxTrans}{Two applications of IH.}

  \rlcase{PrecAxUpDown}{ Adding two more casts yield the right form. }

  \rlcase{PrecAxDownUp}{Removing two casts gives the right form}

  \rlcase{PrecAxIntermed}{Modifying one cast gives the right form}

  \rlcase{PrecAxComposeUp}{Contracting 2 casts into 1 or expanding one cast into two is of the right form.}

  The proofs of (2) and (3) follow the same form as the (1), but for \rrule{PrecAxTrans}
  we use (1) in the case that a $\lambda$ term is produced.

\end{proofEnd}

\begin{theoremEnd}[apxlem]{lemma}[Reduction Preserves $\squbstar$]
  \label{lem:star-red-preserve}
  Suppose $\Gamma |- \g{T_{1}} \squbstar \g{T_{2}}$,
  where   $\g{T_{1}} \stepstostar \g{T'_{1}}$
  and $\g{T_{2}} \stepstostar \g{T'_{2}}$.
  Then
  $\Gamma |- \g{T'_{1}} \squbstar \g{T'_{2}}$.
\end{theoremEnd}
\begin{proofEnd}
  Confluence implies that reduction preserves normal forms,
  and so the precision of the normal forms is preserved
  by stepping.
\end{proofEnd}

  \begin{theoremEnd}[apxlem]{lemma}[Mutual Simulation]
  Consider $\Gamma |- \g{t_{low}} : \g{T}$ and $\Gamma |- \g{t_{high}} : \gT$,
  where $\Gamma |- \g{t_{low}} \squbc \g{t_{high}}$. Then:

  \begin{enumerate}
	\item If $\g{t_{low}} \stepstostar \g{t'_{low}}$, then there must exist some $\g{t'_{high}}$
		  such that $\Gamma |- \g{t'_{low}}\squbc \g{t'_{high}}$.
	\item If $\g{t_{low}} \stepsto \g{t'_{low}}$, then there must exist some $\g{t'_{high}}$
		  such that $\Gamma |- \g{t'_{low}}\squbc \g{t'_{high}}$.
	\item If $\g{t_{low}} = \g{\lambda (x : T) \ldotp \g{t_{body}}}$, then $\g{t_{high}}\stepstostar \g{\lambda (x : T) \ldotp t'_{{high}}}$
		  such that $\Gamma |- \g{t_{body}} \squbc \g{t_{high}}$, or $\g{t_{high}\stepstostar \qm_{\gT}}$
	\item If $\g{t_{low}} \in \set{\g{true}, \g{false}}$, then
		  $\g{t_{high}} \stepstostar \qm_{\g{\bB}}$ or $\g{t_{high}} \stepstostar \g{t_{low}}$.

	\item If $\qm_{\gT}\squbc \g{t'}$ and then $\g{t'} \stepstostar \qm_{\gT}$.

				\end{enumerate}
\end{theoremEnd}
\begin{proofEnd}
	We perform lexicographic induction on the number of steps that $\g{t_{low}}$ takes before normalization,
	and the derivation of $\Gamma |- \g{t_{low}}\squbc \g{t_{high}}$. \je{TODO: check lexicographic, cite SN lemma}.

	Proof of (1): follows from applying (2) by IH to the first step, then (1) by IH on the remaining steps. \je{TODO check WF}

	Proof of (2):
	We proceed by induction on the derivation of $\g{t_1} \squbc \g{t_2}$.

	\rcase{PrecGenUnk}{
	  We have $\g{t_{low}} \squbc \qm_{\g{T_{high}}}$, so we know that $\Gamma |- \g{t_{low}}\psynth \gV$
	  and $\g{T_{high}} \stepstostar \g{\gV}$. Our goal is to find $\g{T'_{high}}$
	  such that $\g{T_{high}} \stepstostar \g{T'_{high}}$ and $\g{t'_{low}} \squbc \qm_{\g{T'_{high}}}$.
	  By \cref{lem:preservation}, we know that $\Gamma |- \g{t'_{low}} : \g{T}$, and by \cref{TODO}
	  we know that $\g{T} \stepstostar \g{V}$. So $\g{t'_{low}}\psynth \gV$.
	  We know that $\g{T_{high}} \stepstostar \g{V}$, so we can choose $\g{T'_{high}} = \g{T_{high}}$
	  (e.g. zero steps)
	  and apply \rrule{PrecGenUnk} to obtain our goal.
		}

		\rlcase{PrecGenErr}{
		  $\err$ only steps contextually, so, the resulting
		  type ascription has the same normal form, so we can choose $\g{t'_{high}} = \g{t_{high}} $.
	}

	\rcase{PrecCongPi, PrecCongLam, PrecCongTag}{
	  The only steps are contextual, so the result follows from our hypothesis.
	  For \rrule{PrecCongPi} we rely on the additional fact that
	  any reduction that $\g{T_{2}}$ can take, $[\cast{T'_{1}}{T_{1}}\gx / \gx]\g{T_{2}}$
	  can also take.
	}

	\lcase{\rrule{PrecCongIf}, contextual step}{Same reasoning as above.}

	\case{\rrule{PrecCongApp}, contextual step}{
	  If we step in $\gT$ or $\g{T'}$ then, the result holds from confluence, along with our IH. \je{TODO formalize}

	   Otherwise, we step in $\g{t_1}$ or $\g{t_{2}}$ and the result follows
	  immediately from our hypothesis.
	}

	\lcase{\rrule{PrecCongCast}, contextual step }{
		Then we have $\cast{T_1}{T_2}\gt \squbc \cast{T_1}{T_2} \g{t'}$.
		If we step in $\g{T_1}$ or $\g{T_2}$, we can take the exact same step in
		$\cast{T_1}{T_2} \g{t'}$, giving us our result.
		If we step in $\gt$, then the result follows from IH.
	  }

	  \case{TODO non-context steps}{TODO}

	  \rlcase{PrecAxRefl}{Trivial.}

	  \rlcase{PrecAxTrans}{
		Then there is some $\g{t_{mid}}$ such that
		$ \Gamma |- \g{t_{low}} \squbc \g{ t_{mid}} $ and $ \Gamma |- \g{t_{mid}} \squbc \g{t_{high}}$,
		  and $\g{t_{low}} \stepsto \g{t'_{low}}$.
		  Then by IH, we have $\g{t'_{mid}} \stepstostar \g{t'_{mid}}$
		  with $\g{t'_{low}} \squbc \g{t'_{mid}}$.
		  Then by part (1) of IH,
		  we have $\g{t_{high}} \stepstostar \g{t'_{high}}$ with $\g{t'_{mid}}\squbc \g{t'_{high}}$.
		  The result then follows by applying \rrule{PrecAxTrans} along with the transitivity of
		  $\stepstostar$.
		}

		\rlcase{PrecAxUpDown}{Then any step we take in $\gt$ we can take in $\gt$ in the RHS.}

		\lcase{\rrule{AxUpDown, PrecAxDownUp, PrecAxIntermed, PrecAxComposeUp}, contextual step}{
		  Either we step in a type, and apply \cref{lem:star-red-preserve}},
		or we step in $\gt$ and can step in both sides.

		\lcase{\rrule{PrecAxCastBot}, contextual step}{Either we step in $\g{T_{1}}$, and can again apply \rrule{PrecAxCastBot},
		  or we step in $\g{T_{2}}$ and also step in $\g{T_{2}}$ in $\g{T_{high}}$.
		}

		\lcase{\rrule{PrecAxEta}, contextual step}{
		  Then we step in one of $\g{T'_{1}}, \g{T'_{2}}, \g{T_{1}}, \g{T_{2}}$ or $\gt$.
		  Whatever step we take, we can also take in $\g{t_{{high}}}$.
		}

		\case{$\g{t_{high}} = \err_{{\gT}}$}{Follows from preservation and \rrule{PrecGenErr}.}

		\case{Step by $\bB\bB$, \textsc{Type} or \textsc{??}}{
		  Choose $\g{t'_{high}} = \g{t_{high}}$.
		  By our premise we have $\cast{T}{T}\gt \squbc \g{t_{high}}$,
		  so by \rrule{PrecAxTrans} and \rrule{PrecAdTrivialCast}
		  we have $\gt \squbc \g{t_{high}}$.
		}

		We have shown our result for all cases where $\g{t_{low}}$ steps to $\g{t'_{low}}$
		contextually, and for all steps for \rrule{PrecGenUnk, PrecGenErr, PrecCongPi, PrecCongLam, PrecCongTag, PrecAxRefl, PrecAxTrans, PrecAxUpDown }.
		We have also covered any case where $\g{t'_{high}} = \err{\gT}$, meaning
		we have covered all cases for \rrule{PrecAxCastBot}.

		This means our remaining proof burden is showing our result for the cases where $\g{t_{1}} \leadsto \g{t'_{1}} \neq \err_{\gT}$,
		and $\g{t_{1}} \squbc \g{t'_{1}}$ by one of
		\rrule{PrecCongApp, PrecCongIf, PrecCongCast, PrecAxDownUp, PrecAxIntermed, PrecAxComposeUpL,PrecAxComposeUpR, PrecAxEta}.

		\case{\rrule{PrecCongApp}, step by $\beta$}{
		  Then
		  $\g{t_{low}} = \g{(\lambda(x : T)\ldotp t_{lowf})\ \g{t_{lowa}}}$,
		  $\g{t_{high}} = \g{t_{highf}}\ \g{t_{higha}}$,
		and $\g{t'_{low}} = [\g{t_{lowa}}/\gx]\g{t_{lowf}}$,
		where $\g{(\lambda(x : T)\ldotp t_{lowf})}\squbc \g{t_{highf}}$
		and $\g{t_{lowa}} \squbc \g{t_{higha}}$.
		By IH (3), we know that $\g{t_{highf}}\stepstostar \g{\lambda (x : T)\ldotp \g{t'_{highf}}}$,
		where $\g{\g{t_{lowf}}\squbc \g{t'_{highf}}}$, or to $\qm_{\g{T'}}$ for some $\g{T'}$.
		}

		\je{TODO: check conditions on type asascription}		\je{TODO: reformulate subst to include cast on one side}

		\case{\rrule{PrecCongApp}, step by $\beta\textbf{\qm}$}{TODO}

		\case{\rrule{PrecCongApp}, step by $\beta\errsym$}{TODO}

		\case{\rrule{PrecCongIf}, step by \textsc{true} or \textsc{false}}{TODO}

		\case{\rrule{PrecCongIf}, step by \textsc{If?}}{TODO}

		\case{\rrule{PrecCongIf}, step by \textsc{If}$\errsym$}{TODO}

		\case{Step by \textsc{?tag}}{TODO}

		\case{Step by \textsc{?tagfun}}{TODO}

		\case{Step by \textsc{?}$\Pi$}{TODO}

		\case{Step by $\Pi\Pi$}{TODO}
	\end{proofEnd}

\begin{theoremEnd}[apxlem]{lemma}[Reduction Preserves Precision]
  Suppose $\Gamma |- \g{t_{1}} : \g{T}$ $\Gamma |- \g{t_{2}} : \gT$ and $\Gamma |- \g{t_{1}} \squbc \g{t_{2}}$.
  If $\g{t_{1}} \stepstostar \g{t'_{1}}$, then there exists $\g{t'_{2}}$ such that
  $\g{t_{2}} \stepstostar \g{t'_{2}}$ and $\Gamma |- \g{t'_{1}} \squbc \g{t'_{2}}$.
\end{theoremEnd}

\begin{theoremEnd}[apxlem]{lemma}[$\eta$-expansion preserves precision]
  Suppose $\Gamma |- \g{v_{1}} : \g{V}$ $\Gamma |- \g{v_{2}} : \gV$ and $\Gamma |- \g{v_{1}} \squbc \g{v_{2}}$.
	If $\Gamma |- \g{v_{1}} \etasteps \g{v'_{1}} :  \gV$ and $\Gamma |- \g{v_{2}} \etasteps \g{v'_{2}} : \gV$,
	then $\Gamma |- \g{v'_{1}} \squbc \g{v'_{2}}$.
\end{theoremEnd}
\begin{proofEnd}
	TODO
\end{proofEnd}

	\begin{theorem}[Normalization Gradual Guarantee]
		\label{thm:norm-guarantee}
		If $\Gamma |- \g{t_{1}} : \gV$, $\Gamma |- \g{t_{2}} : \gV$ and $\Gamma |- \g{t_{1}} \squbc \g{t_{2}}$,
		then
		 
		$\Gamma |- \g{t_{1}} \etasteps \g{v_{1}} : \gV$, $\Gamma |- \g{t_{2}} \etasteps \g{v_{2}} : \gV$,
		and $\Gamma |- \g{v_{1}} \squbc \g{v_{2}}$ for some $\g{v_{1}},\g{v_{2}}$.
	\end{theorem}

\begin{theoremEnd}[apxlem]{lemma}[Precision Value Catch-up]
	Suppose $\Gamma |- \g{t} : \gT$ and $\Gamma |- \g{v} : \gT$,
	where $\Gamma |- \g{t} \squbc \g{v}$.
		Then either $\gt$ is a value, or $\gt \stepstostar \g{t'}$
		such that $\Gamma |- \g{t'} \squbc \gv$.
	\end{theoremEnd}

	\begin{proofEnd}
	TODO
	\end{proofEnd}

The last definition we need before stating the SGG is what precision means for \surflang.
Normally, this is defined by lifting $\gt \sqsubseteq \qm$ across each constructor,
but we must account for the level ascriptions on $\qm$,
which we do by defining precision in terms
of well-typed substitutions.

\begin{theoremEnd}[apxlem]{lemma}[Elaboration Is Preserved Under Substitution]
	TODO
\end{theoremEnd}

\begin{definition}[Precision for \surflang]
	Suppose $\mathcal{D} :: \Gamma |- \echeck{[\g{s'}/ \gx]\g{s}}{\gT}{\gt} $
	and $\mathcal{D'} $ is a sub-derivation of $\mathcal{D}$,
	where $\mathcal{D'} :: \Gamma' \Gamma |- \esynth{\g{s'}}{\g{T'}}{\g{t'}} $
	or $\mathcal{D'} :: \Gamma' \Gamma |- \echeck{\g{s'}}{\g{T'}}{\g{t'}} $.

	We say that ${\Gamma |- [\g{s'}/ \gx]\g{s} \squbs_{1} [\qmat{\ell}/ \gx]\g{s} \g }$
	if ${\Gamma |- \g{T'} : \gType{\ell}}$.

	Then $\_\!\!\squbs\!\!\_$ is the transitive reflexive closure of $\_\!\!\squbs_{1}\!\!\_$.
\end{definition}

We now state the elaboration gradual guarantee,
which relates surface term precision to elaboration synthesis and checking.
The SGG and DGG follow as
 corollaries.
 The proof is omitted for space,
 but works by induction, using \cref{thm:norm-guarantee} to relate types' normalizations,
 and using the axioms of algebraic precision to relate terms with casts.
 The proof never makes use of the assumption that \rrule{To?} does not operate on function types,
so the results apply to \clang's exact semantics (\cref{sec:elab}).

\begin{theoremEnd}[apxlem]{lemma}[Precision Weakening]
	If $\Gamma |- \g{T_{low}} \squbc \g{T_{high}}$, and $\Gamma'(\gx : \g{T_{low}})\Gamma |- \esynth{\gs}{\gT}{\gt}$,
	\\then $([\cast{T_{high}}{T_{low}}\gx/\gx]\Gamma')(\gx : \g{T_{high}})\Gamma |- \esynth{\gs}{\g{T'}}{g{t'}}$
	\\where $([\cast{T_{high}}{T_{low}}\gx/\gx]\Gamma')(\gx : \g{T_{high}})\Gamma |- [\cast{T_{high}}{T_{low}}\gx/\gx] \gT \squbc \g{T'}$
	\\and  $([\cast{T_{high}}{T_{low}}\gx/\gx]\Gamma')(\gx : \g{T_{high}})\Gamma |-\\ \cast{[\cast{T_{high}}{T_{low}}\gx/\gx] \gT}{T'}[\cast{T_{high}}{T_{low}}\gx/\gx] \g{t} \squbc \g{t'}$.

\end{theoremEnd}
\begin{proofEnd}
	By induction on the elab derivation of $\gs$.

	\rcase{EVar}{Then $\Gamma'(\gx : \g{T_{low}})\Gamma |- \esynth{\gy}{\gT}{\gt}$.
		If $\gy$ is in $\Gamma$ the result holds trivially.

		If $\gy$ is in $\Gamma'$ then we have\\
		$([\cast{T_{high}}{T_{low}}\gx/\gx]\Gamma')(\gx : \g{T_{high}})\Gamma |- \esynth{\gy}{[\cast{T_{high}}{T_{low}}\gx/\gx]\Gamma'(\gy)}{\gy}$.
		Then\\ $([\cast{T_{high}}{T_{low}}\gx/\gx]\Gamma')(\gx : \g{T_{high}})\Gamma |-
		[\cast{T_{high}}{T_{low}}\gx/\gx] \Gamma'(\gy) \squbc 	[\cast{T_{high}}{T_{low}}\gx/\gx] \Gamma'(\gy) $
		by \rrule{PrecAxRefl}.
		Finally, we have\\
		$([\cast{T_{high}}{T_{low}}\gx/\gx]\Gamma')(\gx : \g{T_{high}})\Gamma |- \\
		\cast{[\cast{T_{high}}{T_{low}}\gx/\gx] \Gamma'(\gy)}{[\cast{T_{high}}{T_{low}}\gx/\gx]\Gamma'(\gy)} \g{y}
		\squbc \g{y}$ by \rrule{PrecAdIsoCast}.

		If $\gy = \gx$, then by our premise,
		\\ $([\cast{T_{high}}{T_{low}}\gx/\gx]\Gamma')(\gx : \g{T_{high}})\Gamma |- [\cast{T_{high}}{T_{low}}\gx/\gx] \g{T_{low}} \squbc \g{T_{high}}$ by \rrule{PrecAxRefl} since $\gx$ is not free in $\g{T_{low}}$.
		Similarly, we have
		\\  $([\cast{T_{high}}{T_{low}}\gx/\gx]\Gamma')(\gx : \g{T_{high}})\Gamma |-\\ \cast{\g{T_{low}}}{T_{high}}\cast{T_{high}}{T_{low}}\gx \gx \squbc \g{x}$ by \rrule{PrecAxDownUp}.
	}

	\rlcase{EType, EBool, ETF}{Trivial.}

	\rcase{ELam}{TODO}

\end{proofEnd}

\begin{theoremEnd}[apxlem]{lemma}[Full Elaboration Gradual Guarantee]
	Consider \surflang terms $\gs$ and $\g{s'}$, where $\Gamma |- \gs \squbs \g{s'}$.
	Consider $\Gamma'$ such that $\Gamma \squbG \Gamma'$.
	\begin{itemize}

		\item Suppose that $\Gamma |- \echeck{\gs}{\gV}{\gt}$,
					that $\Gamma |- \g{V} \squbc \g{T'}$,
					and $\Gamma |- \g{T'} \etasteps \g{V'}$.
					Then $\Gamma' |- \echeck{\gs}{\g{V'}}{\g{t'}}$
					for some $\g{t'}$ where $\Gamma' |- \cast{V}{V'}\castenv{\Gamma}{\Gamma'}\gt \squbc \g{t'}$.

		\item Suppose $\Gamma |- \esynth{\gs}{\gT}{\gt}$.
					Then either $\g{s'}$ is $\qmat{\ell}$ and $\Gamma' |- \echeck{\qmat{\ell}}{\gV}{\qm_{\gT}}$
					where $\Gamma' |- \castenv{\Gamma}{\Gamma'}\g{T} \etasteps \g{V}$,
					or $\Gamma' |- \esynth{\g{s'}}{\g{T'}}{\gt'}$
					for some $\g{t'}$ and $\g{T'}$ where $\Gamma' |- \cast{T}{T'} \castenv{\Gamma}{\Gamma'} \gt \squbc \g{t'}$
					and $\Gamma' |- \castenv{\Gamma}{\Gamma'}\gT \squbc \g{T'}$.
	\end{itemize}
\end{theoremEnd}
\begin{proofEnd}
	\je{TODO element-wise Gamma comparison}
	By induction on the typing derivation of $\gs$. We assume we have already proved \cref{thm:strong-norm}.

	\case{$\g{s'} = \qmat{\ell}$}{
		Then by our premise,
		$\Gamma |- \gT <= \gType{\ell}$, so $\Gamma' |- \echeck{\qmat{\ell}}{\gT}{\qm_{\gT}}$,
		and by \cref{thm:strong-norm} we have $\Gamma' |- \castenv{\Gamma}{\Gamma'} \gT \etasteps \gV$.
		So we have $\Gamma' |- \castenv{\Gamma}{\Gamma'} \gt \squbc \qm_{\g{V}} $ by \rrule{PrecGenUnk}.
		We assume henceforth that $\g{s'} \neq \qmat{\ell}$.
	}

	\rlcase{EVar}{Then $\castenv{\Gamma}{\Gamma'}\Gamma(\gx) \squbc \Gamma'(\gx)$. For the elaboration, we have
	need to show $\Gamma' |- \cast{\Gamma(\gx)}\Gamma$}

\case{EType, EBool, ETrueFalse}{Then $\g{s'} = \gs $, so the result holds trivially.
}

\case{ELam}{ Then $\g{s} = \g{\lambda x \ldotp \g{s_{2}}}$, $\g{s'} = \g{\lambda x \ldotp \g{s'_{2}}}$,
	and $\g{T} = \g{(x : V_{1}) -> V_{2}}$.
	By our premise we have $(\gx : \g{V_{1}})\Gamma |- \echeck{s_{2}}{V_{2}}{t_{2}} $.
	There are two possible forms for $\g{T'}$: $\qml$ or $\g{(x : T'_{1}) ->T'_{2}}$.

	For $\qml$: IH gives $\Gamma |- \echeck{\g{s'_{2}}{\qml}{t'_{2}}}$
	where $(\gx : \g{V_{1}})\Gamma |- \cast{V_{2}}{\qml}\g{t_{2}} \squbc \g{t'_{2}} $.
	By \rrule{CLam} and \rrule{ELamUnk} we have
	$\Gamma |- \echeck {\lambda x \ldotp \g{s_{2}}}{\qml}{\cast{\qml -> \qml}{\qml}\g{(\lambda x : \qml \ldotp t'_{2})}}$.
	It remains to show that

}

\end{proofEnd}

\begin{theoremEnd}[apxlem]{lemma}[Elaboration Gradual Guarantee]
	Consider \surflang terms $\gs$ and $\g{s'}$, where $\Gamma |- \gs \squbs \g{s'}$.
	\begin{itemize}

		\item If  $\Gamma |- \echeck{\gs}{\gV}{\gt}$ and $\Gamma |- \g{V} \squbc \g{V'}$
					then $\Gamma |- \echeck{\gs}{\g{V'}}{\g{t'}}$
					 where $\Gamma |- \cast{V}{V'}\gt \squbc \g{t'}$.

		\item If $\Gamma |- \esynth{\gs}{\gT}{\gt}$ and $\gs \neq \qml$.
					
					then $\Gamma |- \esynth{\g{s'}}{\g{T'}}{\gt'}$
					 where $\Gamma |- \cast{T}{T'} \gt \squbc \g{t'}$
					and $\Gamma |- \gT \squbc \g{T'}$.
	\end{itemize}
\end{theoremEnd}

\begin{corollary}[SGG for \surflang]
	If $|- \gs : \gS$ and $\gs \squbs \g{s'}$,
	then $ |- \g{s'} : \g{S}$.
	\end{corollary}

	\begin{corollary}[(Semantic) DGG for \clang]
		Suppose $\cdot |- \gt : \gT$ and $\cdot : \g{t'} : \gT$,
		where $\cdot |- \gt \squbc \g{t'}$.
		Then for any context $\gamma : \gT -> \g{\bB} $,
		$\gamma\ \g{t} \stepstostar \g{v}$ and $\gamma\ \g{t'} \stepstostar \g{v'}$,
		where $\cdot |- \g{v} \squbc \g{v'}$.
	\end{corollary}

\section{Translating \clang to \tlang}
\label{sec:translation}

The final important property of \clang's approximate normalization is that each term terminates,
so we know type-checking is decidable. The ultimate result of this section is:

\begin{theoremEnd}[apxproof]{theorem}[Strong Approximate Normalization]
  \label{thm:strong-norm}
  If $\Gamma |- \gt : \gV$, then $\gt \stepstostar \gv$ for some $\gv$. 
\end{theoremEnd}
\begin{proofEnd}
  We rely on the strong normalization of \tlang,
  proceeding by strong induction on the number of reductions in the normalization of $\T{\gt}$.
  If $\T{\gt}$ is normal, then by \cref{lem:value-preserve} so is $\gt$,
  giving us our result.
  Otherwise, suppose $\T{\gt}$ is not normal. If $\gt$ is normal, we have our result.
  Otherwise, $\gt \stepsto \g{t'}$ and $\Gamma |- \g{t'} : \gV$ for some $\g{t'}$.
  If $\g{t'}$ is $\err$, we have our result. Otherwise, by \cref{lem:sim},
  $\T{\gt} \stepstoplus \T{\g{t'}}$. Since $\T{\g{t'}}$ reduces to a normal form
  in strictly fewer steps, our hypothesis and the transitivity of $\stepstostar$
  yield our result.
\end{proofEnd}

We show this using a \textit{syntactic model} in the sense advocated by \citet{10.1145/3018610.3018620}: each \clang term
is translated into a term from a variant of static CIC such that
every \clang reduction step corresponds to at least one step in the static language,
ensuring that no term reduces infinitely.

The syntactic model is useful for more than just strong normalization.
It allows us to
understand the dynamic semantics of \clang through
a well-known static language.
Moreover, it suggests an implementation strategy: terms can be normalized
by first elaborating from \surflang to \clang (\cref{sec:elab}), then translating the \clang term to an existing language,
such as Agda or Idris.

We translate \clang terms to \tlang, a predicative variant of CIC extended with inductive-recursive types.
While inductive-recursive types have been studied extensively in the context of \MLTT,
it is usually presented with a typed normalization relation.
Conversely, CIC has an untyped reduction semantics by which we can easily
show a simulation of \clang's reduction steps.
The model uses a similar technique
as \citet{bertrand:gcic}, but it serves as the foundation for our novel approach to
equality (\cref{sec:equality}).

We use Agda-like syntax for readability.

\subsection{Supporting Definitions}
\label{subsec:trans-support}

\myparagraph{Type Definitions}

To translate from \clang to \tlang, we define and use several types and functions.
In \cref{fig:codes}, we define $\s{GBool}$ to represent \clang booleans,
$\s{GUnk_{\ell}}$ for $\qml$, and $\s{Err}$ for $\err$.
Additionally, we define an inductive-recursive \textit{universe \ala Tarski}: a datatype $\s{Code_{\ell}}$ whose elements
describe the structure of types, and a function $\s{El_{\ell}}$ interpreting those descriptions back into types.

 The idea is that, while \tlang has no type-case operator, $\s{Code}$ is a regular data type that we can
 pattern match against. $\s{El_{\ell}}$ turns these descriptions of types into actual types, but
 because types may depend on values, the definition of the inductive type $\s{Code}$
 refers to the recursive function $\s{El_{\ell}}$ that matches on $\s{Code}$ values
 (thus the name inductive-recursive type).
This technique is standard for dependent type introspection~\citep{10.1145/3018610.3018620,bertrand:gcic}.
As a notational shorthand in \cref{fig:codes}, we assume that $\s{Code_{\ell-1}}$ is empty when $\s\ell$ is $\s{0}$.
Note that $\s{GUnk_{\ell}}$ only depends on the definition of $\s{Code_{\ell}}$ from strictly lower
universe levels: just as $\qm_{\gType{0}}$ cannot contain values from $\gType{\ell}$,
$\s{GUnk_{0}}$ does not embed $\s{Code_{\ell}}$.

\begin{figure}
      \begin{minipage}{0.3\textwidth}
    \scriptsize
  \begin{flalign*}
    &\s{data\ GBool : \sType{}\ where}\\
    &\qquad \s{BUnk : GBool}\\
    &\qquad \s{BErr : GBool}\\
    &\qquad \s{BTrue : GBool}\\
    &\qquad \s{BFalse : GBool}\\
    &\s{data\ GUnk_\ell : \sType{\ell}\ where}\\
    &\qquad\s{UUnk : GUnk_\ell}\\
    &\qquad\s{UErr : GUnk_\ell}\\
    &\qquad\s{UBool : GBool -> GUnk_\ell}\\
    &\qquad\s{UFun : (\bOne -> GUnk_\ell) -> GUnk_\ell}\\
    &\qquad\s{UCode : Code_{\ell-1} -> GUnk_\ell}
\end{flalign*}
      \end{minipage}
      \begin{minipage}{0.3\textwidth}
    \scriptsize
  \begin{flalign*}
    &\s{data\ Code_\ell\ where}\\
    &\qquad \s{CUnk : Code_\ell}\\
    &\qquad \s{CErr : Code_\ell}\\
    &\qquad \s{CType : Code_\ell}\\
    &\qquad \s{CPi : (c : Code_\ell) ->}\\&\qquad\qquad \s{((x : El_\ell\ c) -> Code_\ell) -> Code_\ell }\\
    &\qquad \s{CBool : Code_\ell}\\
    &\s{data\ GErr_\ell : \sType{\ell}\ where}\\
  &\qquad\s{Err : GErr}
\end{flalign*}
      \end{minipage}
      \begin{minipage}{0.3\textwidth}
    \scriptsize
  \begin{flalign*}
    &\s{El_{\ell} : Code_\ell -> \sType{\ell} }\\
    &\s{El_\ell\ CUnk = GUnk_\ell}\\
    &\s{El_\ell\ CErr = GErr_\ell}\\
    &\s{El_\ell\ CType = \s{Code_{\ell-1}}}\\
    &\s{El_\ell\ (CPi\ dom\ cod) =} \\&\qquad \s{ (x : El_\ell\ dom) -> El_\ell\ (cod\ x)}\\
    &\s{El_\ell\ CBool = GBool}
  \end{flalign*}
  \end{minipage}
  \caption{Inductive-recursive definition of codes and their interpretations}
  \label{fig:codes}
\end{figure}

\begin{figure}
  \begin{minipage}{0.3\textwidth}
    \scriptsize
  \begin{flalign*}
    &\s{\top_{\_} : (c : Code_\ell) -> El_\ell}\\
    &\s{\top_{CUnk} = UUnk}\\
    &\s{\top_{CErr} = Err}\\
    &\s{\top_{CType} = CUnk}\\
    &\s{\top_{(CPi\ dom\ cod)} = (x : \top_{dom}) --> \top_{cod\ x}}\\
    &\s{\top_{CBool} = BUnk}\\
    &\overline{\s{\bot_{\_} : (c : Code_\ell) -> El_\ell}}\\
    &\s{\bot_{CUnk} = UErr}\\
    &\s{\bot_{CErr} = Err}\\
    &\s{\bot_{CType} = CErr}\\
    &\s{\bot_{(CPi\ dom\ cod)} = (x : \bot_{dom}) --> \bot_{cod\ x}}\\
    &\s{\bot_{CBool} = BErr}
  \end{flalign*}
\end{minipage}
  \begin{minipage}{0.3\textwidth}
    \scriptsize
  \begin{flalign*}
    &\s{is\bot : (c : Code_\ell) -> El_\ell -> \bB}\\
    &\s{is\bot\ {CUnk}\ UErr = true}\\
    &\s{is\bot\ {CUnk}\ \_ = false}\\
    &\s{is\bot\ {CErr}\ Err = true}\\
    &\s{is\bot\ {CErr}\ \_ = false}\\
    &\s{is\bot\ {CType}\ CErr = true}\\
    &\s{is\bot\ {CType}\ \_ = false}\\
    &\s{is\bot\ {(CPi\ dom\ cod)}\ f = is\bot\ (cod\ \top_{dom}) (f\ \top_{dom}) } \\
    &\s{is\bot\ {CBool}\ BErr = true}\\
    &\s{is\bot\ {CBool}\ \_ = false}
  \end{flalign*}
\end{minipage}
  \begin{minipage}{0.3\textwidth}
    \scriptsize
  \begin{flalign*}
    &\s{gIf\ : (c : Code_{\ell}) -> GBool ->}\\&\qquad\s{
 El_{\ell}\ c -> El_{\ell}\ c -> El_{\ell}\ c}\\
    &\s{gIf\ c\ GTrue\ t_{1}\ t_{2} = t_{1}}\\
    &\s{gIf\ c\ GFalse\ t_{1}\ t_{2} = t_{2}}\\
    &\s{gIf\ c\ BUnk\ t_{1}\ t_{2} = \top_{c}}\\
    &\s{gIf\ c\ BErr\ t_{1}\ t_{2} = \bot_{c}}
  \end{flalign*}
  \end{minipage}
  \caption{Top, bottom, bottom-checking and gradual conditionals}
  \label{fig:top-botcheck}
\end{figure}

\myparagraph{$\qm$, $\err$ and Error Detection}
\clang has $\qm$ and $\err$ for every type.
We account for this by denoting special top and bottom elements
for each code's interpretation.
Additionally, several reduction rules explicitly check if a term is $\err$,
so we need a way to check if a term is the translation of one that is $\eta$-equivalent
to $\err$. We define such functions (\cref{fig:top-botcheck}), named $\s{\top}, \s{\bot}$ and $\s{is\bot}$ respectively.

One subtlety arises when deciding whether a function is equivalent to $\err$.
The term $\err_{\g{(x : T_{1}->T_{2})}}$ is $\eta$-equivalent to
$\g{\lambda x \ldotp \err_{\g{T_{2}}}}$, which eliminates the need to check
if each function  is $\err$ before applying. However, given an arbitrary function, how can
we determine whether it produces $\err$ on all inputs?
The key is monotonicity: \cref{sec:precision} shows that \clang functions
are monotone with respect to their input. So a function is equivalent to $\err$ if and only if
it produces $\err$ when given $\qm$ as input.

\myparagraph{Casts and Conditionals:}
In addition to representing \clang values, we also need
to represent computation.
Applications are represented by applications.
We need not handle $\qm$ or $\err$ specially, since they are $\eta$-expanded in $\s{\top}$ and $\s{\bot}$.
So the \rrule{$\beta$}, \rrule{$\beta$?} and \rrule{$\beta\errsym$} rules are handled.

To handle conditionals, we define a pattern matching function $\s{gIf}$ (\cref{fig:top-botcheck}),
which acts like a normal conditional, but takes a code for the result type,
which is given to $\s{\top}$ or $\s{\bot}$ if the scrutinee is $\s{BUnk}$ or $\s{BErr}$.
The cases correspond to
\rrule{IfTrue}, \rrule{IfFalse}, \rrule{If?} and \rrule{If$\errsym$}.

\clang allows casts between arbitrary types, so we define casts
between the interpretations of arbitrary codes
by pattern matching on the given codes (\cref{fig:cast-defn}).
The cases
directly correspond to the cast reduction rules of \cref{fig:clang-reductions},
although we break \rrule{To?} and \rrule{From?} into separate rules for each tag.

\begin{figure}
  \begin{minipage}{0.45\textwidth}
    \scriptsize
  \begin{flalign*}
    &\s{cast : (c_1\ c_2 : Code_\ell) -> El_\ell\ c_1 -> El_\ell\ c_2}\\
    &\s{cast\ c_{1}\ c_{2}\ x \mid is\bot\ c_{1}\ x = \bot_{c_{2}}} \text{ (\rrule{Cast$\errsym$})} \\
    &\s{cast\ CErr\ c_{2}\ x  = \bot_{c_{2}}} \text{ (\rrule{From$\errsym$}) }\\
    &\s{cast\ c_{1}\ CErr\ x = Err} \text{ (\rrule{To$\errsym$}) }\\
    &\s{cast\ (CPi\ dom\ cod)\ Unk\ f = } \\& \qquad\s{UFun (\lambda x \ldotp cast\ (cod\ \top_{dom}) CUnk (f\ \top_{dom})) }  \text{ (\rrule{?$\Pi$}) }\\
    &\s{cast\ CBool\ Unk\ b = UBool\ b }  \text{ (\rrule{To?}) }\\
    &\s{cast\ CType\ Unk\ c = UCode\ c }  \text{ (\rrule{To?}) }\\
    &\s{cast\ Unk\ (CPi\ dom\ cod)\ (UFun f) = }\\&\qquad\s{  \lambda x \ldotp cast\ CUnk\ (cod\ x)\ (f\ (cast\ dom\ CUnk)) }  \text{ (\rrule{From?}) }
  \end{flalign*}
  \end{minipage}
    \begin{minipage}{0.45\textwidth}
        \scriptsize
      \begin{flalign*}
        &\textit{...continued}\\
  &\s{cast\ Unk\ CBool\ (UBool\ b) =  b }  \text{ (\rrule{From?}) }\\
    &\s{cast\ Unk\ CType\ (UCode\ c) =  c }  \text{ (\rrule{From?}) }\\
    &\s{cast\ (CPi\ dom1\ cod1)\ (CPi\ dom2\ cod2)\ f = }\\&\qquad\s{ \lambda x \ldotp cast\ (cod1\ x')\ (cod2\ x)\ (f\ x')  }  \text{ (\rrule{$\Pi\Pi$}}) \\
    & \qquad \s{ \text{ where } x' = cast\ dom2\ dom1\ x} \\
    &\s{cast\ CType\ CType\ c =  c }  \text{ (\rrule{TypeType}) }\\
    &\s{cast\ CBool\ CBool\ b =  b }  \text{ (\rrule{$\bB\bB$}) }\\
    &\s{cast\ CUnk\ CUnk\ d =  d }  \text{ (\rrule{??}) }\\
    &\s{cast\_\ \_\ \_ = err\ c_2} \text{ \textit{otherwise} (\rrule{tag$\errsym$, ?$\errsym$}) }
  \end{flalign*}
\end{minipage}
\caption{Conversion functions between interpretations of codes}
  \label{fig:cast-defn}
\end{figure}

\subsection{The Translation}
\label{subsec:term-tran}
Our supporting definitions make it straightforward to define the full translation (\cref{fig:translation}).
The translation is compositional, and \clang has enough type ascriptions that the translation need not be type- directed.
Functions are translated to functions, and other types are translated into
the types from \cref{fig:codes}.
We translate $\qm$ and $\err$ using $\s{top}$ and $\s{\bot}$,
and use $\s{cast}$ to translate casts.

\begin{figure}
  \begin{minipage}{0.35\textwidth}
    \scriptsize
  \begin{flalign*}
    & \T{(x : T_{1})->T_{2}} = \s{CPi}\ \T{T_{1}}\ \s{(\lambda x \ldotp} \T{T_{2} } \s{)} \\
  & \T{\cast{T_{1}}{T_{2}}\gt} = \s{cast}\ \T{T_{1}}\ \T{T_{2}}\ \T{t}\\
    & \T{\attagl{\gType{}}\gT} = \s{GCode}\ \T{T}\\
    & \T{\gType{\ell}} = \s{CType}\\
    & \T{if_{T}\ t_{1}\ t_{2}\ t_{3}} = \s{gIf}\ \T{T}\ \T{t_{1}}\ \T{t_{2}}\ \T{t_{3}}
  \end{flalign*}
\end{minipage}
\begin{minipage}{0.3\textwidth}
    \scriptsize
\begin{flalign*}
    & \T{\bB} = \s{CBool}\\
    & \T{true} = \s{BTrue}\\
    & \T{false} = \s{BFalse}\\
    & \T{\qm_{\gT}} = \s{\top_{\T{T}}}\\
    & \T{\err_{\gT}} = \s{\bot_{\T{T}}}
  \end{flalign*}
\end{minipage}
\begin{minipage}{0.3\textwidth}
    \scriptsize
\begin{flalign*}
    & \T{x} = \sx\\
    & \T{\attagl{\Pi}\gt} = \s{GFun}\ \T{t}\\
    & \T{\attagl{\bB}\gt} = \s{GBool}\ \T{t}\\
    & \T{\lambda x : T \ldotp t} = \s{\lambda x \ldotp } \T{t} \\
    & \T{t_{1}\ t_{2}} = \T{t_{1}}\ \T{t_{2}}
   \end{flalign*}
 \end{minipage}
 \caption{Translation from \clang into \tlang}
  \label{fig:translation}
\end{figure}

To show that the result of translation
is terminating, we need to know that it is well typed in \tlang.
The explicitness of our cast calculus makes this easy to show:
each type translates to a well-typed code,
and each term translates to an element of the interpretation of its type's translation.

\begin{theoremEnd}[apxproof]{lemma}[Translated Type Preservation]
  The following hold:
  \begin{itemize}
\item If $\Gamma |- \gV : \gType{\ell}$, then $\T{\Gamma} |- \T{\gV} : \Code_{\s\ell}$
\item If $\Gamma |- \gt : \gV$, then $\T{\Gamma} |- \T{\gt} : \s{El_\ell}\ \T{\gV}$
  \end{itemize}
\end{theoremEnd}
\begin{proofEnd}
 TODO
\end{proofEnd}

The main property we need for the termination of \clang terms is \textit{simulation}:
each \clang step corresponds to one or more \tlang steps.
This bounds the number of steps a \clang term takes,
so combined with progress and preservation, we deduce strong normalization.

\begin{theoremEnd}[apxproof]{lemma}[Simulation]
  \label{lem:sim}
  If $\Gamma |- \gt : \gV $ and $\gt \stepsto \g{t'}$, then
  $\g{t'}$ is $ \err$ or
  $\T{\gt} \stepstoplus \T{\g{t'}}$.
\end{theoremEnd}
\begin{proofEnd}
 TODO
\end{proofEnd}

\subsection{Values and the Back-Translation}
\label{subsec:back-trans}

In addition to showing strong normalization, we can view our translation as a form of compilation.
The translation of a \clang term reduces in a similar enough way to the original term that
we recover its evaluation from the evaluation of the translation.

To achieve this, we define a \textit{back-translation} (\cref{fig:back-trans}) from
\tlang to \clang.
One difficulty in defining such a translation is handling neutral terms,
since not all pattern matches in \tlang correspond to matches in \clang.
However,
we take advantage of how \clang cannot pattern match on $\gType{}$,
so any match on a value of type $\s{Code_{\ell}}$ was produced by $\s{\top}$,
$\s{\bot}$, $\s{cast}$ or $\s{\El}$.
Another challenge is in handling codes: since $\T{  }$ produces calls to $\El_{\ell}$,
evaluating the resulting \tlang term produces a type. So we translate both types
and codes from \tlang into \clang types.

The back-translation is given in \cref{fig:back-trans}.
as a partial function from \tlang values to \clang values.
For readability, we write the neutral forms of pattern matches
in terms of the functions generating those matches, with the understanding
that the actual \tlang values would have the function names replaced with their bodies.
For example, the rule producing $\g{if}$ actually checks that the argument is of the form
$\s{match\ t_{1}\ with \{GTrue => t_{2} ; GFalse => t_{3} ; GUnk => \top_{c} ; GErr => \bot_{c} \}}$.
Likewise, we assume the back-translation has access to a full typing derivation for the \tlang term,
so that functions may be ascribed with their argument types, $\qml$ may be ascribed
with the proper level, etc. For brevity, we omit the derivation from the definition.

\begin{figure}
  \begin{minipage}{0.4\textwidth}
    \scriptsize
  \begin{flalign*}
   & \V{CPi\ v_{c1}\ (\lambda x \ldotp v_{c2})} = \g{(x : \V{v_{c1}}) -> \V{v_{c2}}} \\
   & \V{CPi\ v_{c1}\ N} = \g{(x : \V{v_{c1}}) -> \V{N}\ x} \\
   & \V{(x : V_{1}) -> V_{2}} = \g{(x : \V{c_{1}} ) -> \V{c_{2}} }   \\
      & \V{gIf\ v_{c}\ N\ v_{1}\ v_{2}} = \g{if}_{\V{v_{c}}}\ \V{N}\ \V{v_{1}}\ \V{v_{2}}\\
      & \V{cast\ v_{c1}\ v_{c_{2}}\ v} = \cast{\V{v_{c1}}}{\V{v_{c2}}} \V{v} \\
      & \qquad \textit{ where one of } \s{v_{c1}}, \s{v_{c2}}, \s{v} \textit{ neutral } \\
      & \V{UUnk : GUnk_{\ell}} = \qml\\
      & \V{UErr : GUnk_{\ell}} = \err_{\gType{\ell}}\\
      & \V{UFun\ v : GUnk_{\ell} } = \attagl{\Pi}{\V{v}} \\
      & \V{UBool\ v : GUnk_{\ell} } = \attagl{\bB}{\V{v}} \\
      & \V{UCode\ v : GUnk_{\ell} } = \attagl{\gType{\ell}}{\V{v}} \\
  \end{flalign*}
\end{minipage}
  \begin{minipage}{0.28\textwidth}
    \scriptsize
    \begin{flalign*}
   & \V{x} = \gx \\
   & \V{Code_{\ell}} = {\gType{\ell}}  \\
   & \V{CType} = {\gType{\ell}}  \\
   & \V{CBool} = {\g{\bB}}  \\
   & \V{GBool} = {\g{\bB}}  \\
   & \V{CErr : Code_{\ell}} =  \err_{\gType{\ell}} \\
   & \V{GErr_{\ell}} =  \err_{\gType{\ell}} \\
   & \V{CUnk : Code_{\ell}} =  \qm_{\gType{\ell}} \\
   & \V{GUnk_{\ell}} =  \qm_{\gType{\ell}} \\
  \end{flalign*}
\end{minipage}
  \begin{minipage}{0.27\textwidth}
    \scriptsize
    \begin{flalign*}
      & \V{\lambda (x : V) \ldotp t} = \g{\lambda (x : } \V{V} \g{) \ldotp} \V{v}\\
      & \V{BTrue} = \g{true}\\
      & \V{BFalse} = \g{false}\\
      & \V{BUnk} = \g{\qm_{\g{\bB}}}\\
      & \V{BErr} = \g{\err_{\g{\bB}}}\\
      & \V{\top_{N}} = \qm_{\V{N}}\\
      & \V{\bot_{N}} = \err_{\V{N}}\\
      & \V{N\ v} = \V{N}\ \V{v}\\
  \end{flalign*}
\end{minipage}
\caption{Back-translation from \tlang into \clang}
  \label{fig:back-trans}
\end{figure}

The important property of back-translation is that it respects
\clang's evaluation: translating a value, then back-translating,
yields the original value. As a corollary, we obtain the algorithm
for computing approximate normal forms by executing \tlang terms.

\begin{theoremEnd}[apxproof]{lemma}[Value Back-Translation Correctness]
  \label{lem:back-trans}
  If $\Gamma |- \gv : \g{V'}$ then  
  $\V{\T{\gv}}{\g{V'}} = \gv$.
\end{theoremEnd}
\begin{proofEnd}
  TODO
\end{proofEnd}

\begin{theoremEnd}[apxproof]{corollary}[Norm. By Translation]
  If $\Gamma |- \gt : \gV$, then $\gt \stepstostar \V{\sv}$
  where $\T{\gt} \stepstostar \sv$.
\end{theoremEnd}
\begin{proofEnd}
  From \cref{thm:strong-norm} we have $\gt \stepstostar_{\textsc{Approx}} \gv$,
  and repeatedly applying \cref{lem:sim} gives that  $\T{\gt} \stepstostar \T{\gv}$.
  Then by \cref{lem:back-trans}, $\gv = \V{\T{\gv}}{\gV}$.
\end{proofEnd}

\section{Adding Propositional Equality and Inductive Types}
\label{sec:equality}

\surflang and \clang are too minimal to be practically useful, but they lay the foundation
on which we build a more fully-featured gradual dependently typed language.
To handle arbitrary user-defined inductive type families, the key is specifying
a gradual representation for propositional equality,
since this can be used to represent constructors' constraints on index values.

We present \surfelang and \celang, which extend  \surflang and \clang repsectively
with support for propositional equality.
We add types and constructors for evidence-based \textit{propositional consistency},
which extends propositional equality to accommodate imprecision among gradual dependent types.
We also introduce \textit{explicit composition expressions} for evidence,
so that composing two pieces of evidence is a syntactic construct within \celang itself.
Our previous proofs of safety, strong normalization, and the gradual guarantees
are all easily modified to encompass these
new language features.
Once equality is added, inductive families are implemented in terms of propositional consistency.

\subsection{Overview of Propositional Consistency}

Thus far, we have alluded to dynamic checks that ensure gradual type safety.
But what do those checks actually entail? We affix a tag to values of type $\qm$,
denoting the top-level type constructor.
When casting from $\qm$ to a more precise type, we fail if the stored tag does not match
the target type.

However, for indexed type families, the type constructor does not provide all the information we need.
Knowing that a value is a $\g{Vec}$ does not say if it is safe to convert from $\g{Vec\ Float\ \qm}$
to $\g{Vec\ Float\ 1}$. So we must also track constraints on the type indices.

\myparagraph{Tracking Index Consistency}

Consider how our incorrect $\g{sort}$ from \cref{sec:intro} operates on the list
$\g{[3.3 , 3.3] : Vec\ Float\ 2}$.
In the simply-typed version, $\g{sort}$
partitions the tail into $\g{[]}$ and $\g{[]}$, sorts each partition,
and produces the incorrect result $\g{[3.3]}$.
We can detect this by tracking the equations induced by
constraints on type indices.
The calls $\g{filter\ (>\ 3.3)\ [3.3]}$ and $\g{filter\ (>\ 3.3)\ [3.3]}$  each
produce $\g{[] : Vec\ Float\ 0}$,
which is cast to $\g{Vec\ Float\ ?}$ to match the return type of $\g{filter}$.
Critically, \textit{we remember that the original length was 0}:
$\g{Vec}$ elaborates to a \cilang inductive type, where there is an extra field
storing $\g{0}$ as evidence that $\g{0}$ is consistent with $\qm$.
Then, $\g{Cons\ 3.3\ []} = \g{[3.3] : Vec\ Float\ (1 + \qm)}$,
but we compose this with the evidence on $\g{[]}$ to get $\g{1}$ as witness between
the consistency of $\g{1 + \qm}$ and $\g{1 + 0}$.
The error is discovered when we cast from $\g{Vec\ Float\ (1 + \qm)}$ to $\g{Vec\ Float\ 2}$,
the return type specified by the given signature.
During this cast, we compose the target index $\g{2}$
with $\g{1}$, the evidence on $\g{[3.3]}$,
which fails when the
distinct precise values $\g{1}$ and $\g{2}$ compose to $\err$.

\myparagraph{Abstracting Out Propositional Consistency}
\label{subsec:prop-cst-intro}
While it is easy to compare and compose members of $\g{\bN}$,
not all compositions are so simple.
We need a representation of evidence that captures plausible equalities between
any two gradual values of the same type.
Since inductive families may be indexed by
any type, we need a general way of tracking indices and composing evidence.
This is precisely the purpose of propositional consistency:
just as the static $\s{refl : t \equiv_{T} t}$ witnesses propositional equality of two terms,
in \lang $\grefl{t}{t_{1}}{t_{2}} : \g{t_{1} \pc_{T} t_{2}}$
witnesses that $\g{t_{1}}$ and $\g{t_{2}}$
\textit{could plausibly be equal}, given their imprecision.
$\g{t}$ is a term which is as precise as both $\g{t_{1}}$ and $\g{t_{2}}$.
If $\g{t}$ is not $\err$,
then we cannot immediately conclude that $\g{t_{1}}$ and $\g{t_{2}}$
are inconsistent.

\surfelang has a propositional equality type with the usual constructors,
but this elaborates to propositional consistency, so that the equalities can be tracked
during normalization.
We elaborate $\g{refl}$ with the initial evidence $\g{\rmeet{t_{1}}{t_{2}}{T}}$.
As the notation suggests, this is the precision greatest lower-bound of $\g{t_{1}}$ and $\g{t_{2}}$,
so it is the least precise evidence for the consistency of $\g{t_{1}}$ and $\g{t_{2}}$.

Once we have propositional consistency, we can handle inductive families. Elaboration uses a known technique
where all inductive families are turned into parameterized inductive types, so that
each constructor has the same return type, but takes equality proofs as arguments
to constrain the indices~\citep{dtfpp,Chapman:2010:GAL:1863543.1863547}.

For example, $\g{Vec}$ is represented as

{\small
\begin{flalign*}
  & \g{\textbf{data}\ Vec : (X : \gType{}) -> (n : \bN) -> \gType{}\ \textbf{where}}\\
  & \qquad \g{Nil : n \pc_{\bN} 0 -> Vec\ X\ n}
  \quad \mid \quad \g{Cons : (m : Nat) -> X -> Vec\ X\ m \-> n \pc_{\bN} 1 + m -> Vec\ X\ n}
\end{flalign*}
}

\myparagraph{Dealing with Neutrals}

The subtlety of defining propositional consistency, and the related composition operator,
lies with neutral terms.
When we cast between propositional consistency types, we must compose the evidence
to ensure monotonicity. But the terms we are composing might have variables.
Consider $\g{\rmeet{(\lambda x\ y \ldotp x)}{(\lambda x\ y \ldotp y)}{\bN -> \bN -> \bN} }$.
This expression cannot produce an error, as
we would need to compare variables for equality,
but languages like CIC do not have such reflection capabilities,
so such comparisons cannot be directly included in a syntactic model.
If it does not produce an error, then the expression is evidence that
$\g{\lambda x\ y \ldotp x}$ and $\g{\lambda x\ y \ldotp y}$ are propositionally consistent, despite
being non-equal static terms.

Our approach is to make syntactic consistency stronger than
propositional consistency.
When comparing terms for consistency during type-checking, $\gx$ and $\gy$ must be
inconsistent, so that we conservatively extend CIC.
Conversely, composition is lazy, so neutral terms are consistent with all other terms
of the same type. When values are provided for variables,
we can evaluate further and possibly produce an error.
For our example, $\g{\rmeet{(\lambda x\ y \ldotp x)}{(\lambda x\ y \ldotp y)}{\bN -> \bN -> \bN} }$
reduces to $\g{\lambda x\ y \ldotp (\rmeet{x}{y}{\bN}) }$.

\subsection{\surfilang and \celang: Adding Propositional Consistency}
\label{subsec:eq-syntax}

\surfelang extends \surflang with the usual propositional equality type constructor $\g{==}$ and proof term $\g{refl}$,
plus an elimination principle $\g{subst}$, so the complexity of
propositional consistency is hidden from the surface language.
\celang, conversely, extends \clang with propositional consistency,
to which \surfelang's propositional equality elaborates,
and
an explicit composition operator.
Evidence ought to be viewed as a separate language from terms: evidence represents a slice
of the plausible equalities that could hold between two gradual terms.
Composition is
only needed for terms representing evidence, but all \celang features are present in the evidence language,
so we do not distinguish it from \celang.
{\small
\begin{displaymath}
  \gt \bnfadd \g{t_{1} \pc_{\gT} t_{2}}
  \bnfalt \grefl{t_{1}}{t_{2}}{t_{3}}
  \bnfalt \g{subst} \bnfalt \rmeet{t_{1}}{t_{2}}{T}
\end{displaymath}
}
The introduction form is $\grefl{t_{1}}{t_{2}}{t_{3}}$, where $\g{t_{1}}$ is \textit{evidence}
of consistency between $\g{t_{2}}$ and $\g{t_{3}}$.
The composition $\g{t_{1}}, \g{t_{2}} : \gT$ as $\rmeet{t_{1}}{t_{2}}{T}$
computes a
 \textit{precision lower bound} of $\g{t_{1}}$ and $\g{t_{2}}$,
an invariant we formalize below.
Normals,  neutrals and term heads (omitted) are like in \clang. The term $\rmeet{t_{1}}{t_{2}}{T}$
is neutral if either $\g{t_{1}}$ or $\g{t_{2}}$ is, unless one is $\qm$ or $\err$,
in which case we can reduce.

For typing, the interesting rule is \rrule{CRefl},
where $\grefl{t_{1}}{t_{2}}{t_{3}}$ synthesizes $\g{t_{2} \pc_{T} t_{3}}$
so long as $\g{t_{1}}$
is at least as precise as both $\g{t_{2}}$ and $\g{t_{3}}$,
and all three have the same type.

Apart from that, $\g{=-=}$ and $\g{subst}$ are typed the same
as their static counterparts, and $\rmeet{t_{1}}{t_{2}}{T}$ has
type $\gT$ so long as its inputs do.

\begin{figure}
  \begin{boxedarray}{@{}l@{}}
  \boxed{\Gamma |- \g{t} : \g{T} \quad \textit{(\celang typing, new rules)}}\\
  \begin{inferbox}
    \inferrule[CEq]{\Gamma |- \gT => \gType{\ell} \\\\
      \Gamma |- \g{t_1} <= \gT\\
      \Gamma |- \g{t_2} <= \gT
    }{\Gamma |- \g{t_1 \pc_{\gT} t_2} => \gType{\ell}}

    \inferrule[CRefl]{
      \Gamma |- \g{t_1} => \g{T}\\
      \Gamma |- \g{t_2} <= \g{T}\\
      \Gamma |- \g{t_3} <= \g{T}\\\\
      \Gamma |- \g{t_1} \squbc \g{t_2}\\
      \Gamma |- \g{t_1} \squbc \g{t_3}
    }
    {\Gamma |- \grefl{t_1}{t_2}{t_3} => \g{t_2 \pc_{\gT} t_3}}

    \inferrule[CComp]{
      \Gamma |- \g{t_1} => \gT \\\\
      \Gamma |- \g{t_2} <= \gT
    }
    {\Gamma |- \rmeet{t_1}{t_2}{\gT} => \gT}

    \inferrule[CSubst]{}
    {\Gamma |- \g{subst} => \g{(X : \gType{\ell}) -> (X_P : X -> \gType{\ell'}) -> (y : X) -> (z : X) -> X_P\ y -> y \pc_X z -> X_P\ z }}
  \end{inferbox}
\end{boxedarray}
\end{figure}

\begin{figure}
  \begin{boxedarray}{@{}l@{}}
    \boxed{\g{t_1} \leadsto \g{t_2} \ \textit{(\celang Reduction))}}\
    \sqcap \text{structural rules omitted, }
    \rrule{$\sqcap\errsym$R} \text{ and } \rrule{$\sqcap?$R} \text{ defined symmetrically}
    \\
  \begin{inferbox}
    \inferrule{}{\g{subst\ T\ T_P\ t_1\ t_2\ t_{Pt1}\ \grefl{t_{ev}}{t_1}{t_2}}
      \redsto_{\rrule{J}} \cast{T_P\ t_{ev}}{T_P\ t_2} \cast{T_P\ t_1}{T_P\ t_{ev}} \g{t_{Pt1}}
      \textit{ where } \g{t_{ev}} =-= \g{t_{ev}}
    }

    \inferrule{}{\g{subst\ T\ T_P\ t_1\ t_2\ t_{Pt1}\ \qm_{t_1 \pc_T t_2}}
      \redsto_{\rrule{J?}} \cast{T_P\ (\rmeet{t_1}{t_2}{T})}{T_P\ t_2} \cast{T_P\ t_1}{T_P\ (\rmeet{t_1}{t_2}{T})} \g{t_{Pt1}}
    }

    \vspace*{-0.9em}
    \inferrule{\g{t_{ev} \not{=-=} \g{t_{ev}}} }{\g{subst\ T\ T_P\ t_1\ t_2\ t''\ t_{ev}}
      \redsto_{\rrule{J$\errsym$}} \err_{\g{(T_P\ t_2)}}
    }

    \inferrule{}{\rmeet{(\lambda x : T\ldotp t_1)}{(\lambda x : T\ldotp t_2)}{(x : T)->T'}
      \redsto_{\rrule{$\sqcap$$\lambda$}} \g{\lambda(x : T)\ldotp \rmeet{t_1}{t_2}{T'}}}

    \inferrule{}
    {\cast{t_1 \pc_{T} t_2}{t'_1 \pc_{T'} t'_2}{\grefl{t_{ev}}{t_1}{t_2}}
      \redsto_\rrule{Eq} \grefl{
        \rmeet{t'_1}{ \rmeet{t'_2}{\cast{T}{T'}t_{ev}}{T'} }{T'}
      }{t'_1}{t'_2}
    }

    \inferrule{\tagOf(\g{t_1}) \neq \tagOf(\g{t_2})}{\rmeet{t_1}{t_2}{T} \leadsto_{\rrule{$\sqcap$Head$\errsym$}} \err_{\gT} }

    \inferrule{}{\rmeet{\qm_T}{t}{T} \leadsto_{\rrule{$\sqcap$?L}} \gt}

    \inferrule{}{\rmeet{\err_T}{t}{T} \leadsto_{\rrule{$\sqcap$$\errsym$L}} \err_{\gT}}

    \inferrule{}{\rmeet{\grefl{t}{t_1}{t_2}}{\grefl{t'}{t_1}{t_2}}{t_1 \pc_T t_2}
      \redsto_{\rrule{$\sqcap$Refl}} \grefl{\rmeet{t}{t'}{T}}{t_1}{t_2}}

    \inferrule{\g{T''} = \rmeet{T}{T'}{\gType{\ell}}}
    {\rmeet{t_1 \pc_{T} t_2}{t'_1 \pc_{T'} t'_2}{\gType{\ell}}
      \redsto_{\rrule{$\sqcap$Eq}} \g{ \rmeet{\cast{T}{T''}t_1}{\cast{T'}{T''}t'_1}{T''} \pc_{T''} \rmeet{\cast{T}{T''}t_2}{\cast{T'}{T''}t'_2}{T''} } }

    \inferrule{\g{T_3} = [\cast{\rmeet{T_1}{T'_1}{\gType{\ell}}}{T_1}\gx/\gx]\g{T_2}
    \\ \g{T'_3} = [\cast{\rmeet{T_1}{T'_1}{\gType{\ell}}}{T'_1}\gx/\gx]\g{T'_2}}{
        \rmeet{(x : T_1 -> T_2)}{(x : T'_1)->T'_2}{\gType{\ell}}
        \redsto_{\rrule{$\sqcap$$\Pi$}}
        \g{(x : \rmeet{T_1}{T'_1}{\gType{\ell}}) -> (\rmeet{T_3}{T'_3}{\gType{\ell}})}
      }

  \end{inferbox}
\end{boxedarray}
\caption{Additional Reduction Rules for \celang}
\label{fig:eq-reductions}
\end{figure}

\myparagraph{Semantics}
\label{subsec:eq-semantics}
\celang extends \clang with two new computational forms and a new type family within which we can cast,
so we introduce new reduction rules (\cref{fig:eq-reductions}).
The first reductions we need are for eliminating via $\g{subst}$.
A \rrule{J} reduction transports a value across an equality by casting it through
the evidence type, which ensures monotonicity, as \citet{10.1145/3236768} observed.
Rule \rrule{J} does the same when given $\qm$, but casts through the meet of the start and target type.
So $\qm_{\g{t_{1} \pc_{T} t_{2}}}$ behaves like $\grefl{\rmeet{t_{1}}{t_{2}}{T}}{t_{1}}{t_{2}}$.
In \rrule{J$\errsym$}, applying $\g{subst}$ to $\err$ produces $\err$ at the target type.

Second, we need rules for casting between equality types (\rrule{Eq}),
which rely heavily on the meet.
When we cast $\grefl{t}{t_{1}}{t_{2}}$ to $\g{t'_{1} \pc_{T} t'_{2}}$, there is no guarantee that $\g{t'_{1}}$
and $\g{t'_{2}}$ are consistent or that $\g{t}$ is valid evidence for $\g{t'_{1}}$ and $\g{t'_{2}}$.
So we cast our evidence into the target type,
then compose with the endpoints of the target equality to ensure
we satisfy the precision requirement of \rrule{CRefl}.
So even if $\g{t'_{1}}$ and $\g{t'_{2}}$ are consistent, we will produce $\err$
if they are not consistent with the evidence $\gt$.

To give an example, suppose we have $\grefl{true}{true}{true} : \g{true \pc true}$.
We could cast this to $\g{true} \pc \g{false}$, but then $\grefl{true}{true}{false}$
would not be well typed at this type. So we compose with $\rmeet{true}{false}{\bB}$,
which normalizes to $\grefl{\err_{\bB}}{true}{false}$.

Finally, we have reductions for composition.
Composing with $\err$ produces $\err$ (\rrule{$\sqcap$$\errsym$L},\rrule{$\sqcap$$\errsym$R}),
and $\qm$ acts as an identity (\rrule{$\sqcap$?L},\rrule{$\sqcap$?R}).
Functions compose by composing their bodies,
and $\grefl{}{}{}$ composes by composing the evidence with which it is ascribed.
Terms with different $\tagOf$ values compose to $\err$ in \rrule{$\sqcap$Tag$\errsym$}.
Rules \rrule{$\sqcap\Pi$} and \rrule{$\sqcap$Eq} work like structural rules, but because of type dependencies,
adds casts to preserve typing.
The remaining meet rules (omitted) are structural rules for terms with the same $\tagOf$:
we apply the constructor to the meet of the arguments.

In addition to our reductions, we add two $\eta$-laws.
First, we have $\qm_{\g{t_{1} \pc_{T} t_{2}}} \stepsto_{\eta} \grefl{\rmeet{t_{1}}{t_{2}}{T}}{t_{1}}{t_{2}} $,
which avoids having $\g{subst}$ produce $\qm$ any time it is given $\qm$,
so potential equalities are dynamically tracked even if $\qm$ is provided in place of $\g{refl}$.
Second, we have $\grefl{\err_{T}}{t_{}1}{t_{2}} \stepsto_{\eta} \err_{\g{t_{1} \pc_{T} t_{2}}}$:
if $\err$ is the only term we know is more precise than $\g{t_{1}}$ and $\g{t_{2}}$,
then we have failed to show they could be equal.

\myparagraph{Elaboration}
\label{subsec:eq-elab}
\surfelang elaborates to \celang much like \surflang does to \clang
(\cref{fig:eq-elab-prec-cst}).
Equality is elaborated directly (\rrule{EEq}).
A proof term $\g{refl}$ is elaborated by taking
composing the equated terms.
We also check that the terms are syntactically consistent,
to avoid accepting fully static terms like $\g{refl : x == y}$.
Like with functions, we have an extra rule \rrule{EReflUnk} that generates $\g{refl}$ with $\qm$ evidence
when the target type is $\qm$, since this is consistent with an equality type.

\begin{figure}
  \begin{boxedarray}{@{}l@{}}
    \boxed{\Gamma |- \esynth{\gs}{\gT}{\gt} \mid \Gamma |- \echeck{\gs}{\gT}{\gt}  \quad \textit{(\surfelang elaboration)}}\ \rrule{EEq} \textit{ omitted}\\
    \begin{inferbox}

      \inferrule[ERefl]{
        \Gamma |- \g{t_1} <= \gT \\
        \Gamma |- \g{t_1} \etasteps \g{v_1} \\\\
        \Gamma |- \g{t_2} <= \gT \\
        \Gamma |- \g{t_2} \etasteps \g{v_2} \\
        \g{t_1} =-= \g{t_2}
      }{
        \Gamma |- \echeck{\g{refl}}{\g{t_1 \pc_T t_2}}{\grefl{\rmeet{t_1}{t_2}{T}}{t_1}{t_2} }}

      \inferrule[EReflUnk]{ }{
        \echeck{\g{refl}}{\qml}{\attagl{\g{Eq}}{\grefl{\qm_{\qml}}{\qm_\qml}{\qm_\qml}}}
      }
    \end{inferbox}
    \\
    \small
  \boxed{\Gamma |- \g{t_{1}} \squbc \g{t_{2}} \quad \textit{(\celang precision)}}\qquad
  \boxed{\g{v_{1}} =-= \g{v_{2}} \quad \textit{(\celang consistency)}}\\
  \rrule{CstUnkEq, CstUnkSubst, CstUnkRefl, CstMeetR} \textit{ omitted }\\
  \begin{inferbox}
    \inferrule[PrecCongEq]{
      \Gamma |- \g{T} \squbc \g{T'}\\
      \Gamma |- \cast{T}{T'}\g{t_1} \squbc \g{t'_1}\\\\
      \Gamma |- \cast{T}{T'}\g{t_2} \squbc \g{t'_2}
    }{\Gamma |- \g{t_1 \pc_T t_2} \squbc \g{t'_1 \pc_{T'} t'_2}}

    \inferrule[PrecAxGreatest]{
      \Gamma |- \g{t_3} \squbc \g{t_1}\\\\
      \Gamma |- \g{t_3} \squbc \g{t_2}
    }{\Gamma |- \g{t_3} \squbc \rmeet{t_1}{t_2}{T}}

    \inferrule[PrecCongRefl]{\Gamma |- \g{t} \squbc \g{t'}}{\Gamma |- \grefl{t}{t_1}{t_2} \squbc \grefl{t'}{t_1}{t_2}}

    \inferrule[PrecAxBound]{\g{i} \in \set{\g{1}, \g{2}}}{\Gamma |- \rmeet{t_1}{t_2}{T} \squbc \g{t_i} }

    \inferrule[PrecAxThrough]{
    }{\Gamma |- \cast{T_1}{T_2}\gt \sqeqc \cast{\rmeet{T_1}{T_2}{\gType{\ell}}}{T_2}\cast{T_1}{\rmeet{T_1}{T_2}{\gType{\ell}}}}

    \inferrule[CstEq]{
      \g{v_1} =-= \g{v'_1}\\
      \g{v_2} =-= \g{v'_2}
    }{ \g{v_1 \pc_{V} v_2} =-= \g{v'_1 \pc_{V'} v'_2} }

    \inferrule[CstSubst]{  }{\g{subst} =-= \g{subst}}

    \inferrule[CstRefl]{
      \g{t} =-= \g{t'}
    }
    {\grefl{t}{t_1}{t_2} =-= \grefl{t'}{t'_1}{t'_2}}

    \inferrule[CstMeetL]{
      \g{t_1} =-= \g{t_3}\\
      \g{t_2} =-= \g{t_3}
    }{\rmeet{t_1}{t_2}{T} =-= \g{t_3}}
  \end{inferbox}
\end{boxedarray}
\caption{\surfelang and \celang: New Elaboration, Precision, and Consistency Rules}
\label{fig:eq-elab-prec-cst}
\end{figure}

In \rrule{ERefl}, since the meet was used as the initial evidence of consistency,
if our elaboration is to be type preserving, the meet must always be a lower-bound
of its inputs. For neutral terms, this means we add axioms to precision.

\myparagraph{Precision}
In \cref{fig:eq-elab-prec-cst} we define the laws for algebraic precision.
\rrule{PrecAxBound} establishes that $\rmeet{t_{1}}{t_{2}}{T}$ is a lower bound for
$\g{t_{1}}$ and $\g{t_{2}}$ at type T, and \rrule{PrecAxGreatest} establishes
that it is the greatest such lower bound.
\rrule{PrecAxThrough} establishes that casting between two types
is equivalent to casting \textit{through} their meet,
making \rrule{PrecAxIntermed} redundant.

Our new syntactic forms also require structural rules.
\rrule{PrecCongRefl} compares by evidence ascriptions, and
\rrule{PrecCongEq} compares the equated terms after being cast into
the type of the (potentially) more precise equality, similar to how
\rrule{PrecCongPi} works (\cref{fig:ax-precision-ax}).
From these rules, we can define all the properties that are expected of a
preorder's meet operator.

\begin{figure}
  \begin{boxedarray}{@{}l@{}}
  \boxed{\Gamma |- \g{t_{1}} \squbc \g{t_{2}} \quad \textit{(\celang precision, new admissible rules)}}\\
  \begin{inferbox}
  \mprset {fraction ={\cdot\cdots\cdot}}
    \inferrule[PrecAdMeet]{
      \Gamma |- \g{t_1} \squbc \g{t'_1}\\
      \Gamma |- \g{t_2} \squbc \g{t'_2}
    }{\Gamma |- \rmeet{t_1}{t_2}{T} \squbc \rmeet{t'_1}{t'_2}{T}}

			\inferrule[PrecAdIntermed]{\Gamma |- \g{T_1} \squbstar \g{T_2}}
			{\Gamma |- \cast{T_1}{T'}\cast{T}{T_1}\gt \squbc \cast{T_2}{T'}\cast{T}{T_2}\gt}

      \inferrule[PrecAdCommut]{ }{\Gamma |- \rmeet{t_1}{t_2}{T} \sqeqc \rmeet{t_2}{t_1}{T}}

      \inferrule[PrecAdAbsorb]{\Gamma |- \g{t_1} \squbc \g{t_2} }{\Gamma |- \rmeet{t_1}{t_2}{T} \sqeqc \g{t_1} }

      \inferrule[PrecAdIdem]{ }{\Gamma |- \rmeet{t}{t}{T} \sqeqc \gt}

  \end{inferbox}
\end{boxedarray}
\caption{Precision: Admissible Rules for \celang}
\label{fig:eq-prec-admitted}
\end{figure}

\myparagraph{Consistency}
Consistency must account for our new syntactic forms
(\cref{fig:eq-elab-prec-cst}).
For equality, we add the necessary structural rules,
along with the \rrule{CstUnk} rules for consistency with $\qm$.
For the meet, we need to make sure upward-closure still holds. The precision rules \rrule{PrecAxBound}
and \rrule{PrecAxGreatest} mean that the composition $\rmeet{t_{1}}{t_{2}}{T}$ is only consistent
with terms that are consistent with both $\g{t_{1}}$ and $\g{t_{2}}$ (\rrule{CstMeetL}, \rrule{CstMeetR}).

\myparagraph{Syntactic Model}
\label{subsec:eq-elab}

    Extending our syntactic model to include equality is once again easy.
Just as we wrote a function $\s{cast : (c_{1}\ c_{2} : Code_{\ell}) -> El_{\ell}\ c_{1} -> El_{\ell}\ c_{2} }$,
we write a function
\\$\s{comp : (c : Code_{\ell}) -> (x\ y : El_{\ell}\ c) -> El_{\ell}\ c}$,
which directly implements the reductions for $\rmeet{}{}{}$ in \tlang.
New $\s{Code}$ and $\s{Unk}$ variants are added for $\g{\pc}$,
and the $\s{cast}$ case between equality codes are implemented using $\s{comp}$.
The type and eliminator for propositional equality/consistency are:
{\small
    \begin{flalign*}
      & \s{\textbf{data}\ GEq (c : Code_{\ell}) (x\ y : El_{\ell}\ c) : \sType{}\ \textbf{where} \quad \mid \quad}
      \s{GRefl : (z : El_{\ell}\ c) -> GEq\ c\ x\ y }\\
&\s{gSubst : (c : Code_{\ell}) -> (P : El_{\ell}\ c -> Code_{\ell'}) -> (x\ y : El_{\ell}\ c) -> El_{\ell'}\ (P\ x) -> GEq\ c\ x\ y -> El_{\ell'}\ (P\ y) }
    \end{flalign*}
    }
    We allow any term to act as evidence in $\s{GRefl}$:
    \rrule{CRefl} requires that evidence be more precise than the equated terms, but this restriction is for monotonicity, not safety.
The function $\s{gSubst}$ casts through the evidence in the given $\s{GEq}$ argument
like the \rrule{J} reduction rule.

\subsection{\lang: Realizing Inductive Families with Propositional Consistency}
\label{sec:inductives}

We now have everything we need to extend \surfelang with inductive types to create \lang.
Having added propositional equality, most of the work is done, since
a \textit{family} of types can be turned into a single parameterized
inductive type whose constructors use equality proofs to constrain
the possible values of indices
Likewise, by expressing \celang casts in terms of type tags, our formalism
easily accommodates a fixed set of user-defined types.

Inductive types in CIC are notoriously expansive in their formal notation, so
we focus on giving the intuition behind the formalization.
We refer the reader to \citep{paulinmohring:hal-01094195} or \citep{Dybjer1994} for a standard reference on the formalism behind
static inductive types.
{\small
\begin{displaymath}
 \g{s} \bnfadd \g{C}\ \seq{\gs} \bnfalt \g{D}\ \seq{\gs} \bnfalt \g{ind_{C}\ \gx\ \gs \ldotp}\ \seq{\g{D}\ \seq{\gy} \mapsto \gs} \qquad
 \g{t} \bnfadd \g{C}\ \seq{\gt} \bnfalt \g{D}\ \seq{\gt} \bnfalt \g{ind_{C}\ \gx\ \gt \ldotp}\ \seq{\g{D}\ \seq{\gy} \mapsto \gt}
\end{displaymath}
}
A \lang inductive type is declared by specifying a type constructor $\g{C}$
with (possibly dependent) typing
$\g{C} : \seq{\g{(x : \gS)}} {-> \gType{\ell}}$,
along with some fixed number of data constructors $\g{D}_{1}\ldots \g{D}_{n}$
typed as $\g{D}_{i} : \seq{\g{(y : \gS)}} \g{-> C}\ \seq{\gs}$.
In \cilang, the syntax is the same, but the return of each constructor is $\g{C}\ \seq{\gx}$.
We follow \citet{bertrand:gcic} use an
induction form $\g{ind}$ that pattern-matches on the given value,
but binds a name for self-reference in each branch.
Termination is guaranteed with an external structural recursion check,
that is standard and orthogonal to our work.
Typing $\g{ind}$ as a checking form (omitted) helps reduce the number of type ascriptions
required, compared to the usual CIC presentation.
Likewise, precision and consistency differ only in the addition of new structural rules.

The typing rules for gradual inductives (omitted) are identical to their
static counterparts: the difficulty
is handled by using propositional equality for index constraints.

\myparagraph{Elaboration}

Inductive families are transformed into inductive types with
equality constraints during elaboration via the standard translation.
Any time a type constructor gives a specific expression $\gs$ for an index,
either of a recursive argument in the return type, the expression is replaced with a fresh parameter $\gx$,
and an argument of type $\g{x \pc \gt}$ is added to the constructor. Likewise, the eliminator
for an inductive family elaborates to the eliminator for the corresponding constrained inductive type,
but rewrites the goal type using $\g{subst}$.

The last thing we consider is the elaboration of constructors. We could elaborate $\lang$
constructors directly into $\cilang$ constructors, adding the initial evidence to $\g{refl}$.
However, this is too lazy to obtain the dynamic-checking behavior discussed in our introduction.
In the example we gave above, we could $\g{Nil}$ with evidence $\grefl{0}{0}{\qm}$,
and $\g{Cons\ 3.3\ Nil}$ with evidence $\grefl{1 + \qm}{1 + \qm}{1 + \qm}$.
This is well typed according to the definition of $\g{Vec'}$ using $\pc$,
but it does not eagerly find the error, because we have a length-2 of propositional consistencies
that we have not composed.

We solve this issue by inserting an extra composition into the elaboration of data constructors.
Every elaborated constructor carries evidence of consistency
between the actual index and the return index of the constructor.
Each constructor's elaboration
accesses this evidence for each inductive argument, use $\g{subst}$ to lift it to a propositional consistency
on the return index, and compose it with the index specified by the constructor.
For our $\g{Vec}$ example, $\g{Nil}$ has evidence $\grefl{0}{0}{\qm}$.
 $\g{Cons}$ is used at type $\g{Float -> Vec\ Float\ \qm -> Vec\ Float\ (1 + \qm)}$,
so we elaborate with evidence $\grefl{1+\qm}{1+\qm}{1+\qm}$.
Evidence from $\g{Nil}$ is lifted
by $\g{subst}$ to $\grefl{1}{1 + \qm}{1 + \qm}$, and composed with the initial evidence
to produce $\grefl{1}{1 + \qm}{1 + \qm}$.
Then composition fails when we cast to $\g{Vec\ Float\ 2}$.
In the case where the inductive argument is not
a member of the inductive type but a function producing a member of the inductive type,
we can supply $\qm$ for all arguments bound function image's evidence.

\myparagraph{Semantics}

The interesting part of adding inductive types is handling casts between these types,
but again, the work is done for us by propositional equality. We define the singular cast rule
between inductive types:
$
  \castnog{\g{C}\ \seq{\g{t_{1}}}}{\g{C}\ \seq{\g{t_{2}}}} \g{(D}\ \seq{\g{t)}}
  \redsto_{\rrule{Ctor}}
  \g{D}\ \seq{ \castnog{\seq{[\g{t_{1}} / \gx]}\gT}{\seq{[\g{t_{2}}/\gx]}\gT} \g{t} }
$
  where\ $\g{D} : \seq{\g{(\gy : \gT)}} \g{-> C\ } \seq{\gx}$.
That is, we cast between different parameterizations of an inductive type
by casting each field of a constructor to a type with the new parameters replacing
the old ones. Index conversions are handled by casts on $\pc$ types.

All other casts are handled by the existing rules, which are
generic in the type tags involved. Likewise, the (omitted) rules for reducing $\g{ind_{C}}$ are nearly indentical
to the static form, since elaboration inserts all the necessary casts.
Applying $\g{ind}$ to $\qm$ or $\err$ produces $\qm$ or $\err$ respectively.

Like with $\g{subst}$, we add a check that the term under elimination
is self-consistent, which ensures that if the evidence of index consistency
ever becomes $\err$ for a member of an inductive type, that value behaves as $\err$.

Finally, we add an $\eta$ rule that
${\Gamma |- \g{D}\ \seq{\gt} \stepsto_{\eta} \err_{\g{C}\ \seq{\g{t'}_{i}}}^{i} : \g{C}\ \seq{\g{t'}}}$
whenever
  ${\g{t}_{i} \stepsto_{\eta} \err_{\gT}}$.
So casts between types with incompatible indices
produce $\err$ immediately, unlike GCIC~\citep{bertrand:gcic} where
inconsistencies are only discovered during elimination.

The above semantic features are easily incorporated in our syntactic model,
for any fixed set of inductive types. We add
constructors to $\g{Code}$ and $\g{Unk}$ and add
cases to $\s{cast}$ and $\s{meet}$.
For each inductive type $\g{C}$,
we define a static type $\s{GC}$, with $\g{C}$'s constructors plus $\s{D_{CUnk}}$
and $\s{D_{CErr}}$ denoting $\qm$ and $\err$ respectively.
Since the structure of $\s{GC}$ and $\g{C}$ are the same, $\g{ind_{C}}$ is easily represented
by the elimination of inductive datatypes in \tlang.

\section{Discussion}
\label{sec:discussion}

\myparagraph{Related Work}
\lang builds on long line of work mixing dynamic and static enforcement of specifications,
most prominently GDTL~\citep{Eremondi:2019:ANG:3352468.3341692} and GCIC~\citep{bertrand:gcic}.
\Citet{dynamicDependent} support mixed static and dynamic checking of boolean-valued properties.
Similarly,
\citet{Tanter:2015:GCP:2816707.2816710} develop a system of casts for Coq, using an unsound axiom
to represent type errors. The casts supported subset types \ie a value paired with a proof
that some boolean-valued function returns true for that value, but not general inductive types.

\citet{Osera:2012:DI:2103776.2103779} present \textit{dependent interoperability}
for principled mixing of dependently typed and non-dependently typed programs.
This was extended by \citet{partialTypeEquiv,dagandtabareautanter2018},
who provide a general mechanism for lifting higher-order programs to the dependently-typed setting.
All of these approaches presuppose separate simple and dependent versions of types,
related by boolean-valued predicates. Our composition of evidence provides similar checks,
but by keeping evidence of propositional consistency,
we do not need types to be reformulated in terms of subset types or boolean predicates.

Our technique of representing inductive types by codes dates back to at least \citep{mcbrideOrnamental},
and has been used to produce very minimal dependently-typed core languages~\citep{Chapman:2010:GAL:1863543.1863547}, to model
constructors and eliminators generically~\citep{Diehl:2014:GCE:2633628.2633630}, and to model a dependent type-case~\citep{10.1145/3018610.3018620}. The representation of $\qm$ and $\err$ we use in inductives types
is essentially that of Exceptional Type Theory~\citep{exceptionalTT,Pedrot:2019:RET:3352468.3341712}, where extra constructors are added to account for the additional possibilities for each type.

Our notion of algebraic precision bears some resemblance to the embedding-projection pairs
that \citet{10.1145/3236768} use to provide a semantic account of gradual typing.
However, the embedding-projection property is too restrictive for approximate normalization,
so we are unable to show the same properties that they show in a non-dependently typed setting.

\myparagraph{Limitations and Future Work}
\lang shows that when programming with gradual types and inductive families, the programmer
can have confidence that their compiler will eventually terminate, and that
type errors always reflect true type inconsistencies. Programmers can omit proofs,
yet test their programs, and receive feedback on which equalities did or did not hold during execution,
specifying those equalities as dependent types.

Our work leaves ample room for future developments.
While \lang supports inductive types, its translation supports an arbitrary but fixed set
of inductive types. This is suitable for proofs, but means that the translation can only work
on whole programs. Also, the translation currently does not support exact evaluation:
while we can adapt the translation to target a non-total language and never approximate,
there would be well-typed terms that were not well-typed in the target, since their types
have no exact normal-form.

Our algebraic precision relation compares values of the same type, making it stricter
than that of \citet{bertrand:gcic} or \citet{refinedCriteria}, so there is room to explore
whether algebraic precision can be used to reason in more traditional gradual languages.
Also, we suspect that the reasoning principles provided by algebraic precision could be used
to drive optimizations of gradual types, such as removing unnecessary casts.
Both Abstracting Gradual Typing~\citep{agt} and our work rely on Galois connections as a key property,
which warrants investigation of a deeper connection.
Likewise, while we cannot show a full embedding-projection pair for approximate normalization,
there may be a stronger property about exact semantics that we can show,
which could capture the fact that reducing then increasing precision is never a lossy operation
for exact execution.

\bibliographystyle{ACM-Reference-Format}
\bibliography{myRefs}


\begin{thebibliography}{28}


\ifx \showCODEN    \undefined \def \showCODEN     #1{\unskip}     \fi
\ifx \showDOI      \undefined \def \showDOI       #1{#1}\fi
\ifx \showISBNx    \undefined \def \showISBNx     #1{\unskip}     \fi
\ifx \showISBNxiii \undefined \def \showISBNxiii  #1{\unskip}     \fi
\ifx \showISSN     \undefined \def \showISSN      #1{\unskip}     \fi
\ifx \showLCCN     \undefined \def \showLCCN      #1{\unskip}     \fi
\ifx \shownote     \undefined \def \shownote      #1{#1}          \fi
\ifx \showarticletitle \undefined \def \showarticletitle #1{#1}   \fi
\ifx \showURL      \undefined \def \showURL       {\relax}        \fi
\providecommand\bibfield[2]{#2}
\providecommand\bibinfo[2]{#2}
\providecommand\natexlab[1]{#1}
\providecommand\showeprint[2][]{arXiv:#2}

\bibitem[\protect\citeauthoryear{Anderson and Drossopoulou}{Anderson and
  Drossopoulou}{2003}]%
        {ANDERSON200353}
\bibfield{author}{\bibinfo{person}{Christopher Anderson} {and}
  \bibinfo{person}{Sophia Drossopoulou}.} \bibinfo{year}{2003}\natexlab{}.
\newblock \showarticletitle{BabyJ: From Object Based to Class Based Programming
  via Types}.
\newblock \bibinfo{journal}{\emph{Electronic Notes in Theoretical Computer
  Science}} \bibinfo{volume}{82}, \bibinfo{number}{8} (\bibinfo{year}{2003}),
  \bibinfo{pages}{53--81}.
\newblock
\showISSN{1571-0661}
\urldef\tempurl%
\url{https://doi.org/10.1016/S1571-0661(04)80802-8}
\showDOI{\tempurl}
\newblock
\shownote{WOOD2003, Workshop on Object Oriented Developments (Satellite Event
  of ETAPS 2003).}


\bibitem[\protect\citeauthoryear{Boulier, P\'{e}drot, and Tabareau}{Boulier
  et~al\mbox{.}}{2017}]%
        {10.1145/3018610.3018620}
\bibfield{author}{\bibinfo{person}{Simon Boulier},
  \bibinfo{person}{Pierre-Marie P\'{e}drot}, {and} \bibinfo{person}{Nicolas
  Tabareau}.} \bibinfo{year}{2017}\natexlab{}.
\newblock \showarticletitle{The next 700 Syntactical Models of Type Theory}. In
  \bibinfo{booktitle}{\emph{Proceedings of the 6th ACM SIGPLAN Conference on
  Certified Programs and Proofs}} (Paris, France) \emph{(\bibinfo{series}{CPP
  2017})}. \bibinfo{publisher}{Association for Computing Machinery},
  \bibinfo{address}{New York, NY, USA}, \bibinfo{pages}{182–194}.
\newblock
\showISBNx{9781450347051}
\urldef\tempurl%
\url{https://doi.org/10.1145/3018610.3018620}
\showDOI{\tempurl}


\bibitem[\protect\citeauthoryear{Brady}{Brady}{2013}]%
        {idrisPaper}
\bibfield{author}{\bibinfo{person}{Edwin Brady}.}
  \bibinfo{year}{2013}\natexlab{}.
\newblock \showarticletitle{{Idris}, a general-purpose dependently typed
  programming language: Design and implementation}.
\newblock \bibinfo{journal}{\emph{Journal of Functional Programming}}
  \bibinfo{volume}{23}, \bibinfo{number}{5} (\bibinfo{year}{2013}),
  \bibinfo{pages}{552–593}.
\newblock
\urldef\tempurl%
\url{https://doi.org/10.1017/S095679681300018X}
\showDOI{\tempurl}


\bibitem[\protect\citeauthoryear{Chapman, Dagand, McBride, and Morris}{Chapman
  et~al\mbox{.}}{2010}]%
        {Chapman:2010:GAL:1863543.1863547}
\bibfield{author}{\bibinfo{person}{James Chapman},
  \bibinfo{person}{Pierre-\'{E}variste Dagand}, \bibinfo{person}{Conor
  McBride}, {and} \bibinfo{person}{Peter Morris}.}
  \bibinfo{year}{2010}\natexlab{}.
\newblock \showarticletitle{The Gentle Art of Levitation}. In
  \bibinfo{booktitle}{\emph{Proceedings of the 15th ACM SIGPLAN International
  Conference on Functional Programming}} (Baltimore, Maryland, USA)
  \emph{(\bibinfo{series}{ICFP '10})}. \bibinfo{publisher}{ACM},
  \bibinfo{address}{New York, NY, USA}, \bibinfo{pages}{3--14}.
\newblock
\showISBNx{978-1-60558-794-3}
\urldef\tempurl%
\url{https://doi.org/10.1145/1863543.1863547}
\showDOI{\tempurl}


\bibitem[\protect\citeauthoryear{Coquand}{Coquand}{1986}]%
        {Coquand86}
\bibfield{author}{\bibinfo{person}{Thierry Coquand}.}
  \bibinfo{year}{1986}\natexlab{}.
\newblock \showarticletitle{An Analysis of Girard's Paradox}. In
  \bibinfo{booktitle}{\emph{LICS}}. \bibinfo{publisher}{IEEE Computer Society},
  \bibinfo{pages}{227--236}.
\newblock
\showISBNx{0-8186-0720-3}
\urldef\tempurl%
\url{http://dblp.uni-trier.de/db/conf/lics/lics86.html\#Coquand86}
\showURL{%
\tempurl}


\bibitem[\protect\citeauthoryear{Dagand, Tabareau, and {\'Eric Tanter}}{Dagand
  et~al\mbox{.}}{2016}]%
        {partialTypeEquiv}
\bibfield{author}{\bibinfo{person}{Pierre-{\'E}variste Dagand},
  \bibinfo{person}{Nicolas Tabareau}, {and} \bibinfo{person}{{\'Eric Tanter}}.}
  \bibinfo{year}{2016}\natexlab{}.
\newblock \showarticletitle{Partial Type Equivalences for Verified Dependent
  Interoperability}. In \bibinfo{booktitle}{\emph{Proceedings of the 21st ACM
  SIGPLAN Conference on Functional Programming (ICFP 2016)}}.
  \bibinfo{publisher}{ACM Press}, \bibinfo{address}{Nara, Japan},
  \bibinfo{pages}{298--310}.
\newblock


\bibitem[\protect\citeauthoryear{Dagand, Tabareau, and Tanter}{Dagand
  et~al\mbox{.}}{2018}]%
        {dagandtabareautanter2018}
\bibfield{author}{\bibinfo{person}{Pierre-{\'E}variste Dagand},
  \bibinfo{person}{Nicolas Tabareau}, {and} \bibinfo{person}{{\'E}ric Tanter}.}
  \bibinfo{year}{2018}\natexlab{}.
\newblock \showarticletitle{Foundations of dependent interoperability}.
\newblock \bibinfo{journal}{\emph{Journal of Functional Programming}}
  \bibinfo{volume}{28} (\bibinfo{year}{2018}), \bibinfo{pages}{e9}.
\newblock
\urldef\tempurl%
\url{https://doi.org/10.1017/S0956796818000011}
\showDOI{\tempurl}


\bibitem[\protect\citeauthoryear{Diehl and Sheard}{Diehl and Sheard}{2014}]%
        {Diehl:2014:GCE:2633628.2633630}
\bibfield{author}{\bibinfo{person}{Larry Diehl} {and} \bibinfo{person}{Tim
  Sheard}.} \bibinfo{year}{2014}\natexlab{}.
\newblock \showarticletitle{Generic Constructors and Eliminators from
  Descriptions: Type Theory As a Dependently Typed Internal DSL}. In
  \bibinfo{booktitle}{\emph{Proceedings of the 10th ACM SIGPLAN Workshop on
  Generic Programming}} (Gothenburg, Sweden) \emph{(\bibinfo{series}{WGP
  '14})}. \bibinfo{publisher}{ACM}, \bibinfo{address}{New York, NY, USA},
  \bibinfo{pages}{3--14}.
\newblock
\showISBNx{978-1-4503-3042-8}
\urldef\tempurl%
\url{https://doi.org/10.1145/2633628.2633630}
\showDOI{\tempurl}


\bibitem[\protect\citeauthoryear{Dybjer}{Dybjer}{1994}]%
        {Dybjer1994}
\bibfield{author}{\bibinfo{person}{Peter Dybjer}.}
  \bibinfo{year}{1994}\natexlab{}.
\newblock \showarticletitle{Inductive families}.
\newblock \bibinfo{journal}{\emph{Formal Aspects of Computing}}
  \bibinfo{volume}{6}, \bibinfo{number}{4} (\bibinfo{date}{01 Jul}
  \bibinfo{year}{1994}), \bibinfo{pages}{440--465}.
\newblock
\showISSN{1433-299X}
\urldef\tempurl%
\url{https://doi.org/10.1007/BF01211308}
\showDOI{\tempurl}


\bibitem[\protect\citeauthoryear{Dybjer and Setzer}{Dybjer and Setzer}{2003}]%
        {DYBJER20031}
\bibfield{author}{\bibinfo{person}{Peter Dybjer} {and} \bibinfo{person}{Anton
  Setzer}.} \bibinfo{year}{2003}\natexlab{}.
\newblock \showarticletitle{Induction–recursion and initial algebras}.
\newblock \bibinfo{journal}{\emph{Annals of Pure and Applied Logic}}
  \bibinfo{volume}{124}, \bibinfo{number}{1} (\bibinfo{year}{2003}),
  \bibinfo{pages}{1 -- 47}.
\newblock
\showISSN{0168-0072}
\urldef\tempurl%
\url{https://doi.org/10.1016/S0168-0072(02)00096-9}
\showDOI{\tempurl}


\bibitem[\protect\citeauthoryear{Eremondi, Tanter, and Garcia}{Eremondi
  et~al\mbox{.}}{2019}]%
        {Eremondi:2019:ANG:3352468.3341692}
\bibfield{author}{\bibinfo{person}{Joseph Eremondi}, \bibinfo{person}{\'{E}ric
  Tanter}, {and} \bibinfo{person}{Ronald Garcia}.}
  \bibinfo{year}{2019}\natexlab{}.
\newblock \showarticletitle{Approximate Normalization for Gradual Dependent
  Types}.
\newblock \bibinfo{journal}{\emph{Proc. ACM Program. Lang.}}
  \bibinfo{volume}{3}, \bibinfo{number}{ICFP}, Article \bibinfo{articleno}{88}
  (\bibinfo{date}{July} \bibinfo{year}{2019}), \bibinfo{numpages}{30}~pages.
\newblock
\showISSN{2475-1421}
\urldef\tempurl%
\url{https://doi.org/10.1145/3341692}
\showDOI{\tempurl}


\bibitem[\protect\citeauthoryear{Garcia, Clark, and Tanter}{Garcia
  et~al\mbox{.}}{2016}]%
        {agt}
\bibfield{author}{\bibinfo{person}{Ronald Garcia}, \bibinfo{person}{Alison~M.
  Clark}, {and} \bibinfo{person}{\'{E}ric Tanter}.}
  \bibinfo{year}{2016}\natexlab{}.
\newblock \showarticletitle{Abstracting Gradual Typing}. In
  \bibinfo{booktitle}{\emph{Proceedings of the 43rd Annual ACM SIGPLAN-SIGACT
  Symposium on Principles of Programming Languages}} (St. Petersburg, FL, USA)
  \emph{(\bibinfo{series}{POPL '16})}. \bibinfo{publisher}{ACM},
  \bibinfo{address}{New York, NY, USA}, \bibinfo{pages}{429--442}.
\newblock
\showISBNx{978-1-4503-3549-2}
\urldef\tempurl%
\url{https://doi.org/10.1145/2837614.2837670}
\showDOI{\tempurl}


\bibitem[\protect\citeauthoryear{Lennon-Bertrand, Maillard, Tabareau, and Éric
  Tanter}{Lennon-Bertrand et~al\mbox{.}}{2020}]%
        {bertrand:gcic}
\bibfield{author}{\bibinfo{person}{Meven Lennon-Bertrand},
  \bibinfo{person}{Kenji Maillard}, \bibinfo{person}{Nicolas Tabareau}, {and}
  \bibinfo{person}{Éric Tanter}.} \bibinfo{year}{2020}\natexlab{}.
\newblock \bibinfo{title}{Gradualizing the Calculus of Inductive
  Constructions}.
\newblock
\newblock
\showeprint[arxiv]{2011.10618}~[cs.PL]


\bibitem[\protect\citeauthoryear{McBride}{McBride}{2000}]%
        {dtfpp}
\bibfield{author}{\bibinfo{person}{Conor McBride}.}
  \bibinfo{year}{2000}\natexlab{}.
\newblock \emph{\bibinfo{title}{Dependently typed functional programs and their
  proofs}}.
\newblock \bibinfo{thesistype}{Ph.D. Dissertation}. \bibinfo{school}{University
  of Edinburgh, {UK}}.
\newblock
\urldef\tempurl%
\url{http://hdl.handle.net/1842/374}
\showURL{%
\tempurl}


\bibitem[\protect\citeauthoryear{McBride}{McBride}{2011}]%
        {mcbrideOrnamental}
\bibfield{author}{\bibinfo{person}{Conor McBride}.}
  \bibinfo{year}{2011}\natexlab{}.
\newblock \bibinfo{title}{Ornamental algebras, algebraic ornaments}.
\newblock
\newblock
\urldef\tempurl%
\url{http://plv.mpi-sws.org/plerg/papers/mcbride-ornaments-2up.pdf}
\showURL{%
\tempurl}


\bibitem[\protect\citeauthoryear{New and Ahmed}{New and Ahmed}{2018}]%
        {10.1145/3236768}
\bibfield{author}{\bibinfo{person}{Max~S. New} {and} \bibinfo{person}{Amal
  Ahmed}.} \bibinfo{year}{2018}\natexlab{}.
\newblock \showarticletitle{Graduality from Embedding-Projection Pairs}.
\newblock \bibinfo{journal}{\emph{Proc. ACM Program. Lang.}}
  \bibinfo{volume}{2}, \bibinfo{number}{ICFP}, Article \bibinfo{articleno}{73}
  (\bibinfo{date}{July} \bibinfo{year}{2018}), \bibinfo{numpages}{30}~pages.
\newblock
\urldef\tempurl%
\url{https://doi.org/10.1145/3236768}
\showDOI{\tempurl}


\bibitem[\protect\citeauthoryear{Norell}{Norell}{2009}]%
        {agdaPaper}
\bibfield{author}{\bibinfo{person}{Ulf Norell}.}
  \bibinfo{year}{2009}\natexlab{}.
\newblock \showarticletitle{Dependently Typed Programming in {Agda}}. In
  \bibinfo{booktitle}{\emph{Proceedings of the 4th International Workshop on
  Types in Language Design and Implementation}} (Savannah, GA, USA)
  \emph{(\bibinfo{series}{TLDI '09})}. \bibinfo{publisher}{ACM},
  \bibinfo{address}{New York, NY, USA}, \bibinfo{pages}{1--2}.
\newblock
\showISBNx{978-1-60558-420-1}
\urldef\tempurl%
\url{https://doi.org/10.1145/1481861.1481862}
\showDOI{\tempurl}


\bibitem[\protect\citeauthoryear{Osera, Sj\"{o}berg, and Zdancewic}{Osera
  et~al\mbox{.}}{2012}]%
        {Osera:2012:DI:2103776.2103779}
\bibfield{author}{\bibinfo{person}{Peter-Michael Osera},
  \bibinfo{person}{Vilhelm Sj\"{o}berg}, {and} \bibinfo{person}{Steve
  Zdancewic}.} \bibinfo{year}{2012}\natexlab{}.
\newblock \showarticletitle{Dependent Interoperability}. In
  \bibinfo{booktitle}{\emph{Proceedings of the Sixth Workshop on Programming
  Languages Meets Program Verification}} (Philadelphia, Pennsylvania, USA)
  \emph{(\bibinfo{series}{PLPV '12})}. \bibinfo{publisher}{ACM},
  \bibinfo{address}{New York, NY, USA}, \bibinfo{pages}{3--14}.
\newblock
\showISBNx{978-1-4503-1125-0}
\urldef\tempurl%
\url{https://doi.org/10.1145/2103776.2103779}
\showDOI{\tempurl}


\bibitem[\protect\citeauthoryear{Ou, Tan, Mandelbaum, and Walker}{Ou
  et~al\mbox{.}}{2004}]%
        {dynamicDependent}
\bibfield{author}{\bibinfo{person}{Xinming Ou}, \bibinfo{person}{Gang Tan},
  \bibinfo{person}{Yitzhak Mandelbaum}, {and} \bibinfo{person}{David Walker}.}
  \bibinfo{year}{2004}\natexlab{}.
\newblock \showarticletitle{Dynamic Typing with Dependent Types}. In
  \bibinfo{booktitle}{\emph{Exploring New Frontiers of Theoretical
  Informatics}}, \bibfield{editor}{\bibinfo{person}{Jean-Jacques Levy},
  \bibinfo{person}{Ernst~W. Mayr}, {and} \bibinfo{person}{John~C. Mitchell}}
  (Eds.). \bibinfo{publisher}{Springer US}, \bibinfo{address}{Boston, MA},
  \bibinfo{pages}{437--450}.
\newblock
\showISBNx{978-1-4020-8141-5}


\bibitem[\protect\citeauthoryear{Paulin-Mohring}{Paulin-Mohring}{2015}]%
        {paulinmohring:hal-01094195}
\bibfield{author}{\bibinfo{person}{Christine Paulin-Mohring}.}
  \bibinfo{year}{2015}\natexlab{}.
\newblock \showarticletitle{{Introduction to the Calculus of Inductive
  Constructions}}.
\newblock In \bibinfo{booktitle}{\emph{{All about Proofs, Proofs for All}}},
  \bibfield{editor}{\bibinfo{person}{Bruno~Woltzenlogel Paleo} {and}
  \bibinfo{person}{David Delahaye}} (Eds.). \bibinfo{series}{Studies in Logic
  (Mathematical logic and foundations)}, Vol.~\bibinfo{volume}{55}.
  \bibinfo{publisher}{{College Publications}}.
\newblock
\urldef\tempurl%
\url{https://hal.inria.fr/hal-01094195}
\showURL{%
\tempurl}


\bibitem[\protect\citeauthoryear{P{\'e}drot and Tabareau}{P{\'e}drot and
  Tabareau}{2018}]%
        {exceptionalTT}
\bibfield{author}{\bibinfo{person}{Pierre-Marie P{\'e}drot} {and}
  \bibinfo{person}{Nicolas Tabareau}.} \bibinfo{year}{2018}\natexlab{}.
\newblock \showarticletitle{Failure is Not an Option}. In
  \bibinfo{booktitle}{\emph{Programming Languages and Systems}},
  \bibfield{editor}{\bibinfo{person}{Amal Ahmed}} (Ed.).
  \bibinfo{publisher}{Springer International Publishing},
  \bibinfo{address}{Cham}, \bibinfo{pages}{245--271}.
\newblock
\showISBNx{978-3-319-89884-1}


\bibitem[\protect\citeauthoryear{P{\'e}drot, Tabareau, Fehrmann, and
  Tanter}{P{\'e}drot et~al\mbox{.}}{2019}]%
        {Pedrot:2019:RET:3352468.3341712}
\bibfield{author}{\bibinfo{person}{Pierre-Marie P{\'e}drot},
  \bibinfo{person}{Nicolas Tabareau}, \bibinfo{person}{Hans~Jacob Fehrmann},
  {and} \bibinfo{person}{\'{E}ric Tanter}.} \bibinfo{year}{2019}\natexlab{}.
\newblock \showarticletitle{A Reasonably Exceptional Type Theory}.
\newblock \bibinfo{journal}{\emph{Proc. ACM Program. Lang.}}
  \bibinfo{volume}{3}, \bibinfo{number}{ICFP}, Article \bibinfo{articleno}{108}
  (\bibinfo{date}{July} \bibinfo{year}{2019}), \bibinfo{numpages}{29}~pages.
\newblock
\showISSN{2475-1421}
\urldef\tempurl%
\url{https://doi.org/10.1145/3341712}
\showDOI{\tempurl}


\bibitem[\protect\citeauthoryear{Pierce and Turner}{Pierce and Turner}{2000}]%
        {localTypeInference}
\bibfield{author}{\bibinfo{person}{Benjamin~C. Pierce} {and}
  \bibinfo{person}{David~N. Turner}.} \bibinfo{year}{2000}\natexlab{}.
\newblock \showarticletitle{Local Type Inference}.
\newblock \bibinfo{journal}{\emph{ACM Trans. Program. Lang. Syst.}}
  \bibinfo{volume}{22}, \bibinfo{number}{1} (\bibinfo{date}{Jan.}
  \bibinfo{year}{2000}), \bibinfo{pages}{1--44}.
\newblock
\showISSN{0164-0925}
\urldef\tempurl%
\url{https://doi.org/10.1145/345099.345100}
\showDOI{\tempurl}


\bibitem[\protect\citeauthoryear{{Ringer}, {Palmskog}, {Sergey}, {Gligoric},
  and {Tatlock}}{{Ringer} et~al\mbox{.}}{2019}]%
        {qedAtLarge}
\bibfield{author}{\bibinfo{person}{T. {Ringer}}, \bibinfo{person}{K.
  {Palmskog}}, \bibinfo{person}{I. {Sergey}}, \bibinfo{person}{M. {Gligoric}},
  {and} \bibinfo{person}{Z. {Tatlock}}.} \bibinfo{year}{2019}\natexlab{}.
\newblock \bibinfo{booktitle}{\emph{QED at Large: A Survey of Engineering of
  Formally Verified Software}}.
\newblock \bibinfo{publisher}{now}.
\newblock
\showISBNx{null}
\showISSN{null}
\urldef\tempurl%
\url{https://ieeexplore.ieee.org/document/8824174}
\showURL{%
\tempurl}


\bibitem[\protect\citeauthoryear{Siek and Taha}{Siek and Taha}{2006}]%
        {gradualTypeInitial}
\bibfield{author}{\bibinfo{person}{Jeremy~G. Siek} {and} \bibinfo{person}{Walid
  Taha}.} \bibinfo{year}{2006}\natexlab{}.
\newblock \showarticletitle{Gradual Typing for Functional Languages}. In
  \bibinfo{booktitle}{\emph{Scheme and Functional Programming Workshop}}.
  \bibinfo{pages}{81--92}.
\newblock


\bibitem[\protect\citeauthoryear{Siek, Vitousek, Cimini, and Boyland}{Siek
  et~al\mbox{.}}{2015}]%
        {refinedCriteria}
\bibfield{author}{\bibinfo{person}{Jeremy~G. Siek}, \bibinfo{person}{Michael~M.
  Vitousek}, \bibinfo{person}{Matteo Cimini}, {and} \bibinfo{person}{John~Tang
  Boyland}.} \bibinfo{year}{2015}\natexlab{}.
\newblock \showarticletitle{{Refined Criteria for Gradual Typing}}. In
  \bibinfo{booktitle}{\emph{1st Summit on Advances in Programming Languages
  (SNAPL 2015)}} \emph{(\bibinfo{series}{Leibniz International Proceedings in
  Informatics (LIPIcs)}, Vol.~\bibinfo{volume}{32})},
  \bibfield{editor}{\bibinfo{person}{Thomas Ball}, \bibinfo{person}{Rastislav
  Bodik}, \bibinfo{person}{Shriram Krishnamurthi}, \bibinfo{person}{Benjamin~S.
  Lerner}, {and} \bibinfo{person}{Greg Morrisett}} (Eds.).
  \bibinfo{publisher}{Schloss Dagstuhl--Leibniz-Zentrum fuer Informatik},
  \bibinfo{address}{Dagstuhl, Germany}, \bibinfo{pages}{274--293}.
\newblock
\showISBNx{978-3-939897-80-4}
\showISSN{1868-8969}
\urldef\tempurl%
\url{https://doi.org/10.4230/LIPIcs.SNAPL.2015.274}
\showDOI{\tempurl}


\bibitem[\protect\citeauthoryear{Siek and Wadler}{Siek and Wadler}{2010}]%
        {Siek:2009:TWB:1570506.1570511}
\bibfield{author}{\bibinfo{person}{Jeremy~G. Siek} {and}
  \bibinfo{person}{Philip Wadler}.} \bibinfo{year}{2010}\natexlab{}.
\newblock \showarticletitle{Threesomes, with and Without Blame}. In
  \bibinfo{booktitle}{\emph{Proceedings of the 37th Annual ACM SIGPLAN-SIGACT
  Symposium on Principles of Programming Languages}} (Madrid, Spain)
  \emph{(\bibinfo{series}{POPL '10})}. \bibinfo{publisher}{ACM},
  \bibinfo{address}{New York, NY, USA}, \bibinfo{pages}{365--376}.
\newblock
\showISBNx{978-1-60558-479-9}
\urldef\tempurl%
\url{https://doi.org/10.1145/1706299.1706342}
\showDOI{\tempurl}


\bibitem[\protect\citeauthoryear{Tanter and Tabareau}{Tanter and
  Tabareau}{2015}]%
        {Tanter:2015:GCP:2816707.2816710}
\bibfield{author}{\bibinfo{person}{\'{E}ric Tanter} {and}
  \bibinfo{person}{Nicolas Tabareau}.} \bibinfo{year}{2015}\natexlab{}.
\newblock \showarticletitle{Gradual Certified Programming in Coq}. In
  \bibinfo{booktitle}{\emph{Proceedings of the 11th Symposium on Dynamic
  Languages}} (Pittsburgh, PA, USA) \emph{(\bibinfo{series}{DLS 2015})}.
  \bibinfo{publisher}{ACM}, \bibinfo{address}{New York, NY, USA},
  \bibinfo{pages}{26--40}.
\newblock
\showISBNx{978-1-4503-3690-1}
\urldef\tempurl%
\url{https://doi.org/10.1145/2816707.2816710}
\showDOI{\tempurl}


\end{thebibliography}

\end{document}